\documentclass[12pt,preprint]{aastex}
\shorttitle{Line-Strengths in Central Coma}
\shortauthors{Matkovi\'c et al.}

\begin{document}


\title{Kinematic Properties and Stellar Populations of Faint
  Early-Type Galaxies.  II.  Line-Strength Measurements of
  Central Coma Galaxies}


%\author{A. Matkovi\'c \altaffilmark{1}}
\author{A. Matkovi\'c }
\affil{Astronomy Department, University of Florida, P.O. Box 112055,
Gainesville, FL 32611, USA}

\author{R. Guzm\'an}
\affil{Astronomy Department, University of Florida, P.O. Box 112055,
Gainesville, FL 32611, USA}

\author{P. S\'anchez-Bl\'azquez}
\affil{Centre for Astrophysics, University of Central Lancashire,
PR1 2HE, Preston, UK}

\author{J. Gorgas and N. Cardiel}
\affil{Departamento de Astrof\'isica, Facultad de F\'isicas,
Universidad Complutense de Madrid, Ciudad Universitaria, 28040
Madrid, Spain}

\author{N. Gruel}
\affil{Astronomy Department, University of Florida, P.O. Box 112055,
Gainesville, FL 32611, USA}


\begin{abstract}

  We present line-strength measurements for 74 early-type galaxies in
  the core of the Coma cluster reaching down to velocity dispersions,
  $\sigma$, of 30 km s$^{-1}$. The index-$\sigma$ relations for our
  sample, including galaxies with $\sigma<100$ km s$^{-1}$
  (low-$\sigma$), differ in shape depending on which index is used.
  We notice two types of relations for the metallic indices: one
  showing a break in the slope around $~\sim 100$ km s$^{-1}$, and
  another group with strong linear relations between an index and
  $\log \sigma$.  We find no connection between the behavior of
  index-$\sigma$ relations with either $\alpha$- or Fe-peak elements.
  However, we find indications that the relations are tighter for
  indices which do not depend on the micro-turbulent velocities of
  stellar atmospheres.  We confirm previous results that low-$\sigma$
  galaxies including dE/dS0s are on average younger, less metal rich,
  and have lower [$\alpha$/Fe] in comparison to E/S0s.  Our data show
  that these trends derived for high-$\sigma$ galaxies extend down to
  dE/dS0s. This is a factor of $\sim 2$ lower in $\sigma$ than
  previously published work. We confirm that the observed
  anti-correlation between age and metallicity for high-$\sigma$
  galaxies is consistent with the effects of correlated errors.
  Low-$\sigma$ galaxies also show a similar relation between age and
  metallicity as a result of correlated errors.  However, they are
  offset from this relationship so that, on average, they are less
  metal rich and younger than their high-$\sigma$ counterparts.

%  The index-$\sigma$ relations for our Coma galaxy sample, including
%  low-$\sigma$ ($\sigma < 100$ km s$^{-1}$) galaxies, differ in shape,
%  depending on which index is used.  We notice three different types
%  of relations: {\it i)} a discontinuity in the index sigma
%  relationship between low- and high-$\sigma$ galaxies, {\it ii)} a
%  continuous trend with a weak relation and large scatter, and {\it
%  iii)} robust linear relation between an index and $\log \sigma$.  We
%  find no connection between the behavior of index-$\sigma$ relation
%  with either $\alpha$ or Fe-peak elements.  However, we find
%  indications that the relations are tighter for indices which do not
%  depend on the micro-turbulent velocities of the underlying stars. 

\end{abstract}

%%%%%%%%%%%%%%%%%%%%%%%%%%%%%%%%%%%%%%%%%%%%%%%%%%%%%%%%%%%%%%%%%%%%%%%%%%%%%


%%%%%%%%%%%%%%%%%%%%%%%%%%%%%%%%%%%%%%%%%%%%%%%%%%%%%%%%%%%%%%%%%%%%%%%%%%%%%
%%%%%%%%%%%%%%%%%%%%%%%%%%%%%%%%%%%%%%%%%%%%%%%%%%%%%%%%%%%%%%%%%%%%%%%%%%%%%
\section{INTRODUCTION}

The ``classical'' view of the formation of massive early-type
galaxies is that they formed relatively fast at early redshifts,
and that they evolved passively. This is supported by tight
scaling relations that these galaxies exhibit. However, the
question of how their low mass counterparts, dwarf early-type
(dEs/dS0s) galaxies, formed remains unanswered.

One of the ways to investigate the star formation histories of
early-type galaxies is through their line-strength indices which,
when combined with stellar population models (SPM), yield ages and
metallicities.  However, the process of deriving star formation
histories of galaxies by comparing line-strengths with models is
complicated by the degeneracy between age and metallicity
\citep{Worthey94b}, abundance ratio differences between the
calibration stars used to calculate the models and the galaxy
spectra, and strong Balmer lines which can be caused by either
young stellar populations or an extended horizontal branch.

The most recent studies have incorporated iterative procedures and/or
simultaneous fitting of as many indices as possible while deriving
ages, metallicities and relations between the line strengths and
velocity dispersions ($\sigma$) of early-type galaxies
\citep{Proctor04a, Thomas05, Nelan05, Denicolo05b, Sanchez06a,
  Sanchez06b}.  Despite these improved techniques and larger
samples extending to lower luminosities, we still do not have a clear
picture of the star formation histories of early-type galaxies,
especially at the low mass end.

It is well known that lower-luminosity early-type galaxies show a
wider range in age than their more luminous counterparts
\citep[e.g.][]{Caldwell83b, Bender93, Worthey97, Kuntschner98,
  Poggianti01, Caldwell03}.  Some studies find that the lower mass
galaxies also display younger ages \citep{Poggianti01, Caldwell03,
  Proctor04a, Thomas05, Nelan05}, while others do not find this
 relation, or find that it depends on environment \citep{Trager00a,
  Kuntschner01, Sanchez06a}.  Similarly a number of studies show that
lower mass galaxies have lower metallicities \citep{Brodie91,
  Poggianti01, Kuntschner01, Mehlert03, Proctor04a, Nelan05, Thomas05,
  Sanchez06b, Bernardi06}.  Further, these galaxies display lower
(closer to solar) abundance of $\alpha$-elements than the luminous
elliptical galaxies (Es) which have an overabundance of
$\alpha$-elements when compared to the values in the solar
neighborhood \citep{Gorgas97, Jorgensen97, Trager00a, Nelan05, Thomas05,
  Denicolo05b, Bernardi06, Sanchez06b}.  The lower values for the
abundance ratio, $\alpha$/Fe, for low-mass galaxies suggests that
these galaxies have had more extended star formation histories than
their massive counterparts.

The main goal of this project is to characterize internal
kinematics and stellar populations of dE/dS0 galaxies as a
function of cluster environment.  In  \citet[][hereafter Paper
I]{Matkovic05} we described the internal kinematics of these
galaxies in the dense central region of the Coma cluster.  Here
(paper II) we investigate the properties of the underlying stellar
populations of these galaxies.  The following papers will examine
the same properties of these galaxies in a region SW of the
cluster center, just outside the virial core, and will investigate how the
location within the cluster affects these properties. Currently,
there are only a few other studies which include dE/dS0 galaxies
in their samples.  While these studies reach $\sigma
\approx 50$ km s$^{-1}$ \citep{Poggianti01, Caldwell03, Nelan05,
Sanchez06a}, we present a statistically representative number of
dE/dS0s in Coma reaching $\sigma = 30$ km s$^{-1}$.  In this
paper, we investigate trends between the Lick/IDS line-strengths
and $\sigma$ and determine mean luminosity-weighted ages,
metallicities and $\alpha$/Fe ratios for a sample of 74 early-type
galaxies in the central part of the Coma cluster.  In $\S 2$ we
describe our sample selection and our spectroscopic data; $\S 3$
describes the absorption line strength measurements; in $\S 4$ we
compare our measurements with other authors; in $\S 5$ we
investigate the relations between different indices and $\sigma$;
$\S 6$ describes the ages, metallicities and $\alpha$-ratios.

%%%%%%%%%%%%%%%%%%%%%%%%%%%%%%%%%%%%%%%%%%%%%%%%%%%%%%%%%%%%%%%%%%%%%%%%%%%%%
%%%%%%%%%%%%%%%%%%%%%%%%%%%%%%%%%%%%%%%%%%%%%%%%%%%%%%%%%%%%%%%%%%%%%%%%%%%%%

\section{SAMPLE SELECTION AND SPECTROSCOPIC DATA}

We observed spectra of bright and faint early-type galaxies in two
different environments within the Coma cluster, the central $20
\arcmin \times 20 \arcmin$ and a SW region, just outside the
virial core ($\sim 1\degr$).  In this paper we discuss
measurements of line-strength indices for 74 early-type galaxies
(out of which 36 are dE/dS0) in the center of the Coma cluster,
while the subsequent papers will also address a region outside the
cluster virial core.  We provided a full description of the sample
selection and spectroscopic observations in Paper I. Here, we only
present a summary of our observations.

For the dE/dS0 sample selection we utilized photometry from WIYN's
Mini-Mosaic imager and the Isaac Newton Telescope Wide Field
Camera. We determined the following cutoffs from color-magnitude
and color-color diagrams: $0.2 < (U-B) < 0.6$ mag, $1.3<(B-R)<
1.5$ mag and $M_{B} \geqslant -17.3$ mag \citep*[][]{Ferguson94}
assuming a distance modulus for the Coma cluster of 35.078, $d=99$
Mpc.

We obtained the follow-up spectroscopic data with the Hydra
multi-fiber spectrograph on WIYN 3.5 m telescope.  The instrumental
resolution was FWHM $=1.91$ \AA~which was sampled at $\sim 0.705$
\AA~px$^{-1}$.  We used IRAF's `dohydra' package for data reduction.
%RRRRRRRRRRRRRRRRRRRRRRRRR
The diameter of the blue cable fiber was $3.1 \arcsec$ which corresponds to
$\sim 1.45$ kpc at the distance of the Coma cluster ($d=99$ Mpc).  
%RRRRRRRRRRRRRRRRRRRRRRRRR

In Paper I we determined that 100\% of the observed galaxies in
our sample (down to m$_B \approx 19.2$ mag) are consistent with
being members of the Coma cluster as their recession velocities
range between 4,000 and 10,000 km s$^{-1}$ \citep{Colless96}.  The
luminosity of the sample spans $-20.3 \lesssim M_B \lesssim -15.8$
mag while the velocity dispersions are $30 \lesssim \sigma
\lesssim 260$ km s$^{-1}$.  All the spectra have S/N$\geqslant 15$
per pixel (Paper I, also see Appendix \ref{snr} for S/N
determination).

Throughout this paper we refer to galaxies fainter than M$_B
\approx -18$ and with $\sigma < 100$ km $s^{-1}$ as `faint', or
low-$\sigma$ galaxies.  This group includes the 36 dE/dS0 galaxies
and 6 intermediate early-type galaxies.  The remaining 32 objects
with $\sigma \geqslant 100$ km $s^{-1}$ we refer to as bright, or
high-$\sigma$ galaxies.

%%%%%%%%%%%%%%%%%%%%%%%%%%%%%%%%%
%We also classified galaxies in our sample depending on their
%bulge-to-total ratios, B/T (see Paper I for more details).  There are
%30 galaxies with bulge-dominated luminosity profiles, 16 with
%bulge+single exponential component, 19 with a single exponential
%component, while no morphology was available for the
%remaining 9 galaxies.  The luminosity profiles are from
%\citet{Gutierrez04} and GG03.
%%%%%%%%%%%%%%%%%%%%%%%%%%%%%%%%%

In order to determine the luminosity-weighted ages and
metallicities for the early-type galaxies in the Coma cluster we
measure their spectral line-strength indices and compare them with
stellar population models by \citet[][hereafter TMB03]{TMB03}.

%%%%%%%%%%%%%%%%%%%%%%%%%%%%%%%%%%%%%%%%%%%%%%%%%%%%%%%%%%%%%%%%%%%%%%%%%%%%%
%%%%%%%%%%%%%%%%%%%%%%%%%%%%%%%%%%%%%%%%%%%%%%%%%%%%%%%%%%%%%%%%%%%%%%%%%%%%%
\section{ABSORPTION-LINE STRENGTH INDEX MEASUREMENTS}
%*************************************************************************
%\subsection{Line-Strength Measurements}
\label{index measurements}

We use the Lick/IDS system of indices \citep[originally defined
by][]{Burstein84, Gorgas93, Worthey94a, Trager98} as it covers a wide
range in optical absorption-lines among which are some of the most
prominent features of early-type galaxies.  Furthermore, the Lick/IDS
system is also a basis for an extensive collection of stellar
synthesis models, allowing one to derive ages and metallicities of
galaxies.

We measured the line-strengths for our 74 early-type galaxies with the
software REDUCEME \citep{Cardiel99}, task INDEX.  This software allows
for careful determination of uncertainties in the index measurements
as it uses Monte Carlo simulations (see Appendix \ref{errors} for more
%RRRRRRRRRRRRRRRRRRRRRRRRR
details).  We were able to measure the following Lick/IDS indices:
%RRRRRRRRRRRRRRRRRRRRRRRRR
Ca4227, G4300, Fe4383, Ca4455, Fe4531, C$_2$4668, H$\beta$, Fe5015,
Mg$_1$, Mg$_2$, Mg$_b$, Fe5270, and Fe5335 \citep{Trager98}; and their
extensions to higher order Balmer lines H$\gamma_A$ and H$\gamma_F$
\citep{Worthey97}.  We also include the [MgFe]$\arcmin$ (as defined by
TMB03) and $\langle$Fe$\rangle$ \citep*{Gonzalez93} indices since they
closely measure metallicity and are common in the literature allowing
for an easy comparison.  We use these two indices with stellar
population models to predict luminosity weighted ages, metallicities
and the [$\alpha$/Fe]-ratios.

The standard procedure for matching the Lick system of
line-strength indices is to re-sample the data to the resolution
of Lick/IDS and to correct for the systematic differences between
the indices of standard stars which were also observed by Lick.
Additionally, one needs to correct the nebular emission and apply
velocity dispersion and aperture corrections.  We were not able to
fully transform our data into the Lick/IDS system because we did
not observe stars in common with the original stellar library (see
below).  Below, we describe the procedure we followed:

\begin{itemize}
\item[{\it i)}] {\it Re-sampling to Lick/IDS resolution}\\
%RRRRRRRRRRRRRRRRRRRRRRRRR
  We adjusted the spectral resolution of our data  
%RRRRRRRRRRRRRRRRRRRRRRRRR re-sampled our data 
  to match the resolution of the Lick/IDS $\sim 7-10$ \AA~(FWHM)
  depending on the wavelength.  This was done by broadening each
  galaxy spectrum by the quadratic difference between the Lick
  resolution and that of the observed galaxy.
  \begin{equation}
    \label{resampling}
      {\sigma_{broad}=\sqrt{\sigma_{Lick}^2-\sigma_{gal}^2-\sigma_{Instr}^2}}
  \end{equation}
  where $\sigma_{broad}$ is the amount to broaden the galaxy by (in km
  s$^{-1}$), $\sigma_{Lick}$ is Lick/IDS resolution for a particular
  index \citep*[see Table 5 of][]{Sanchez06a}, $\sigma_{gal}$ the
  velocity dispersion of the galaxy and $\sigma_{Instr}$ represents the
  instrumental resolution of our data. Note that
  $\sigma_{obs}^2=\sigma_{gal}^2+\sigma_{Instr}^2$.

\item [{\it ii)}] {\it Flux calibration}\\ Since we lacked
  observations of flux calibration stars in our sample, we used the
  flux calibration of \citet*{Sanchez06a} who have 8 galaxies in
  common with our sample.  First, we obtained the response curves for
  the 8 matching galaxies by dividing each galaxy spectrum by that of
  the flux-calibrated galaxy.  Then, we fitted a polynomial function
  to each of the response curves and created a mean flux calibration
  curve.  This curve is used to flux calibrate the line-strengths,
  while the individual response curves are later used to calculate the
  flux related uncertainties in the index measurements (see Appendix
  \ref{errors}).

\item [{\it iii)}] {\it Offsets to Lick/IDS standards}\\
  We did not observe stars in common with Lick/IDS.  Furthermore, we
  only have 4 galaxies also observed by Lick/IDS \citep{Trager98}
  which is insufficient for determining any offsets between our data
  and the Lick/IDS system.  We are, however, able to compare our data
  to other data sets in the literature (see Section \ref{comparison}).

%RRRRRRRRRRRRRRRRRRRRRRRRRRRRRRRRRRRRRR
\item [{\it iv)}] {\it Velocity dispersion corrections}\\ The velocity
  dispersion measurements of our sample of Coma cluster galaxies are
  described in Paper I.  The velocity dispersion corrections are
  usually done by, first, broadening an individual galaxy spectrum to
  an effective resolution of $\sqrt{(\sigma_{Lick}^2 +
    \sigma_{gal}^2)}$. Second, the index measurements are corrected
  for the extra broadening due to velocity dispersion of the galaxy
  ($\sigma_{gal}$) via polynomial fitting.  In our case, $\sigma_{gal}
  < \sigma_{Lick}$ for all but 4 galaxies in our sample.  Therefore,
  we are able to apply the velocity dispersion corrections by
  broadening our spectra by $\sqrt{(\sigma_{Lick}^2-\sigma_{gal}^2)}$
  and then measuring the indices.  This approach also avoids
  introducing a source of uncertainty associated with velocity
  dispersion corrections which may affect the derived $I-\sigma$
  relations \citep{Kelson06}.

%RRRRRRRRRRRRRRRRRRRRRRRRRRRRRRRRRRRRRR
%it was not necessary to apply the velocity dispersion corrections.  We
%  excluded those galaxies when their velocity dispersion was larger
%  than the resolution of the Lick/IDS index in question.  This
%  approach also avoids introducing a source of uncertainty associated
%  with velocity dispersion corrections which may affect the derived
%  $I-\sigma$ relations \citep{Kelson06}.

%RRRRRRRRRRRRRRRRRRRRRRRRRRRRRRRRRRRRRRRRRRRRRRRRRRRRRRRRRRRRRRRR
\item [{\it v)}] {\it Aperture corrections}\\ The galaxies in our
  sample are typically smaller than the $3 \arcsec$ aperture of HYDRA
  fibers \citep*{GG03}.  Therefore, we found no need for aperture
  corrections. In fact, for most of our data, we are sampling the
  entire galaxy, rather than only its central parts.  Since dE
  galaxies have flat velocity dispersion gradients
  \citep[e.g.][]{Pedraz98}, there should not be a significant change
  between the central and global measurements of these galaxies.
  Although we do not apply aperture corrections, we do recognize that
  a few massive galaxies in this sample are likely to be larger than
  the HYDRA-spectrograph fibers in which case we are sampling their
  central regions only.
%RRRRRRRRRRRRRRRRRRRRRRRRRRRRRRRRRRRRRRRRRRRRRRRRRRRRRRRRRRRRRRRR

%RRRRRRRRRRRRRRRRRRRRRRRRRRRRRRRRRRRRRRRRRRRRRRRRRRRRRRRRRRRRRRRR
\item [{\it vi)}] {\it Emission corrections}\\ It is well known that
  some early-type galaxies have nebular emission which can contaminate
  measurements of certain line-strength indices.  For example, nebular
  emission in the H$\beta$ feature would cause the H$\beta$
  line-strength index to appear lesser in value which in turn would
  lead to derivation of an older age for a given galaxy.  We selected
  our sample via color-magnitude and color-color diagrams to minimize
  the presence of emission. We also checked the spectra for weak
  emission lines by: co-adding all the galaxies to enhance the S/N of
  any weak emission, and dividing each galaxy by it's template
  \citep{Kuntschner01}.  Finally, we used a method similar to that of
  \citet*{Hammer01} and \citet{Kuntschner02} to determine whether any
  galaxies in our sample contained emission.  From each galaxy
  spectrum we subtracted it's optimal template (a spectrum of a linear
  combination of template stars optimized to fit each galaxy spectrum,
  see Paper I for more details).  This method revealed that 2 galaxies
  in our sample had H$\beta$ and OIII emission.  We exclude these two
  galaxies, GMP 3733 and GMP 2516, from our analysis.  We note that
  the low number of dE/dS0 galaxies with emission is consistent with
  other studies in the Coma cluster \citep*{Sanchez06a,Smith08} and
  with dwarf galaxies having their gas removed when they enter the
  cluster environment.
%RRRRRRRRRRRRRRRRRRRRRRRRRRRRRRRRRRRRRRRRRRRRRRRRRRRRRRRRRRRRRRRR
\end{itemize}

A sample of our measurements is presented in Tables
\ref{mastertable1} and \ref{mastertable2}, while the full data set
is in electronic format.  These indices are at the Lick/IDS
resolution, so they can be easily compared to other sources in the
literature.  The error in each index measurement includes
uncertainties associated to the flux calibration and to the photon
noise.  For a more detailed description of the uncertainty
measurements, please see Appendix \ref{errors}.

%%%%%%%%%%%%%%%%%%%%%%%%%%%%%%%%%%%%%%%%%%%%%%%%%%%%%%%%%%%%%%%%%%%%%%%%%%%%%
%%%%%%%%%%%%%%%%%%%%%%%%%%%%%%%%%%%%%%%%%%%%%%%%%%%%%%%%%%%%%%%%%%%%%%%%%%%%%
\section{COMPARISON WITH LITERATURE}
\label{comparison}

Converting our data to the Lick/IDS system resolution allows a direct
comparison to other data sets in the literature.  We compare our line
strength measurements to that of \citet*[][hereafter
P01]{Poggianti01}, \citep*[][hereafter NFPS]{Nelan05} and
\citep*[][hereafter S08]{Smith08}. Although these
studies contain a significantly larger number of galaxies, our data is
complementary to these samples as it includes a larger number of low
mass galaxies ($\sigma \lesssim 100$ km s$^{-1}$) with velocity
dispersion measurements reaching as low as $\sigma \approx 30$ km
s$^{-1}$ in Coma.

\begin{figure}
\includegraphics[width=6.5in]{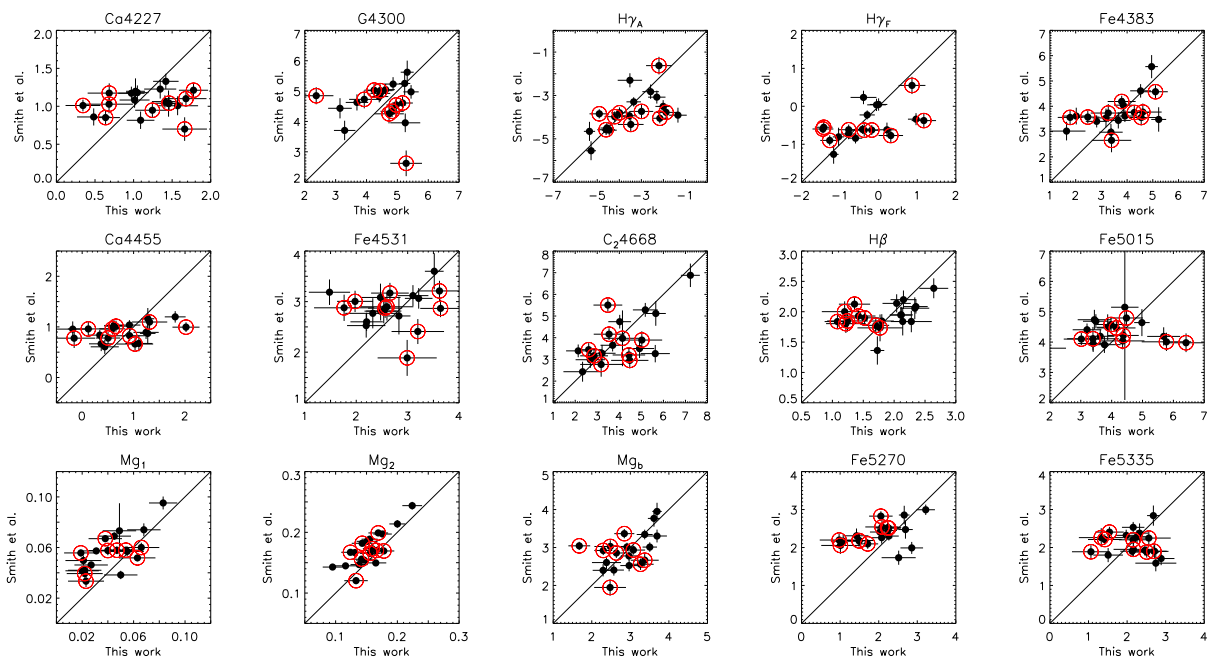}
%  \centerline{\psfig{figure=f1.eps,width=7.0truein}}
%  \centerline{\psfig{figure=comp_Hecto_paper.eps,width=7.0truein}}
  \caption{Comparison of our sample with \citet*{Smith08}.
    The lines have a slope of 1.0 and
    indicate a one-to-one agreement between the two data sets.
    Galaxies which do not fit on the model grids, the ``off-grid''
    galaxies, are labeled as red open circles and are discussed in
    $\S$ \ref{offgridSect}.
\label{compHectofig}}
\end{figure}

\begin{figure}
\includegraphics[width=6.5in]{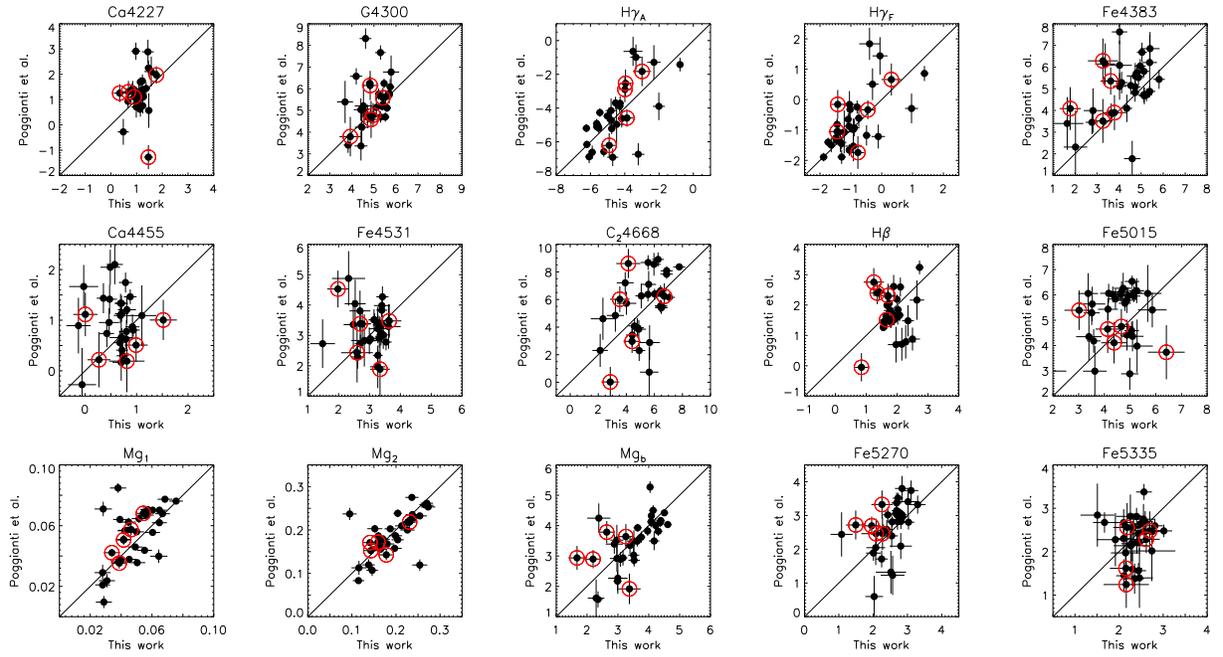}
%  \centerline{\psfig{figure=f2.eps,width=7.0truein}}
%  \centerline{\psfig{figure=comp_mob3.eps,width=7.0truein}}
  \caption{Comparison of our sample with \citet*{Poggianti01}
    represented as solid circles. Symbols and lines are the same as in Figure \ref{compHectofig}.
    \label{compMobfig}}
\end{figure}

\begin{figure}
\includegraphics[width=6.5in]{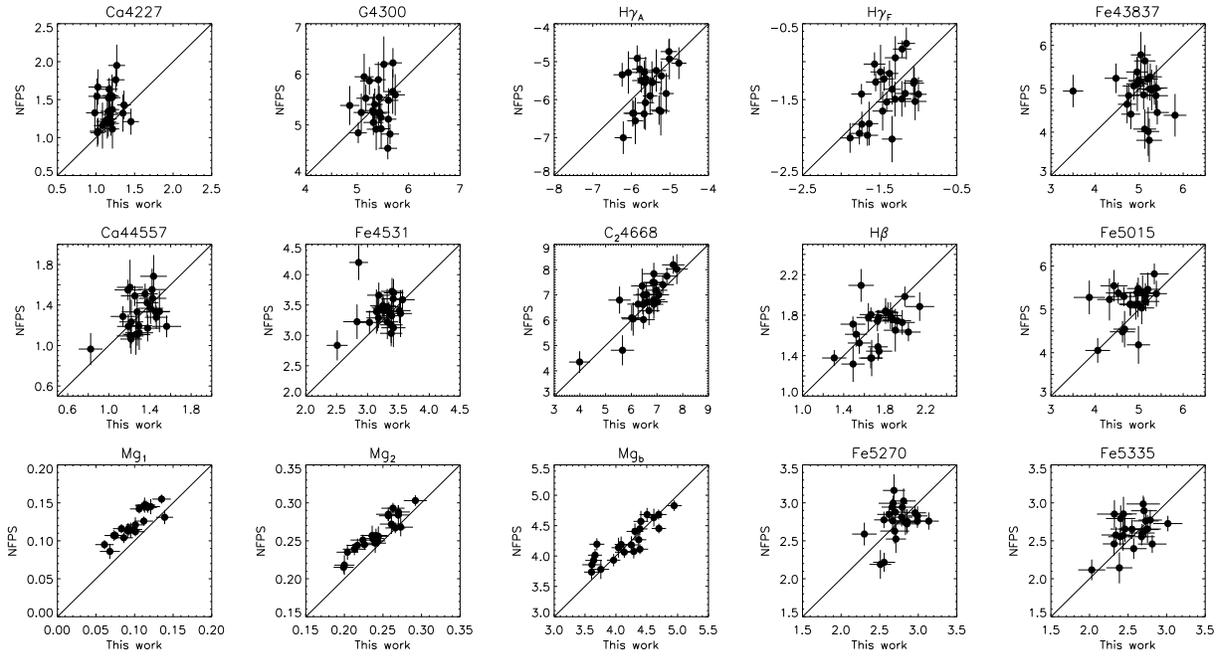}
%  \centerline{\psfig{figure=f3.eps,width=7.0truein}}
%  \centerline{\psfig{figure=comp_NFPS.eps,width=7.0truein}}
    \caption{Comparison of our sample with \citet*{Nelan05}.
      Symbols and lines are the same as in Figure \ref{compHectofig}.
      \label{compNFPSfig}}
\end{figure}

The comparison between the three literature samples and our data
are shown in Figures \ref{compHectofig}, \ref{compMobfig}, and
\ref{compNFPSfig}.  The offsets are presented in Table
\ref{offsetstabl}.  The mean offsets are defined as a difference
between the index measurement in this paper and the index
measurement in P01, S08, or NFPS (N05) $\langle \Delta I \rangle =
I_{here}-I_{other}$.  We also calculated the error in the mean
offset, standard deviation, and the standard deviation expected by
the errors.

%--------------------------------------------------------------------
%Comparison with Smith
%--------------------------------------------------------------------

The offsets between our sample and S08 are small and not significant
for the majority of the indices.  The indices for which the offsets
are significant are Mg$_1$, Mg$_2$ and Fe5270 with mean offsets of
$-0.014 \pm 0.003$, $-0.018 \pm 0.004$ and $-0.334 \pm 0.127$,
respectively.  However, the scatter between the S08 sample and ours is
large and cannot be explained by the errors.  The indices Ca4227,
G4300, H$\gamma_A$, H$\gamma_F$, and Ca4455 each have a scatter $\sim
2$ times larger than the scatter expected by the errors, while the
scatter is $\sim 1.3$ times larger than the scatter due to the errors
for majority of the other indices.  The large scatter in the
comparison between the S08 index measurements and ours is dominated by
a group of galaxies which are systematically offset in the H$\beta$
plot (red open circles in Figure \ref{compHectofig}).  Because our
H$\beta$ and [MgFe]$\arcmin$ indices for these galaxies cannot be
reproduced by the stellar population models, we exclude them from
further analysis and revisit this issue in $\S$ \ref{offgridSect}.
%RRRRRRRRRRRRRRRRRRRRRRRRRRRRRRRRRRRRRRRRRRRRRRRRRRRRRRRRRRRRRRRR
However, once these galaxies are excluded from the analysis, we no
longer have a statistically significant number of galaxies in common
with S08 sample for comparison.
%RRRRRRRRRRRRRRRRRRRRRRRRRRRRRRRRRRRRRRRRRRRRRRRRRRRRRRRRRRRRRRRR

%--------------------------------------------------------------------
%  Comparison with Poggianti
%--------------------------------------------------------------------
We find that the mean offsets between our index measurements and those
of \citet*{Poggianti01} are small and in most cases
insignificant. However, Ca4455, Fe5015, and Mg$_1$ have non-negligible
offsets, $-0.259 \pm 0.116$, $-0.652 \pm 0.208$ and $-0.012 \pm
0.006$, respectively.  We note that some indices may be offset because
our indices are not fully transformed to the Lick/IDS system like the
P01 sample.  Similarly to the S08 data, the scatter in the indices
when comparing our data with P01 is large for most indices, except for
Mg$_1$ and Mg$_2$.  Furthermore, the large scatter between these two
data sets cannot be explained by the errors.  The discrepancy between
the measured scatter and the scatter due to the errors is the largest
for G4300, H$\gamma_A$, H$\gamma_F$, and Mg$_1$.

%--------------------------------------------------------------------
%  Comparison with NFPS
%--------------------------------------------------------------------
On the other hand, the NFPS index measurements are in good agreement
with ours.  Among the atomic indices only Ca4227, Fe4531 and Fe5015
have significant offsets: $-0.207 \pm 0.047$, $-0.152 \pm 0.066$ and
$-0.313 \pm 0.102$, respectively, while the rest of the indices are
consistent between the two studies with a small scatter.  Most
noticeably, our and the NFPS indices are in very good agreement in the
Mg-indices after a systematic offset in Mg$_1$ of $-0.023 \pm 0.002$
and in Mg$_2$ of $-0.017 \pm 0.002$ is applied.  The NFPS data are not
flux calibrated which would mostly affect the molecular indices like
the Mg$_1$ and Mg$_2$ and may explain why there is an offset between
the data sets for these two indices.  However, the scatter in these
two molecular indices is remarkably small (0.01 mag for both).  The
scatter in the remaining indices is also consistent within the errors
in both studies.

%RRRRRRRRRRRRRRRRRRRRRRRRRRRRRRRRRRRRRRRRRRRRRRRRRRRRRRRRRRRRRRRR
We have demonstrated that our index measurements are in good agreement
with those of NFPS, which is one of the highest quality data available
in the Coma cluster so far.  Considering our internal and external
error analysis, we conclude that our data set has a similar enough
quality and our error measurements are well estimated.
%In the following sections we only include the NFPS data set for
%comparison purposes, since the \citet*{Poggianti01} and
%\citet*{Smith08} do not include velocity dispersions. Additionally,
%our index measurements are in better agreement with NFPS.  
In the following sections we only include the NFPS data set for
comparison purposes, since this sample also includes velocity dispersions.
%and is in good agreement with our data.  
The NFPS spectra were not flux calibrated, and for further analysis we
apply an offset to the NFPS data to be consistent with ours according
to the average values listed in Table \ref{offsetstabl}.
%RRRRRRRRRRRRRRRRRRRRRRRRRRRRRRRRRRRRRRRRRRRRRRRRRRRRRRRRRRRRRRRR

%%%%%%%%%%%%%%%%%%%%%%%%%%%%%%%%%%%%%%%%%%%%%%%%%%%%%%%%%%%%%%%%%%%%%%%%%%%%%
%%%%%%%%%%%%%%%%%%%%%%%%%%%%%%%%%%%%%%%%%%%%%%%%%%%%%%%%%%%%%%%%%%%%%%%%%%%%%
\section{INDEX -- VELOCITY DISPERSION RELATIONS}
\label{i-sigma}

In the following paragraphs we investigate the relationships
between different line-strength indices and velocity dispersion
($\sigma$). We compare these relations between the low- and
high-$\sigma$ galaxy samples, and discuss possible parameters that
drive the index-$\sigma$ relations (hereafter $I-\sigma$).

\subsection{Results of $I-\sigma$ Relations}

\begin{figure}
\includegraphics[height=5.5in]{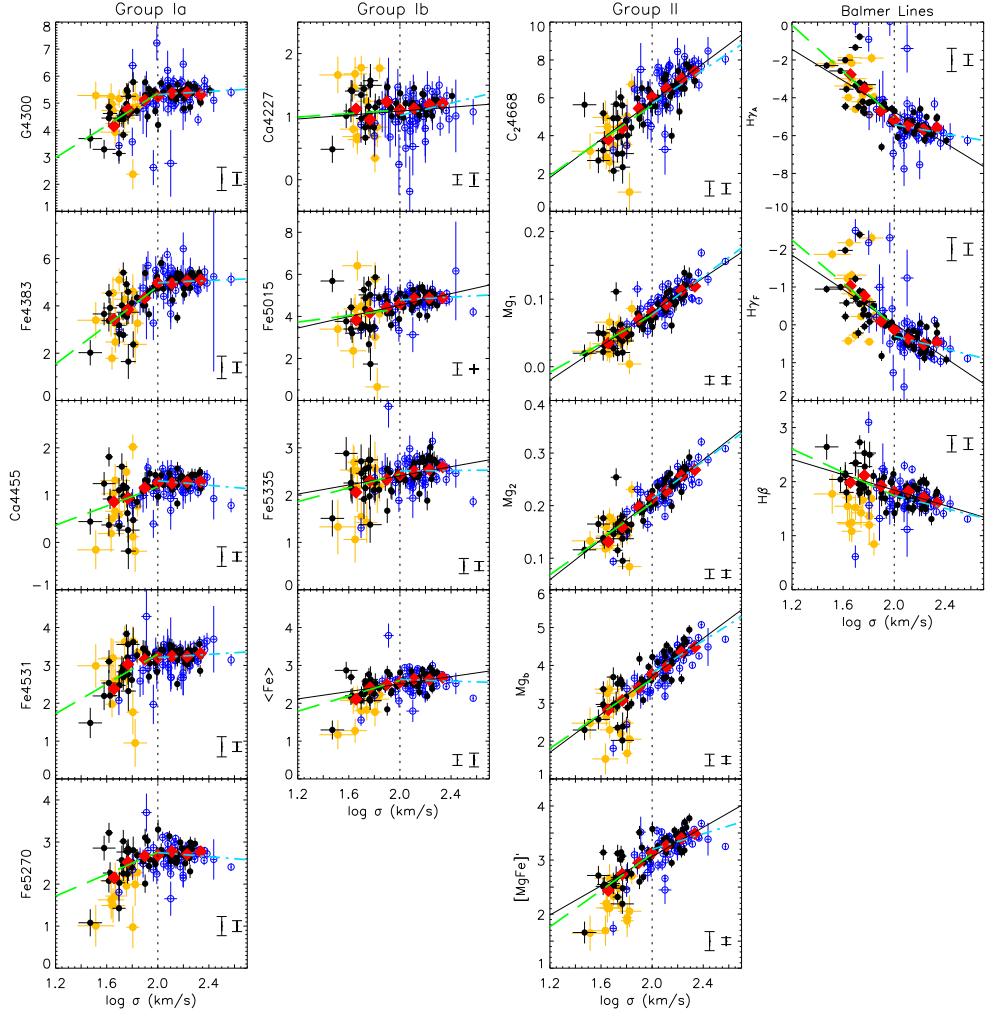}
%from /scratch/matkovic/dEs_gEs/lick_index/nfps/index_sigma_2Groups_Oct2_2008.pro
%\centerline{\psfig{figure=index_sigma_3colHg.eps,height=9truein}}
%  \centerline{\psfig{figure=f4.eps,width=\textwidth}}
%  \centerline{\psfig{figure=index_sigma_4colHg.eps,width=\textwidth}}
  \caption{Index -- $\log \sigma$ plots.  Data from the NFPS sample are marked by blue open circles, and our sample with black filled circles and yellow circles for galaxies not fitted by models.  We binned the data by $\sigma$ where each bin contains an equal number of objects.  The
red diamonds are weighted average values of each bin.  The vertical
dotted line corresponds to $\sigma = 100$ km s$^{-1}$.  Group Ia and Ib represent the set of
$I-\sigma$ relations that show evidence for a break in the slope,
group II $I-\sigma$ with strong linear relations, and the last column contains
Balmer lines.  In the bottom right corner of $I-\sigma$ plots in Group
Ia we show the 1-sigma scatter for low-$\sigma$ on the left and
high-$\sigma$ galaxies on the right.  In the case of Group Ib and
Balmer lines these lines correspond to the intrinsic scatter for
low- and high-$\sigma$ respectively.  For Group II, we show the
intrinsic and observed scatter, respectively, in the bottom right
corner.  The black lines represent linear fits for all the galaxies in
the individual $I-\sigma$ figures, while the green dashed and light
blue dash-dotted lines mark linear fits to low- and high-$\sigma$
galaxies respectively.  Note that the ``off-grid'' galaxies were not
included in the bins and linear regression calculations.
\label{index-sigma-fig}}
\end{figure}

We examine relations between 15 Lick/IDS indices, plus
$\langle$Fe$\rangle$ and [MgFe]$\arcmin$, and the central velocity
dispersion of our galaxies in Figure \ref{index-sigma-fig}.  In this
plot, we include galaxies from our sample and galaxies in the Coma
%RRRRRRRRRRRRRRRRRRRRRRRRRRRRRRRRRRRRRRRRRRRRRRRRRRRRRRRRR
cluster from NFPS.  Objects in common have averaged index
values and velocity dispersions.  We combined the two data sets and
%RRRRRRRRRRRRRRRRRRRRRRRRRRRRRRRRRRRRRRRRRRRRRRRRRRRRRRRRR
binned index measurements for galaxies of similar velocity dispersion.
All the bins have the same narrow interval in $\log \sigma$ of 0.118
dex.  The red diamonds represent the average value of each bin
weighted by the uncertainties in the index measurements.

%RRRRRRRRRRRRRRRRRRRRRRRRRRRRRRRRRRRRRRRRRRRRRRRRRRRRRRRRR
%Different types of behaviours emarge between the indices and velocity
%dispersions for these early-type galaxies in the Coma cluster (Figure
%\ref{index-sigma-fig}). The $I-\sigma$ relations cannot all be
%described by a linear fit.  To more easily describe
%the data, we grouped the $I-\sigma$ trends according to the
%steepness of their slope and the behaviour of the low-$\sigma$
%galaxies with respect to the high-$\sigma$ ones.  We plotted the
%Balmer indices in a separate column in the figure.
%
%The three types of relations are:
%\begin{itemize}
%\item [{\bf I)}]{$I-\sigma$ which cannot be described by a linear
%fit. For these indices, there is a clear offset between the low-
%and high-$sigma$ galaxies, and the slope of the high-$\sigma$
%galaxies is flat.}
%
%\item [{\bf II)}]{$I-\sigma$ with a shallow slope and notable
%scatter especially for low-$\sigma$ galaxies.}
%
%\item [{\bf III)}]{a group of indices with strong linear
%correlations and steep slope with $\log \sigma$.}
%\end{itemize}

Two different types of behaviors emerge between the indices and
velocity dispersions for these early-type galaxies in the Coma cluster
(Figure \ref{index-sigma-fig}).  One group of indices shows evidence
for a break in the slope $~\sim 100$ km s$^{-1}$, {\bf group I}, while
another group exhibits strong linear relations with $\log \sigma$,
{\bf group II}.  We further divide group I into 2 sub-groups where the
evidence for a break in the slope is stronger for group Ia than it is
for group Ib.  

We plot these groups in the different columns of Figure
\ref{index-sigma-fig}.  To more easily describe the difference between
the groups of $I-\sigma$ trends, the plots also include separate
linear fits to low- and high-$\sigma$ galaxies (Table
\ref{linear-table}), even though the correlation coefficients for
these fits are low for group I.  This figure also contains 13
low-$\sigma$ galaxies, marked as yellow filled circles, which lie
outside the model grids and have been excluded from the analysis
(discussed in $\S$ \ref{offgridSect}).  We refer to these objects as
``off-grid'' galaxies.

%their H$\beta$ measurements are systematically offset from
%the \citet*{Smith08} data, and they potentially have spurious
%measurements for some indices (discussed in $\S$
%\ref{offgridSect}).  We exclude these galaxies from further
%analysis and refer to them as ``off-grid'' galaxies.

%--------------------------------------------------------------------
\paragraph{Group I: }

Galaxies with $\sigma \geqslant 100$ km s$^{-1}$ exhibit flat
$I-\sigma$ relations in group I.  On the other hand, these relations
show evidence for a break in their slope for the low-$\sigma$
galaxies.  This break is more evident for indices in group Ia than in
Ib.  Both sub-groups show a significantly larger scatter of index
values in the low-$\sigma$ regime than for the high-$\sigma$ galaxies.
In Table \ref{stats-table}, we calculated the standard deviation of
the low- and high-$\sigma$ galaxies for group Ia (also shown in Figure
\ref{index-sigma-fig}).  The scatter of index values is $\sim 2$ times
larger for the low-$\sigma$ galaxies than it is for the high-$\sigma$
ones.  Additionally, the mean value of an index differs for low- and
high-$\sigma$ galaxies by 7--27 \% for this sub-group.  We also
performed the Kolmogorov-Smirnov test (KS) to determine whether the
two groups of galaxies come from the same distribution.
% (we used IDL/KSTWO from IDL Astronomy User's Library).
In the sam table, we show the KS probability that the two
sets are drawn from the same distribution.  Since the values of the KS
probability are quite small, we conclude that low- and high-$\sigma$
galaxies may indeed come from different populations.  We also show the
linear fits for this sub-group, although their Spearman Rank
correlation coefficients (see Table \ref{linear-table}) are small and
these relations are not statistically significant.

The middle panel of Figure \ref{index-sigma-fig} shows the Ib
sub-group of $I-\sigma$ relations, which also exhibits evidence for
the break in the slope around $~\sim 100$ km s$^{-1}$.  However, the
break is not as obvious as it is for group Ia since there are a number
of low-$\sigma$ galaxies whose index values are equal or higher than
the high-$\sigma$ galaxies.  For this reason we perform linear fits
for the entire $\sigma$ range together.  We take into account both the
errors in the index and $\sigma$ and calculate the intrinsic scatter
($\sigma_{Intr}$) and the scatter predicted by the errors
($\sigma_{Err}$) for both low- and high-$\sigma$ galaxies (Table
\ref{fitexytabl2}).  The spearman Rank correlation coefficients,
$\rho$, are low (lower than 0.5) for this sub-group indicating weak or
null $I-\sigma$ correlations, especially for Ca4227.  Galaxies with
$\sigma < 100$ km s$^{-1}$ have a larger intrinsic scatter\footnote
{\begin{displaymath}
{\sigma_{Intr} = \sqrt{ \frac {\sum_{i=1}^n
(\textrm{Index$_i$}-(\textrm{intercept}+\textrm{slope} \cdot
\sigma_i))^2}{N-2}- \frac {\sum_{i=1}^n
\delta_{\textrm{Index}_i}^2+\textrm{slope}^2 \cdot
\delta_{\sigma_i}^2} {N} } }
\end{displaymath}} 
than their more massive counterparts except for Ca4227 and
$\langle$Fe$\rangle$ indices (see lower right-hand corner of Figure
\ref{index-sigma-fig}).  However, the larger intrinsic scatter for
high-$\sigma$ galaxies in the Ca4227 index is dominated by galaxies
from the NFPS data.  Ca4227 is the only index for which the intrinsic
scatter is smaller than the scatter due to the errors for both low-
and high-$\sigma$ galaxies.  The remaining indices in group Ib display
a larger intrinsic scatter than the scatter due to errors ($\sim
1.2-2.9$ times) for both low- and high-$\sigma$ sub-samples, except
for Fe5015 for which the intrinsic scatter is zero.  The KS test on
the residuals between the indices and their respective linear fits
implies that only Fe5335 shows a high enough
probability that the low- and high-$\sigma$ galaxies are drawn from
the same population.

\paragraph{Group II: }

In the third column of Figure \ref{index-sigma-fig} we show the
metallic indices which display a tight relation with $\sigma$.  This
is quantified in Table \ref{fitexytabl3}, where we calculated the
Spearman-Rank correlation coefficients, and we include the slopes and
intercepts of linear fits to these relations.  We show the intrinsic
scatter and the scatter expected by the errors in the bottom right
hand corner of Figure \ref{index-sigma-fig} for each of these indices.
For all indices in this group the intrinsic scatter is larger than the
scatter due to errors, although the two are very close in value for
Mg$_1$ and Mg$_2$.  Moreover, the correlation coefficients for all the
indices in this group imply a robust relation with $\sigma$.  Many
studies find that the early-type galaxies in clusters show a tight
relation between line-strength index Mg$_2$ and their central velocity
dispersion and find similar slopes for this relation.  Our slope is
consistent with that of NFPS within 1 standard deviation (Table
\ref{fitexytabl3}).
%--------------------------------------------------------------------

\paragraph{Balmer Lines: }

The Balmer lines H$\gamma_A$, H$\gamma_F$ and H$\beta$ exhibit
negative slopes in their $I-\sigma$ relations.  Their Spearman Rank
correlation coefficients imply a strong relation for H$\gamma_F$,
while the relations for H$\gamma_A$ and H$\beta$ have slightly lower
%coefficients, $\rho=-0.665$ and $-0.502$ respectively (Table
coefficients, $\rho=-0.663$ and $-0.488$ respectively (Table
\ref{fitexytabl3}).  In case of H$\beta$, we see a hint for an
asymmetric scatter for low-$\sigma$ galaxies where more galaxies have
a lower H$\beta$ index.
%RRRRRRRRRRRRRRRRRRRRRRRRRRRRRRRRRRRRRRRRR
% the scatter for low-$\sigma$ galaxies is
%asymmetric with more galaxies having a lower H$\beta$ index.  
%RRRRRRRRRRRRRRRRRRRRRRRRRRRRRRRRRRRRRRRRR
Together, our and the NFPS sample have a standard deviation from the
linear fit of 0.306 with 23 galaxies having a lower value of their
H$\beta$ index, compared to the scatter of 0.243 for 18 galaxies with
a higher H$\beta$ index.  The calculations exclude the galaxies which
lie off the model grids.

%%%%%%%%%%%%%%%%%%%%%%%%%%%%%%%%%%%%%%%%%%%%%%%%%%%%%%%%%%%%%%%%%%%%%%%%%%%%%
\subsection{Discussion of Index--$\sigma$ relations}

Historically, most studies have concentrated on relations between
Mg$_2$, a metallicity indicator, and H$\beta$, an age indicator, with
$\log \sigma$, a measure of mass, \citep[to name a
few:][]{Terlevich81, Dressler84, Guzman92, Bender93, Jorgensen96,
  Kuntschner01}.  Recent studies, however, show that these relations
are more complex than originally thought.  For instance, Mg$_2 -
\sigma$ may be significantly dependent on both age ($\sim 15 \%$) and
relative abundances of heavy elements ($\sim 20-30 \%$)
\citep*{Mehlert03,Thomas05}.  While H$\beta-\sigma$ mainly depends on
age, it also changes with metallicity and chemical composition
\citep{Sanchez06a}.

In this study we present, for the first time, the Mg$_2-\sigma$,
H$\beta-\sigma$ and relations between other Lick indices with $\sigma$
down to 30 km s$^{-1}$ for a homogeneous sample of early-type
galaxies.  Our galaxies reside in one of the densest environments in
the nearby universe: the center of the Coma cluster.  We show that the
Mg$_2-\sigma$ relation spans the entire range of 30--260 km s$^{-1}$
with a small scatter. We also investigate relations between the
H$\beta$, H$\gamma$ and the metallicity sensitive line strengths with
$\sigma$.  We discuss the H$\beta-\sigma$ and Mg$_2 - \sigma$
individually, while we examine the remaining indices together.

%-------------------------------------------------------------------------
\subsubsection{H$\beta$ vs. $\sigma$}

Most studies find that the line-strength H$\beta$ and the central
velocity dispersion of early-type galaxies are anti-correlated.  This
relation seems to hold in different environments
\citep*[e.g.][]{Fisher95, Jorgensen97, Trager98, Kuntschner00,
  Caldwell03, Nelan05, Sanchez06a} and is usually interpreted as an
interplay between age and mass.  However, recent evidence that the
H$\beta-\sigma$ relation is weak or flat in the Coma cluster
\citep*{Mehlert03, Sanchez06a} may allude towards its dependence
on the environment.  Furthermore, there is some evidence that the
age variation of galaxies in clusters is not large and the
H$\beta-\sigma$ relation may mostly be driven by metallicity
\citet[in Fornax][]{Kuntschner98}, or by both variations in global
metallicity and relative abundance of different heavy elements for
the galaxies in the Coma cluster \citep{Sanchez06a}.

We confirm that the early-type galaxies in the core of the Coma
cluster show a weak anti-correlation between their H$\beta$ line
strength and velocity dispersions.  
%RRRRRRRRRRRRRRRRRRRRRRRRRRRRRRRRRRRRRRRRRRRRRRRRRRRRRRRRRRRRR
%Our result is similar to that of \citet[][hereafter CRC03]{Caldwell03}
%who also find a large asymmetric scatter for the low-$\sigma$
%galaxies. However, the CRC03 data show an opposite effect in
%asymmetry where more galaxies have higher Balmer line-strengths, while
%we find that more galaxies have lower value of their H$\beta$ index.
%It is possible that this effect is environmental, since CRC03 sample
%includes galaxies lower density environments (Virgo, the field and
%lower density environments) than ours.
Furthermore, we find a hint towards an asymmetric scatter for the
low-$\sigma$ galaxies where more galaxies have higher Balmer
line-strengths (see last part of Section \ref{i-sigma}).
Similarly, \citet[][hereafter CRC03]{Caldwell03} also find a large
asymmetric scatter for the low-$\sigma$ galaxies, although in the
opposite direction.  It is possible that this effect is environmental,
since CRC03 sample includes galaxies in lower density environments
(Virgo, the field, and lower density environments) than ours.
%RRRRRRRRRRRRRRRRRRRRRRRRRRRRRRRRRRRRRRRRRRRRRRRRRRRRRRRRRRRRR
In either case, we extend the H$\beta - \sigma$ relation to
low-$\sigma$ galaxies down to 30 km s$^{-1}$ in $\sigma$, or by 0.2
dex when compared to CRC03.  We confirm that H$\beta$ and $\sigma$ are
anti-correlated, and we find a hint of an asymmetric scatter in the
low-$\sigma$ regime.

%-------------------------------------------------------------------------
\subsubsection{Mg$_2$ vs.~$\sigma$}

Perhaps the most studied $I-\sigma$ relation is the one of
magnesium, in particular, the Mg$_2$ index \citep*[just to name a
few studies:][]{Terlevich81, Gorgas90, Guzman92, Bender93,
Bernardi98,
  Colless99, Jorgensen99, Concannon00, Kuntschner00, Poggianti01,
  Proctor02, Worthey03, Mehlert03, Sanchez06a}.  The tight relation
between Mg$_2$ and $\sigma$ has been interpreted as evidence that all
elliptical galaxies have a low dispersion in age \citep{Bender93,
  Bernardi98}.  Furthermore, the parameter driving this relation has
been under much debate.  Originally, studies argued that the Mg$_2 -
\sigma$ relation depended mostly on metallicity \citep{Forbes98,
  Terlevich99}, while age and relative abundances of different heavy
elements have recently been proposed to also influence this relation
\citep*{Trager98, Jorgensen99, Trager00b, Kuntschner01, Poggianti01a,
  Mehlert03, Caldwell03, Thomas05, Sanchez06a}.

Our Mg$_2 - \sigma$ relation is consistent with other studies in both
the slope (Table \ref{fitexytabl3}) and the low intrinsic scatter.
Although, we note a slightly larger dispersion around the line for
low-$\sigma$ galaxies, for the first time, we confirm that this
relation is robust for the entire range (30--250 km s$^{-1}$) in
$\sigma$.

%-------------------------------------------------------------------------
\subsubsection{Index-$\sigma$ trends in general}

A surprising result in this paper are the different shapes of
$I-\sigma$ relations once we include galaxies with $\sigma < 100$ km
s$^{-1}$.  Most studies find linear $I-\sigma$ relations for
early-type galaxies.  However, these samples are limited to galaxies
with $\sigma \gtrsim 50$ km s$^{-1}$.  Although our Coma cluster data
is not larger than most other studies, it is unique as our sample
contains $\sim 40$ galaxies with $30 \leqslant \sigma < 100$ km
s$^{-1}$ (see Paper I).

%RRRRRRRRRRRRRRRRRRRRRRRRRRRRRRRRRRRRRRRRRRRRRRRRRRRRRRRRR
%The different shapes of $I-\sigma$ trends found in this work are a
%direct consequence of the dramatically larger scatter for low-$\sigma$
%galaxies.  

We find that for the majority of indices, the low-$\sigma$ galaxies
exhibit a larger scatter in the $I-\sigma$ relations than the
high-$\sigma$ galaxies.  \citet*{Concannon00} also found that the
scatter in
%RRRRRRRRRRRRRRRRRRRRRRRRRRRRRRRRRRRRRRRRRRRRRRRRRRRRRRRRR
index-$\sigma$ relations is larger for low mass galaxies.  Their
interpretation of this result for the H$\beta - \sigma$ relation is
that the low mass galaxies have experienced a more varied star
formation history and have a larger spread in age.  Similarly,
\citet*{Sanchez06a} show that the scatter in the $I-\sigma$ relations
is mainly a consequence of the element abundances varying with age.
We investigated whether the shapes of the $I-\sigma$ relations are
related to the variations in individual element abundances using
\citet{Tripi95}, \citet{TMB03} and \citet{Korn05}.

Indices in the left column (group Ia) of Figure \ref{index-sigma-fig}
have strong Fe-dependence in common, except for Fe4531 and G4300 which
mostly depend on Ti and to a lesser extent on Fe.  Group Ib contains
indices which depend on both, $\alpha$-peak elements and Fe.  In group
II we again find a mixture of elements driving the indices.  Carbon
and $\alpha$-peak elements, Mg and O in particular, do appear to
influence most indices in this column, with the exception of
[MgFe]$\arcmin$ which does not depend much on the [$\alpha$/Fe] ratio
\citep*{TMB03}.  Finally, H$\gamma_{A,F}$ indices depend on the
[$\alpha$/Fe] ratio although this dependency diminishes with
increasing metallicity \citep{Korn05}, while the H$\beta$ index is
moderately influenced by elemental abundance ratios.  If the abundance
ratios are what drives the relation between the higher order Balmer
lines and $\sigma$, then there is a possibility that the larger
scatter of these indices toward the low-$\sigma$ galaxies is caused by
the decreased metallicity.  In conclusion, we do not find any clear
correlations between the shape of the $I-\sigma$ relations with the
element abundance driving the indices, neither with $\alpha$- nor
Fe-peak elements.

\citet*{Poggianti01} find that the slopes of the index-magnitude
relations can be explained by trends between the age and metallicity
with luminosity.  These relations are an alternate form of the
$I-\sigma$ relations, since magnitude and $\sigma$ are related via the
Faber-Jackson relation.  \citet*{Sanchez06a} find that variations in
these two parameters are not sufficient to explain the $I-\sigma$
slopes.  They conclude that a likely explanation for the different
$I-\sigma$ relations could be the relative abundance of elements
[Mg/Fe], [C/Fe], and [N/Fe], as already speculated by other authors
\citep*{Worthey92, Greggio97, Jorgensen97, Kuntschner00, Trager00b,
TMB02, Mehlert03, Thomas05}.  More specifically, \citet*{Sanchez06a}
show that the $I-\sigma$ slopes are best reproduced when the
$\alpha$-peak elements change more than the Fe-peak elements and, the
[Mg/Fe] and [N/Fe] ratios change more than the rest of the alpha
elements with $\sigma$.  We find no clear evidence in support of these
results.

The answer to the different shapes of $I-\sigma$ relations may lie in
the finding that C$_2$4668, Mg$_1$, Mg$_2$ and Mg$_b$, {\bf all} in
group II (exhibiting robust linear relations), are independent of the
micro-turbulent velocity of stellar atmospheres \citep*{Tripi95}. This
is unusual, since most other indices depend on this parameter.  In
fact, according to \citet*{Tripi95}, changing the micro-turbulent
velocity of stellar atmospheres just by 1 km s$^{-1}$ causes changes
in the indices which are more significant than if one were to double
the metal abundance. Hence, the shape and/or tightness of the
$I-\sigma$ relations for metallic lines may be determined by how much
an index depends on the micro-turbulent velocity of the underlying
stellar atmospheres.

%*************************************************************************
%*************************************************************************
\section{AGES, METALLICITIES AND ELEMENT ABUNDANCES OF EARLY-TYPE GALAXIES}
\label{models}

One of the main goals of this paper is to derive ages and
metallicities of faint early-type galaxies in the central region
of the Coma cluster.  This is possible through evolutionary
stellar synthesis models.  We use index-index diagrams to compare
the observed line-strengths with the stellar population models
(SPM) of \citet*[][hereafter TMK04]{TMK04}, an extension of TMB03,
to derive the ages, metallicities and abundance ratios for our
sample of galaxies.

The models that we use are based on the evolutionary population
synthesis code from \citet{Maraston98}.  They account for element
ratio changes based on the response functions from \citet{Korn05}
via a method similar to the one introduced by \citet{Trager00a}.
The TMB03 models span a range in age between 1 and 15 Gyr, total
metallicity, [Z/H], from $-2.25$ to 0.65, and the $\alpha$-ratio
values of $0.0-0.5$.  Using an age-sensitive index vs. a
metallicity-sensitive index with the models allows for a
derivation of ages, metallicities and $\alpha$-ratios of galaxies.

We use H$\beta$ as the main indicator of age.  This line-strength
is only marginally sensitive to the $\alpha$/Fe ratio, while the
higher order Balmer lines are significantly affected by
[$\alpha$/Fe] at super-solar metallicities (TMK04).  Furthermore,
H$\beta$ is a prominent feature in the spectra of our galaxies. As
a metallicity gauge we use the [MgFe]$\arcmin$ index as defined by
TMB03, albeit it is also dependent on age.  This index is
sensitive to the overall metallicity and, similarly to H$\beta$,
depends little on the [$\alpha$/Fe] ratio.

%%%%%%%%%%%%%%%%%%%%%%%%%%%%%%%%%%%%%%%%%%%%%%%%%%%%%%%%%%%%%%%%%%
%  Hbeta-[MgFe] figure
%%%%%%%%%%%%%%%%%%%%%%%%%%%%%%%%%%%%%%%%%%%%%%%%%%%%%%%%%%%%%%%%%%

We use a combination of the H$\beta$--[MgFe]$\arcmin$, and
$\langle$Fe$\rangle$--Mg$_b$ indices to determine the ages,
metallicities and the $\alpha$-ratios for our Coma galaxies.  However,
before investigating the relations between age, metallicity and
[$\alpha$/Fe] we note that, due to the tilt of the model grids and the
given errors in individual indices, the errors in derived ages and
metallicities are likely to be correlated \citep*{Kuntschner01,
  Terlevich02}.  In order to reduce the error in the line-strength
indices (and therefore the correlated errors in the derived
parameters) we have obtained an average value of each index for
galaxies with similar velocity dispersions (for $\sigma$ bins, see
Table \ref{tablBins}).  These average values binned by velocity
dispersion represent ``average'' or ``binned'' galaxies.

\begin{figure}
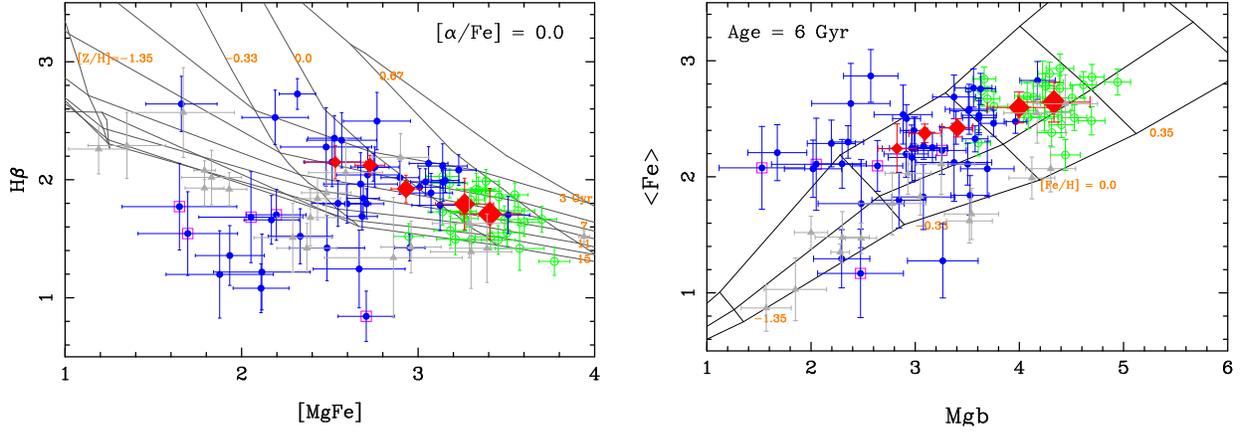

\centering
\includegraphics[angle=-90,width=3.1truein]{f5a.eps}
\hspace{0.1in}
\includegraphics[angle=-90,width=3.1truein]{f5b.eps}
\caption{[MgbFe]$\arcmin$ vs. H$\beta$ and Mgb
  vs. $\langle$Fe$\rangle$ plots for age of 6 Gyr plot.  The low- and
  high-$\sigma$ galaxies are represented by the blue solid and green
  open circles, respectively.  We mark the nucleated dEs \citep{GG03}
  with purple open squares.  The large red solid diamonds denote the
  galaxies binned by velocity dispersion where each bin is of an equal
  interval in $\log \sigma$.  The larger diamonds represent the more
  massive galaxies with larger $\sigma$.  The bins do not include
  galaxies which lie outside the model grids.  For comparison, we also
  include Globular Clusters from \citet{Cenarro07} as the grey
  triangles. \label{modelfigs}}
\end{figure}

As a precursory step, we plot our galaxies on top of the model
grids in Figure \ref{modelfigs}.  In the left panel of this
figure, we fixed the $\alpha$-ratio to the solar value according
to the findings of \citet*{Gorgas97}, so that we can determine the
ages and metallicities. While in the $\langle$Fe$\rangle$--Mg$_b$
panel, the age is set to 6 Gyr, as this is an average age of our
low-$\sigma$ galaxies and corresponds to the ages derived from our
paper I.  Once the age is at a fixed value, we can determine the
metallicities and the [$\alpha$/Fe].  We marked the low- and
high-$\sigma$ galaxies with different symbols and also included
the ``average galaxies'' for which the indices are binned by
velocity dispersion.

The Coma cluster galaxies in our sample exhibit a wide range in
both their ages and metallicities.  Further, the more massive
galaxies have, on average, metallicities equal to or larger than
solar, while the low-$\sigma$ galaxies (the three smallest
diamonds) have on average, subsolar metallicities and younger
ages.  Similar results of wide age and metallicity ranges have
already been noted by other authors in the literature for both the
Coma cluster \citep*{Jorgensen99, Poggianti01, Mehlert03, Nelan05,
Sanchez06b}, and for the lower density environments
\citep*{Caldwell93, Jorgensen97,
  Trager98, Trager00a, Nelan05, Sanchez06b, Bernardi06}.

Our sample seems to split around [MgFe]$\arcmin \sim 3$, or more
precisely around the solar metallicity.  This is in agreement with
\citet*{Poggianti01} who find that their faint Coma cluster galaxies
are divided into two groups, one being metal-rich and the other one
metal-poor.  In our sample, galaxies with the super-solar
metallicities are predominantly high-$\sigma$ early-types.  A group of
low-$\sigma$ galaxies is also present in this regime and these
galaxies are on average younger than the high-$\sigma$ galaxies.  In
contrast, all the other low-$\sigma$ galaxies exhibit sub-solar
metallicities.  This result may imply two different formation
mechanisms for low-$\sigma$ early-type galaxies within the Coma
cluster.  Alternatively, galaxies entering the cluster environment at
different epochs would be stripped from their gas at different
evolutionary stages, possibly explaining the metallicity differences.
An investigation of these parameters and their dependence on the
position within the cluster and the cluster environment is a topic of
our future papers.

On the other hand, \citet{Mehlert03} and \citet{Thomas05} find
that their samples of early-type galaxies split into two
subclasses at H$\beta \sim 2$ \AA~where the younger subclass has
solar or higher metallicities on the H$\beta$--[MgFe]$\arcmin$
plot.  \citet{Mehlert03} find that the `young clump' (their Figure
4) is dominated by S0 galaxies, rather than Es. \citet{Thomas05}
attribute this division to either younger stellar populations or
blue horizontal branch stars.  Here too, we can argue that such a
division exists for the low-$\sigma$ galaxies in our sample, but
not for the more massive Es.  We denote a group of galaxies with
old ages and low metallicities, while the remaining low-$\sigma$
galaxies in our sample have intermediate ages and a large range in
metallicity.  This suggests that some low mass early-type galaxies
harbor younger stellar populations, while the others are old and
metal poor.  We also found no correlation with morphology for this
result.

%%%%%%%%%%%%%%%%%%%%%%%%%%%%%%%%%%%%%%%%%%%%%%%%%%%%%%%%%%%%%%%%%%
%  <Fe>-Mgb figure
%%%%%%%%%%%%%%%%%%%%%%%%%%%%%%%%%%%%%%%%%%%%%%%%%%%%%%%%%%%%%%%%%%

The right panel of Figure \ref{modelfigs} shows a relation of the
Mg$b$ and $\langle$Fe$\rangle$ indices overlaid with models.  When
the age is fixed, it is possible to derive the [$\alpha$/Fe]
ratios for these galaxies.  Similar to the [MgFe]$\arcmin$ vs.
H$\beta$ plot, there is a division in the sample between galaxies
around the solar metallicity in this figure. Majority of galaxies
with super-solar metallicities are high-$\sigma$ galaxies.  They
cluster around [$\alpha$/Fe]$=0.3$ which is consistent with the
well-known overabundance of Mg among Es, i.e., a depression of Fe
with respect to the solar values \citep*{Trager00b}.  Although the
low-$\sigma$ galaxies show a wider range in $\alpha$--ratios
(0.0--0.5) than their more massive counterparts, the majority of
low-$\sigma$ galaxies, with the exception of a few objects, have
low [$\alpha$/Fe]$ \lesssim 0.2$.  This result indicates that the
low-mass galaxies have had a more extended star formation history
\citep{Gorgas97}

%%%%%%%%%%%%%%%%%%%%%%%%%%%%%%%%%%%%%%%%%%%%%%%%%%%%%%%%%%%%%%%%%%%%%%%%%%
\subsection{Off-Grid Galaxies}
\label{offgridSect}

\begin{figure}
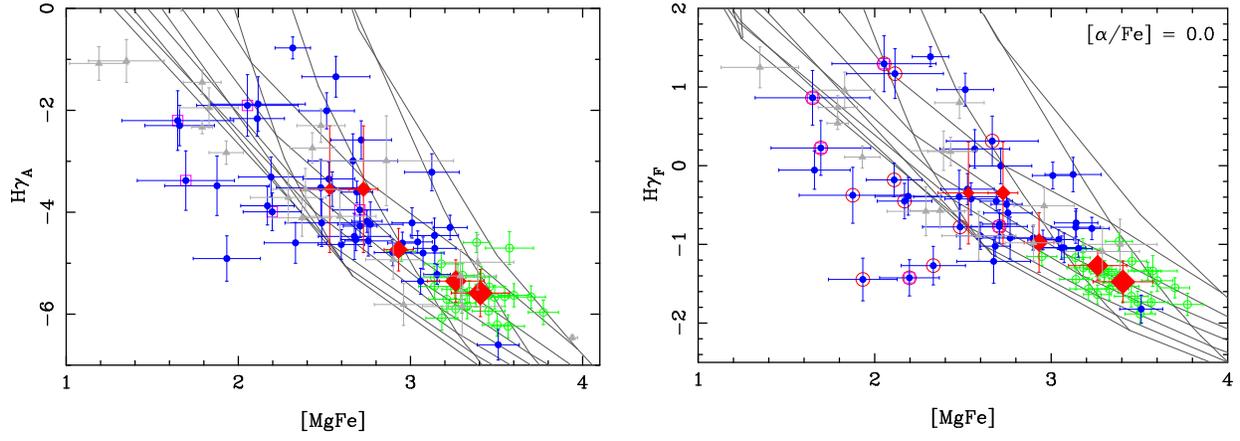

\centering
\includegraphics[angle=-90,width=3.1truein]{f6a.eps}
%HgA_tMgFe.eps
\hspace{0.1in}
\includegraphics[angle=-90,width=3.1truein]{f6b.eps}
\caption{[MgbFe]$\arcmin$ vs. H$\gamma_A$ and H$\gamma_F$ plots.  The
  low- and high-$\sigma$ galaxies are represented by the blue solid
  and green open circles, respectively and Globular Clusters by grey
  triangles.  We mark galaxies which are not fitted by the models from
  Figure \ref{modelfigs} with red open circles.  \label{Hgfig}}
\end{figure}

Our sample contains 13 dE/dS0 galaxies that are not fitted by the
models when using H$\beta-$[MgFe]$\arcmin$ indices (bottom left corner
of Figure \ref{modelfigs}).  To see whether we can recover their ages,
we plot H$\gamma_A$ and H$\gamma_F$ with [MgFe]$\arcmin$ in Figure
\ref{Hgfig}.  Even when we use a different Balmer line strength,
H$\gamma$, 8 out of 13 galaxies still lie off the model grids.  All of
these galaxies are low mass with $30 \leqslant \sigma < 70$ km
s$^{-1}$.  In the following paragraphs we perform a number of tests to
determine whether these galaxies truly have such low values of the
H$\beta$ index.

\begin{figure}
\includegraphics[width=6.5in]{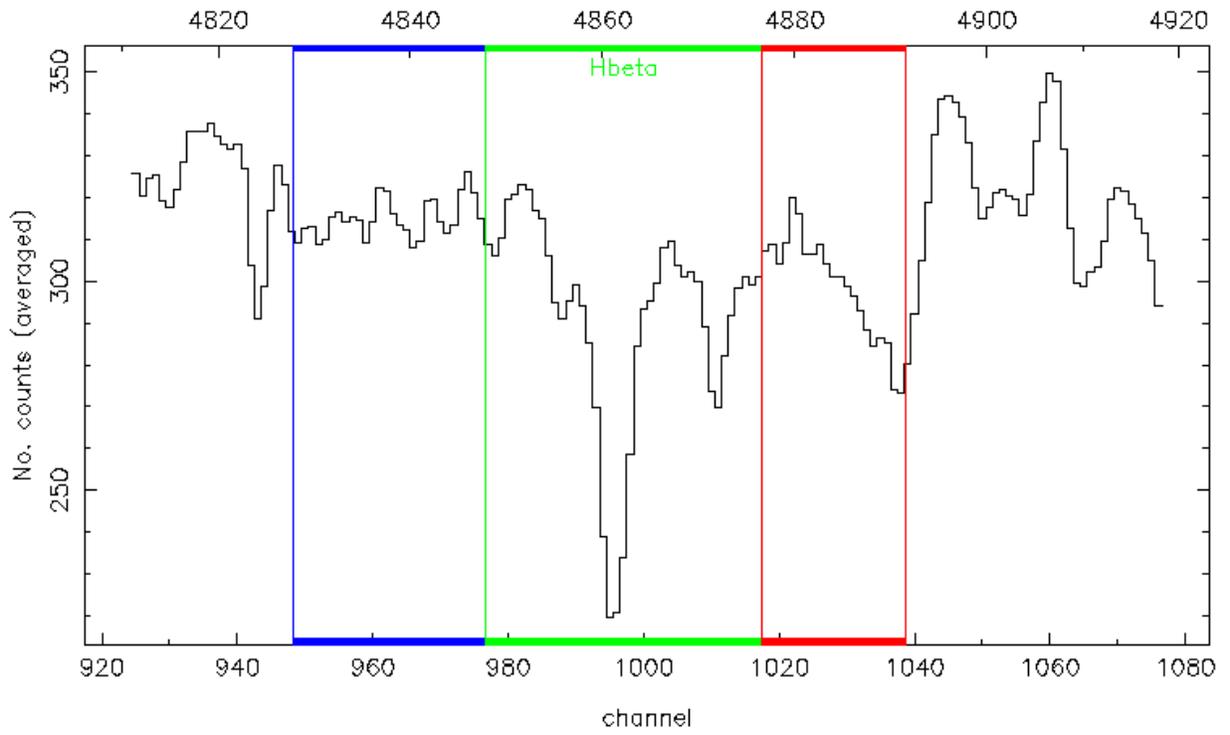}
%  \centerline{\psfig{figure=f6.eps,width=4truein}}
%  \centerline{\psfig{figure=dE_13outliers_paper.ps,width=4truein}}
  \caption{H$\beta$ of the off-grid galaxies combined.  We stacked
    the spectra of 13 galaxies which did not fit on the models for
    [MgFe]$\arcmin$ vs. H$\beta$ plot.
    \label{hbeta}}
\end{figure}

First, we checked for a possibility of nebular emission in the
H$\beta$ feature, as it would make this line strength appear weaker,
i.e. yielding older ages.  Aside from the test that we have already
performed by dividing each spectra with it's optimal template (see
$\S$ \ref{index measurements}), we also stacked the spectra of these
galaxies together (since they have a small range in $\sigma$) at the
original resolution (FWHM = 1.9 \AA) and checked for any possible
emission in the H$\beta$ absorption feature (Figure \ref{hbeta}) which
was not detected previously.  At this resolution, we do not find any
contamination of the off-grid galaxies by nebular emission.

%RRRRRRRRRRRRRRRRRRRRRRRRRRRRRRRRRRRRRRRRRRRRRRRRRRRRRRRRR
Second, we considered possible sky subtraction and scattered light
problems.  If scattered light was the cause of the low H$\beta$ index
values, not only Hbeta but the rest of the indices should be affected
as well.  Further, wrongly subtracted sky levels should also lead to
asymmetrical residuals at the locations of the bright sky
lines. However, this was not the case in our spectra, and we exclude
sky subtraction and the scattered light correction as causes of the
low H$\beta$ index values.
%RRRRRRRRRRRRRRRRRRRRRRRRRRRRRRRRRRRRRRRRRRRRRRRRRRRRRRRRR

Third, we checked whether these off-grid galaxies were consistent
with the position of globular clusters (hereafter GCs) on the
[MgFe]$\arcmin$ vs. H$\beta$ plot.  If the models would extend to
these galaxies, they would correspond to very old and very metal-poor
objects similar to GCs or they could also be ``primordial'' as
suggested by \citet{Rakos04}.  Figure \ref{modelfigs} shows a
possibility that some of the off-grid galaxies are consistent with
GCs, while a number of these galaxies lie in a region even older and
more metal poor than GC.

\begin{figure}
\includegraphics[]{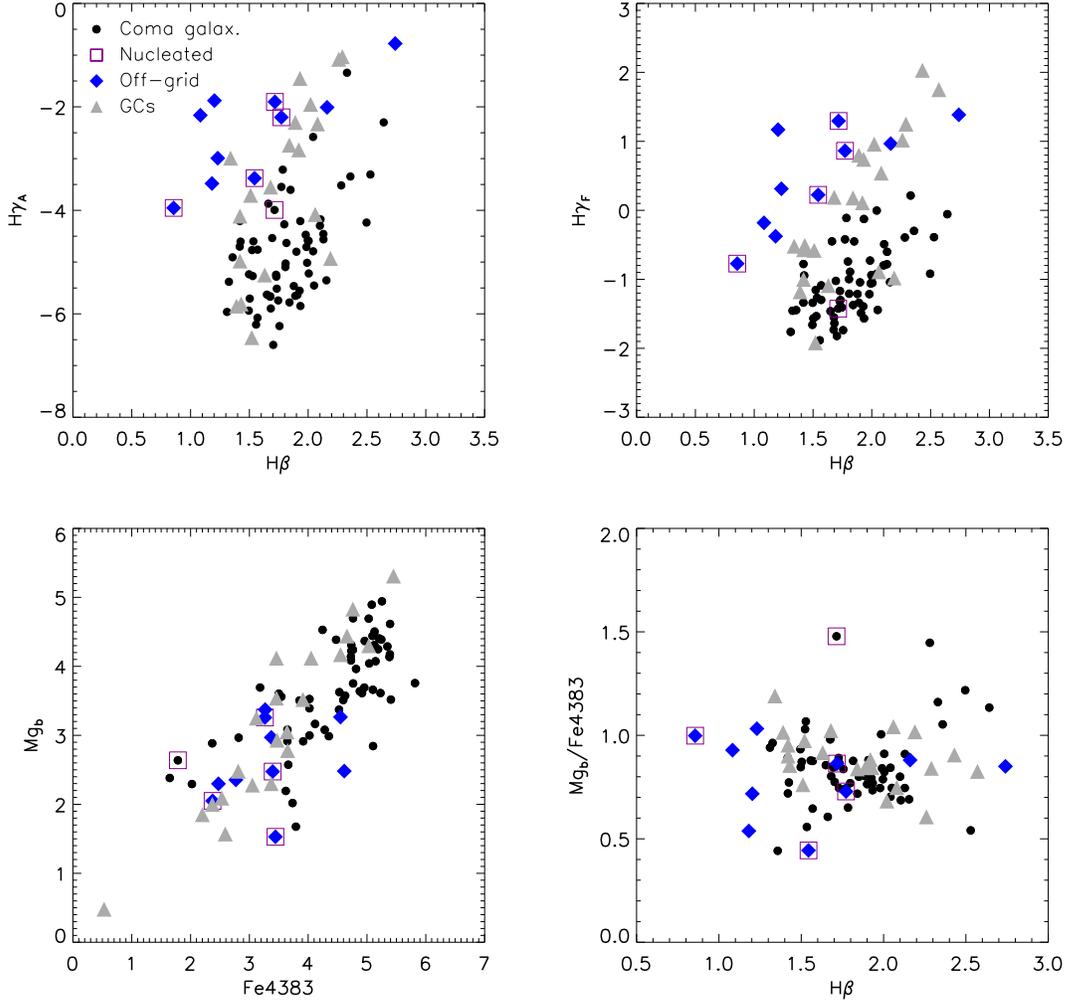}
%  \centerline{\psfig{figure=f7.eps}}
%  \centerline{\psfig{figure=outlying_GCs_paper.eps}}
  \caption{Testing the off-grid galaxies.  The top left and right
    plots show H$\beta$ index vs. the higher order Balmer lines
    H$\gamma_A$ and H$\gamma_F$.  The bottom plots investigate whether
    there are any inconsistencies in the [Mg/Fe] between the off-grid
    galaxies with the other galaxies in the sample and GCs
    \citep{Cenarro07}.  The black filled circles represent the Coma
    galaxies from our sample where the blue diamonds are off-grid
    galaxies.  Nucleated dEs are marked by purple open squares and the
    GCs by grey triangles.
    \label{testoffgridfig}}
\end{figure}

We also checked whether the H$\beta$/H$\gamma_{A,F}$ ratios for the
off-grid galaxies are consistent with the other galaxies in our sample
and with GCs from \citet{Cenarro07}.  This is shown in Figure
\ref{testoffgridfig}.  The off-grid galaxies deviate noticeably from
the H$\gamma_A$ and H$\gamma_F$ with H$\beta$ plots when compared to
other Coma galaxies in our sample and the GCs.  Interestingly, the
off-grid galaxies show no deviations in the [Mg/Fe] plot.  This points
toward a possibility of some problems in the measurements of the
Balmer lines for these off-grid galaxies, which is not necessarily
true for the other indices.

%RRRRRRRRRRRRRRRRRRRRRRRRRRRRRRRRRRRRRRRRRRRRRRRRRRRRRRRRR

Finally, we were able to compare our spectra with those of S08 (also
private communication with Russell Smith).  The comparison showed that
the spectra of the off-grid galaxies likely suffered from some
spurious high-resolution frequency patterns found only in the Balmer
continuum bands used to measure the index.  However, this is not the
case for the rest of our galaxies which have good quality.  To err on
the safe side we exclude these galaxies from the analysis in this
paper.

%In conclusion, we exclude these galaxies from the
%analysis in this paper.
%RRRRRRRRRRRRRRRRRRRRRRRRRRRRRRRRRRRRRRRRRRRRRRRRRRRRRRRRR
%The final test for the off-grid galaxies was to compare our index
%measurements and spectra to that of \citet*[][private communication
%with Russell Smith]{Smith08}.  Indeed, the off-grid galaxies clearly
%deviate in the plot of H$\beta$ measured here and in \citet*{Smith08}.
%We have been unable to find the cause for such differences.
%Conservatively, we exclude these galaxies from the analysis in this
%paper.

%-----------------------------------------------------------------
%%%%%%%%%%%%%%%%%%%%%%%%%%%%%%%%%%%%%%%%%%%%%%%%%%%%%%%%%%%%%%%%%%

\subsection{Method of  Deriving SPM Parameters}

We use an iterative procedure similar to \citet{Thomas05} to derive
the ages, metallicities and $\alpha$-ratios for our sample of
early-type galaxies in the Coma cluster.  This procedure consists of,
first, determining the age and metallicity of each galaxy by
interpolating the model grids for the [MgFe]$\arcmin-$H$\beta$ plot,
at a given $\alpha$-ratio.  This particular combination of indices has
low sensitivity to abundance ratios and is, therefore well suited for
determining the other two SPM parameters, age and metallicity.  Then,
we fix the age as it was derived in the first step, and we derive the
[$\alpha$/Fe] and metallicity with $\langle$Fe$\rangle-$Mg$_b$ index
combination.  This two-step procedure is repeated until the
metallicities derived from [MgFe]$\arcmin-$H$\beta$ and
$\langle$Fe$\rangle-$Mg$_b$ match well (i.e. better than $15 \%$
difference).

We used the same iterative procedure for deriving the error
ellipse in age, metallicity and $\alpha$-ratio values for each
galaxy.  We treated the extremes of the index uncertainties as
individual values. Conservatively, we chose the highest value of
the error ellipse for the uncertainty in the SPM parameter.  The
ages, metallicities and $\alpha$-ratios together with their
respective errors are shown in Tables \ref{mastertable1} and
\ref{mastertable2}, while these parameters for the binned galaxies
are in Table \ref{tablBins}.

% range in SPMs
%            >100              <100
% age:       3-15               2-12
% met: -0.211 _ 0.378     -0.714 _ 0.26
% alf:    0.07-0.44         0.005-0.481

Our data of early-type Coma galaxies span a wide range in all SPM
parameters.  The low-$\sigma$ galaxies have on average: lower ages,
$6.3 \pm 0.6$ vs. $9.4 \pm 0.7$ Gyr for high-$\sigma$ galaxies; lower
metallicities, $-0.082 \pm 0.042$ vs. $0.121 \pm 0.028$ dex;
and slightly lower [$\alpha$/Fe], $0.18 \pm 0.02$ vs. $0.23 \pm
0.02$, closer to the solar value of 0.0.  Here, we excluded the
galaxies which lie off the model grids as discussed in $\S$
\ref{offgridSect}.

%Our data of early-type Coma galaxies span a wide range in all SPM
%parameters.  The low-$\sigma$ galaxies have on average: lower ages,
%$6.1 \pm 0.1$ vs. $8.9 \pm 0.1$ Gyr for high-$\sigma$ galaxies; lower
%metallicities, $-0.050 \pm 0.003$ vs. $1.131 \pm 0.004$ dex;
%%[Z/H]$_\sun$ {\bf PSB: remove the subscript here..,
%%  [Z/H]$_\sun$=0 :) I would put [Z/H]=-0.07 vs. 0.12 dex; };
%and slightly lower [$\alpha$/Fe], $0.173 \pm 0.002$ vs. $0.238 \pm
%0.002$, closer to the solar value of 0.0.  Here, we excluded the
%galaxies which lie off the model grids as discussed in $\S$
%\ref{offgridSect}.

%*************************************************************************
%%%%%%%%%%%%%%%%%%%%%%%%%%%%%%%%%%%%%%%%%%%%%%%%%%%%%%%%%%%%%%%%%%%%%%%%%
%  Model Parameters vs. sigma
%%%%%%%%%%%%%%%%%%%%%%%%%%%%%%%%%%%%%%%%%%%%%%%%%%%%%%%%%%%%%%%%%%%%%%%%%
\section{MODEL PARAMETERS VS. $\sigma$}

In the index-index figures overlaid with models and mentioned in
the previous section, we notice a trend with mass for our
galaxies. Henceforth, we plot the model parameters, age,
metallicity and [$\alpha$/Fe] vs. $\log \sigma$ for the 5 velocity
dispersion bins, (Figure \ref{age_met_sig}).  We also include the
individual galaxies in these plots although we perform the linear
least-squares regression (see Table \ref{tablage_met_sig}) for the
binned data only.  The bin values for the age, metallicity and
[$\alpha$/Fe] were calculated by averaging these parameters for
individual galaxies within each bin.  The errors in the
model-derived average parameters were determined by taking the
standard deviation of the individual parameter values within each
bin.
% and then dividing by the square root of the number of galaxies
%in the bin.  
Although not statistically significant, trends emerge
between the age, metallicity and [$\alpha$/Fe] with $\sigma$, and
we compare them to the same relations from
\citet{Nelan05}.\footnote{We excluded one galaxy from the linear
regression (GMP 2585) since this galaxy is clearly an outlier in
both metallicity- and [$\alpha$/Fe]--$\sigma$ plots and it has an
unusually low metallicity and a high value of [$\alpha$/Fe].}

%%%%%%%%%%%%%%%%%%%%%%%%%%%%%%%%%%%%%%%%%%%%%%%%%%%%%%%%%%%%%%%%%%%%%%%%%%%%
\subsection{Age--$\sigma$}
%------------------------------------------------------------------

\begin{figure}
%%% version with errors as standard dev. of age, met, a/fe and sigma:
\includegraphics[height=6.5in]{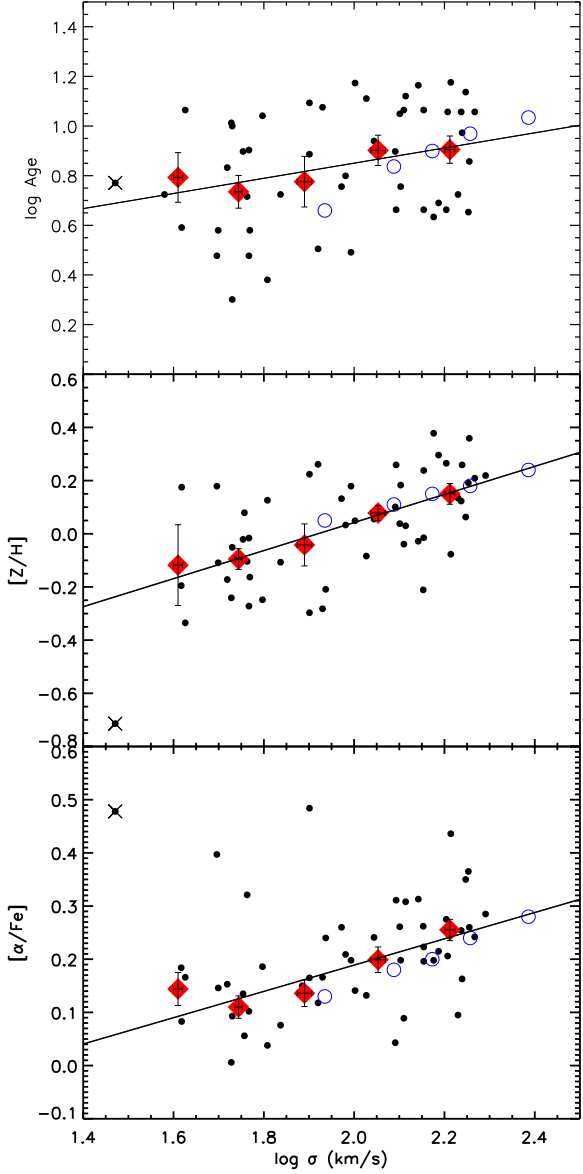}
%%% version with elliptical errors is: age_met_alpha_sigmaEll.ps
%  \centerline{\psfig{figure=age_met_alpha_sigma_stdev.ps,height=6truein}}
%  \centerline{\psfig{figure=f8.eps,height=6truein}}
%  \centerline{\psfig{figure=age_met_alpha_sigma_avgSPM.eps,height=6truein}}
  \caption{$\log \sigma$ versus Age, Metallicity and [$\alpha$/Fe]
    (from top to bottom). The small filled circles are individual
    galaxies in our data set, red diamonds represent ``average''
    galaxies whose indices were binned by velocity dispersion prior to
    deriving the SPM parameters, and the blue open circles represent
    the NFPS data without any offsets applied.
    \label{age_met_sig}}
\end{figure}

The top panel of Figure \ref{age_met_sig} shows the relation between
$\log$ age and $\log \sigma$.  The linear fit between these two
parameters is uncertain due to the large errors in age.  However, both
the binned and the individual galaxies in this figure provide clear
evidence for a trend between age and $\sigma$ where the low-$\sigma$
galaxies display younger ages.

Within the errors, our age--$\log \sigma$ trend is consistent with
\citet{Nelan05}, although there are some differences.  \citet{Nelan05}
find that the age--$\log \sigma$ relation steepens for the low-mass
galaxies.  We do not find this effect, since the age--$\log \sigma$
for these galaxies levels off at $\sim 4$ Gyr in our case.

Whether the age-$\sigma$ relation exists for early-type galaxies or
not is still an unresolved issue in the literature.
\citet*{Jorgensen99}, \citet*{Kuntschner01}, \citet*{Mehlert03},
\citet*{Thomas05}, and \citet*{Sanchez06b} do not find a relation
between these parameters, although results from \citet*{Sanchez06b}
yield a relation for galaxies in low density environments.  However,
at least a trend between age and $\sigma$ is found in the samples of
\citet*{Concannon00}, \citet*{Poggianti01}, \citet*{Caldwell03},
\citet*{Proctor04a}, \citet*{Nelan05}, and \citet*{Bernardi06}.
Additionally, \citet*{Nelan05} derive age-$\sigma$ relations for other
sources in the literature and find them to be in agreement with their
data.

Although we find an age-$\sigma$ trend for our sample of Coma
early-type galaxies, the uncertainties in the age measurements are
large and we cannot confirm a relation between these two
parameters. Nonetheless, we observe that the low-$\sigma$ galaxies
exhibit, on average, younger ages than their more massive
counterparts.

In paper I, we used the scatter in the Color-$\sigma$ relation and
the evolutionary stellar population synthesis models of
\citet{Bruzual03} to estimate the formation epoch for our Coma
galaxies.  We found that, if we assume a strong coordination in
the formation epoch of galaxies in the Coma cluster, most of these
galaxies would have formed about 6 Gyr ago and within a scatter of
1 Gyr.  The results from our previous paper are consistent with
those shown in the age--$\sigma$ panel of Figure
\ref{age_met_sig}.  We find that the average age of low-$\sigma$
galaxies is $\sim 6$ Gyr and the scatter in the $\log \sigma-
\log$ age relation implies a scatter in formation epoch of $\sim
1.5$ Gyr.

%%%%%%%%%%%%%%%%%%%%%%%%%%%%%%%%%%%%%%%%%%%%%%%%%%%%%%%%%%%%%%%%%%%%%%%%%%%%
\subsection{Metallicity--$\sigma$}

We also find a trend between metallicity and $\log \sigma$ (shown in
the middle panel of Figure \ref{age_met_sig}).  The high-$\sigma$
early-type galaxies tend to be more metal-rich than the low-$\sigma$
galaxies which also exhibit a larger range in their metallicities.

Our derived slope, [Z/H] $\propto \sigma^{0.53 \pm 0.11}$, is in
agreement with a number of studies which find a fairly robust
metallicity--$\sigma$ relation \citep*{Kuntschner01, Mehlert03,
  Nelan05, Thomas05, Sanchez06b, Bernardi06}.  Furthermore, our
metallicity--$\sigma$ trend is in good agreement with that of
\citet*{Nelan05} which is also shown in Figure \ref{age_met_sig}.  We
extend this trend to galaxies with $\sigma = 30$ km s$^{-1}$ with no
evidence for a change of slope or offset.

% Slopes:  (values form Nelan05)
%  sigma^0.42 {Thomas05}
%       ^0.76 {Trager00a}
%       ^0.53 {Nelan05}
%       ^0.55  we

%%%%%%%%%%%%%%%%%%%%%%%%%%%%%%%%%%%%%%%%%%%%%%%%%%%%%%%%%%%%%%%%%%%%%%%%%%%%
\subsection{$[\alpha$/Fe$]-\sigma$}

The [$\alpha$/Fe] also increases with increasing $\sigma$ (bottom
panel in Figure \ref{age_met_sig}).  The low mass early-type galaxies
exhibit $\alpha$-ratios closer to the values in the solar
neighborhood, although the trend is suggestive of [$\alpha$/Fe] $> 0$
even at the lowest $\sigma$.  The higher mass galaxies have an
overabundance of [$\alpha$/Fe].

A relation between [$\alpha$/Fe] and $\sigma$ has already been noted
by a number of authors in the literature \citep*{Trager00a,
Kuntschner01, Proctor02, Mehlert03,Thomas05, Bernardi06}, although
conflicting with \citet*{Proctor04a}.  In accordance to the former
studies, we find that the $\alpha$-ratio increases with increasing
velocity dispersion.  However, due to large uncertainties, we can only
confirm a trend and not a correlation between these parameters.

Our [$\alpha$/Fe]--$\sigma$ slope of $0.25 \pm 0.05$ is consistent
with the slopes derived by other authors who find $\sim 0.3$
\citep{Trager00a, Thomas05, Nelan05}.  We also note that there are
3 objects in the [$\alpha$/Fe]--$\sigma$ plot with $\alpha$-ratios
that are quite large (GMP 2306, GMP 3855 and GMP 3780).

%Our [$\alpha$/Fe]--$\sigma$ slope of $0.25 \pm 0.05$ is shallower than
%the slopes derived by other authors who find $\sim 0.3$
%\citep{Trager00a, Thomas05, Nelan05}.  However, we note that there are
%3 objects in the [$\alpha$/Fe]--$\sigma$ plot with $\alpha$-ratios
%that are quite large.  These ``outlying'' points are affecting the
%average values of [$\alpha$/Fe] at a given $\sigma$ toward smaller
%values and, therefore the slope.  If we were to exclude these three
%galaxies (GMP 2306, GMP 3855 and GMP 3780), our slope would be $0.22
%\pm 0.06$ which is consistent with other authors.

% Slopes:  (values form Nelan05)
%  sigma^0.36 {Thomas05}
%       ^0.33 {Trager00a}
%       ^0.31 {Nelan05}
%       ^0.34  we

In general, super-solar $\alpha$-ratios denote that galaxies formed
quickly, i.e., on short star formation time-scales, and at high
redshifts \citep*{Matteucci94}.  Therefore, an [$\alpha$/Fe]--$\sigma$
trend suggests that the low-$\sigma$ galaxies had more extended star
formation histories than their massive counterparts where new stars
formed from already metal-enriched environment \citep*{Gorgas97}.  In
fact, a couple of mechanisms explaining the extended star formation
histories of the low mass galaxies already exist.  One involves UV
background radiation which can extend the duration of star formation
by suppressing cooling more effectively in low mass galaxies
\citep{Kawata01}.  While the other uses a combination of cooling, star
formation, energy feedback, and chemical evolution to extend the star
formation history of these galaxies \citep*{Chiosi02}.
%-----------------------------------------------------------------------

%*************************************************************************

\section{TRENDS BETWEEN THE AGE, METALLICITY AND $\alpha$--RATIO}

\begin{figure}
  % \centerline{\psfig{figure=age_vs_z_vs_alph_sigma.ps,width=6truein}}
  \includegraphics[width=6.5in]{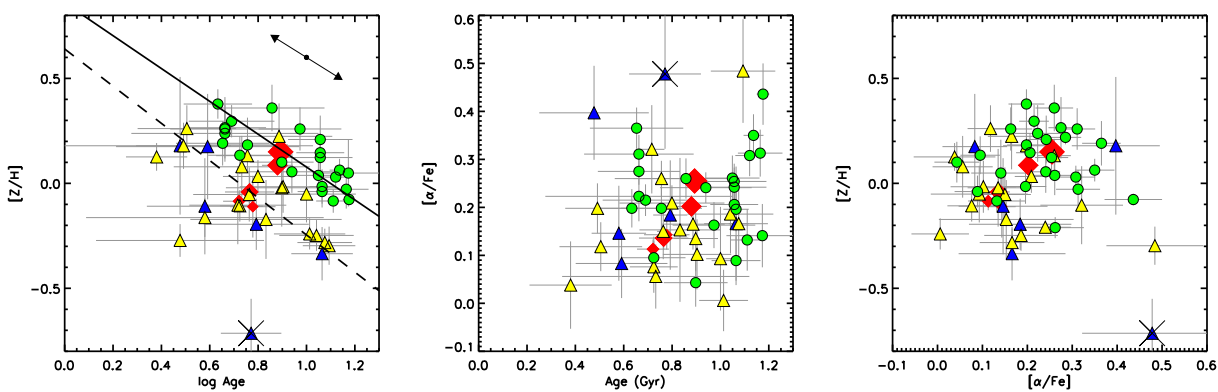}

%  \centerline{\psfig{figure=f9.eps,width=\textwidth}}
%  \centerline{\psfig{figure=age_vs_z_vs_alph_sigma_slopes.ps,width=\textwidth}}
  \caption {Metallicity vs. age, [$\alpha$/Fe] vs. age, and
    metallicity vs. [$\alpha$/Fe] for individual galaxies.  The green
    circles represent galaxies with $100 \leqslant \sigma < 185$ km/s;
    the yellow triangles mark the galaxies with $50 \leqslant \sigma <
    100$; and the blue triangles are galaxies with $30 \leqslant
    \sigma < 50$.  The linear fit (left panel) for the high-$\sigma$
    galaxies is marked with a solid line, while the fit to the
    low-$\sigma$ galaxies ($\sigma < 100$ km s$^{-1}$) is marked by a
    dashed line.  We excluded the galaxy with the lowest metallicity
    from the linear regression (it is marked with a cross).  The
    arrows in the top right corner of the age-metallicity plot
    represent the average correlated error ellipse for age and
    metallicity.  The arrows in the age-metallicity plot represent 1
    $\sigma$ correlated error.
    \label{figAZA}}
\end{figure}

In Figure \ref{figAZA} we investigate relations between metallicity
and age, [$\alpha$/Fe] and age, and metallicity and [$\alpha$/Fe] for
our sample of Coma cluster early-type galaxies.  We do not find any
relations between the model parameters for all the galaxies in our
sample.  However, when we examine the difference between the high- and
low-$\sigma$ galaxies, we find trends in the age-metallicity and the
metallicity--[$\alpha$/Fe] plots.

We do not find any correlations between age and [$\alpha$/Fe], nor
between metallicity and [$\alpha$/Fe] even at a fixed velocity
dispersion.  However, we do note a weak tendency for the high-$\sigma$
galaxies to have higher $\alpha$-ratios and to be more metal rich than
the low-$\sigma$ galaxies.  This effect is stronger in the
\citet{Michielsen07} sample (their Figure 7) whose data extend to
lower metallicities and are in lower density environment than our Coma
cluster galaxies.  Assuming that our and the \citet{Michielsen07}
sample span a similar range in $\sigma$, and that the
metallicity-abundance trends are not due to correlated errors, there
is a possibility that this effect, too is due to the environment.

The relation between the age and metallicity at a given $\sigma$
was first discussed by \citet{Trager00b}.  We study the
possibility of such relations for our low- and high-$\sigma$
galaxies.  The Spearman Rank coefficients (Table
\ref{tablage_met}) imply that the relations between the age and
the metallicity exist for the two $\sigma$-sub-samples, while no
correlations were found for either the age--[$\alpha$/Fe] nor the
metallicity--[$\alpha$/Fe]. Note that the one galaxy, marked by a
cross in Figure \ref{figAZA}, which we excluded from the Spearman
Rank test and linear regression is a galaxy with the lowest
$\sigma = 30$ km s$^{-1}$ in our sample.

Existence of an age-metallicity relation where galaxies with younger
ages tend to be more metal rich, implies that these young galaxies
have had multiple star formation episodes and that they had to form
their stars from already enriched gas.  This relation has been noted
in numerous works both in low density environments and clusters
\citep[to name a few]{Trager98, Jorgensen99, Kuntschner00,
  Kuntschner01, Poggianti01, Terlevich02, Sanchez06b}.  Some studies,
however, suggest that this relation is a consequence of correlated
errors \citep*{Trager98, Trager00b, Ferreras99, Kuntschner01}.
While \citet*{Poggianti01} observe an age-metallicity relation at
all magnitudes in the Coma cluster, \citet*{Sanchez06b} do not
find this relation in Coma and argue that their apparent trend is
due to correlated errors, although they also note a possibility
that their sample is biased toward high-$\sigma$ galaxies, making
the age-metallicity appear flat.

High-$\sigma$ galaxies follow the same relation as the
age--metallicity relation found by \citet{Trager00b}.  To
investigate this, we estimated an average correlated error for our
galaxies from the error ellipse derived with the iterative process
as described in $\S$ \ref{models}.  The direction and size of this
correlated error are shown in the top right corner of Figure
\ref{figAZA}.  As it can be seen, the direction of the correlated
errors coincides with the slope of the age--metallicity relation.
Furthermore, the size of the errors is consistent with the extent
of the distribution of age and metallicity values.  This implies
that the the age--metallicity correlation at a given velocity
dispersion may be simply the result of correlated errors in both
parameters!

The linear regression between the ages and metallicities for the
low- and high-$\sigma$ galaxies shows that the slopes for the two
sub-samples are effectively the same, but offset with different
zero points.  At a given age, the high-$\sigma$ galaxies are more
metal rich by a factor of $\sim 2$ than the low-$\sigma$ galaxies.
Similarly, at a given metallicity, the low-$\sigma$ galaxies are
$\sim 3$ Gyr younger than their more massive counterparts.  This
result also compares well with \citet{Michielsen07} whose data
sample contains dE galaxies from the Virgo cluster and the field.
Both data sets are well anchored to the sample of massive Es from
\citet{Sanchez06b}. However, there are differences in the
distribution of galaxies between our and their sample in the
age-metallicity diagram (their Figure 7). The Coma cluster
galaxies exhibit a similar range of ages to the
\citet{Michielsen07} dEs, but they have a smaller range in
metallicity.  This points towards an environmental dependence of
the age-metallicity relation since the central region of the Coma
cluster is one of the densest regions in the local universe.
Unfortunately, we cannot say this with certainty without knowing
how low in $\sigma$ the \citet{Michielsen07} sample goes, since
the one galaxy in our sample with very low metallicity on the
age-metallicity plot is a galaxy with the lowest $\sigma=30$ km
s$^{-1}$.  The effect of finding the wider range in metallicity in
the lower-density environments than in the core of the Coma
cluster, thus, may be purely due to sampling galaxies with lower
velocity dispersions.

%%%%%%%%%%%%%%%%%%%%%%%%%%%%%%%%%%%%%%%%%%%%%%%%%%%%%%%%%%%%%%%%%%%%%%%%%%
%%%%%%%%%%%%%%%%%%%%%%%%%%%%%%%%%%%%%%%%%%%%%%%%%%%%%%%%%%%%%%%%%%
\section{Conclusions}
%%%%%%%%%%%%%%%%%%%%%%%%%%%%%%%%%%%%%%%%%%%%%%%%%%%%%%%%%%%%%%%%%%%%%%%%%%

In this paper we study the properties of the underlying stellar
populations of faint early-type galaxies in the core of the Coma
cluster.  Our sample is one of the largest homogeneous samples of
cluster dE/dS0 galaxies to date with velocity dispersions measurable
down to $\sigma=30$ km s$^{-1}$.  We present relations between 15 line
strength indices with $\sigma$ and extend them to the dE/dS0 galaxies.
We confirm that, when we include these faint early-type galaxies, the
Mg$_2$ relation and the H$\beta-\sigma$ anti-correlation correspond
well with other sources in the literature.

We find evidence for two types of behaviors between the metallic
indices and $\sigma$'s.  One set of indices (C$_2$4668, Mg$_1$,
Mg$_2$, Mg$_b$, [MgFe]$\arcmin$) exhibit tight linear relations
with $\sigma$, which was also shown by other studies.  The second
group of $I-\sigma$ relations show a break in their slope where this
break is more evident in a sub-set including Ca4227, Fe5015, Fe5335,
$\langle$Fe$\rangle$ than a sub-set with G4300, Fe4383, Ca4455, Fe4531
and Fe5270.
%We find evidence for three different types of behaviors between the
%metallic indices and $\sigma$'s.  One set of indices (C$_2$4668,
%Mg$_1$, Mg$_2$, Mg$_b$, [MgFe]$\arcmin$) exhibit fairly tight linear
%relations with $\sigma$ and steep slope, as shown by other studies.
%The second set of $I-\sigma$ (Ca4227, Fe5015, Fe5335 and
%$\langle$Fe$\rangle$) also shows linear relations, but with shallow
%slopes and notable scatter, while the third group of indices (G4300,
%Fe4383, Ca4455, Fe4531 and Fe5270) displays non-linear relations with
%$\sigma$ once low-$\sigma$ galaxies are included.

We also find that the relations between the Balmer lines and $\sigma$
have negative slopes and are fairly robust, with H$\beta-\sigma$ having the
weakest correlation coefficient.  We find a hint of an asymmetric
scatter in the H$\beta-\sigma$ relation with more galaxies having a
lower H$\beta$ index.  Since the asymmetry is in the opposite
direction from the one found in the lower density environments
\citep*{Concannon00}, it is possible that this effect depends on the
environment.

Although majority of the indices are influenced by the overall
metallicity, we also investigated whether each $I-\sigma$ group of
indices is driven by the $\alpha$- or Fe-peak elements.  We found no
connection between these elements and the occurrence of the break in
the slope for the $I-\sigma$ trends.  However, the non-Balmer indices
with tight $I-\sigma$ relations are all independent of micro-turbulent
velocity of stellar atmospheres.  This may be a main factor which
determines the shape and/or tightness of the $I-\sigma$ relations for
metal-dependent line strengths.

We use the stellar population models to derive ages, metallicities and
[$\alpha$/Fe] for our Coma cluster galaxies.  We find a wide range in
all the SPM parameters where the galaxies with super-solar
metallicities are dominated by the high-$\sigma$ galaxies, while the
low-$\sigma$ galaxies are on average younger, have lower metallicities
and their $\alpha$-ratios scatter around the solar value.  This
implies that the low-$\sigma$ galaxies had some residual star
formation in their recent history, and that their star formation
histories are more extended than they are for the high-$\sigma$
galaxies.  These results are also confirmed by the trends we find
between the age, metallicity and [$\alpha$/Fe] with $\sigma$.

We find that the age-metallicity anti-correlation is most likely due
to correlated errors.  We were able to compare our results with those
of \citet{Michielsen07} who observed dE galaxies in the Virgo cluster
and the field.  Our Coma cluster galaxies seem to have a smaller
range in metallicity when compared to the \citet{Michielsen07} data
set.  Therefore, there is a possibility of an environmental effect on
the metallicity range for dE/dS0 galaxies, unless the
\citet{Michielsen07} data include galaxies with lower velocity
dispersions.

%%%%%%%%%%%%%%%%%%%%%%%%%%%%%%%%%%%%%%%%%%%%%%%%%%%%%%%%%%%%%%
\acknowledgments
%%%%%%%%%%%%%%%%%%%%%%%%%%%%%%%%%%%%%%%%%%%%%%%%%%%%%%%%%%%%%%
We thank J. Cenarro for his help on investigating the spectra for
potential problems and for providing the GC data.  We are also
grateful to Russell Smith for providing the data from the Hectospec
Coma cluster survey, double checking our measurements and error
estimates.  R.~G. gratefully acknowledges University of Yale for the
awarded nights at WIYN telescope. P.~S\'anchez-Bl\'azquez acknowledges
the support by a Marie Curie Intra-European Fellowship within the 6th
European Community Framework Programme.  We also thank the anonymous
referee whose suggestions have improved this work.
%%%%%%%%%%%%%%%%%%%%%%%%%%%%%%%%%%%%%%%%%%%%%%%%%%%%%%%%%%%%%%%%%%%%
\appendix

\section{APPENDIX \\ UNCERTAINTY MEASUREMENTS}
%%%%%%%%%%%%%%%%%%%%%%%%%%%%%%%%%%%%%%%%%%%%%%%%%%%%%%%%%%%%%%%%%%%%%%%%%%%%%
%\section{Uncertainty Measurements}
\label{errors}

We created the ``error spectra'' which allowed us to measure the
uncertainties in our line-strength measurements.  This process
consisted of multiple steps.  First, we produced an optimal
template for each galaxy by combining the 6 template stars (same
procedure as in Paper I).  The template stars are primarily G and
K spectral type and are well matched with the dominating stellar
population of early-type galaxies.  Then, we shifted the optimal
template of each galaxy to the rest frame and broadened this
spectra and that of the galaxy to the resolution of Lick/IDS.
Optimal template is a spectrum of a linear combination of template
stars optimized to fit each galaxy spectrum, (see Paper I for more
details).  We then found a polynomial fit (we chose the 7$^{th}$
order polynomial) to the galaxy's and the template's black body
curve.  This allowed us to calculate the residuals between the
galaxy spectrum and the ``model galaxy'' in the following way:
\begin{equation}
\label{residuals}
{\textrm{R}=\textrm{G}-\left(\frac{\textrm{T}}{\textrm{P$_T$}}
\times {\textrm{P$_G$}}\right)}
\end{equation}
where R represents the residuals, G the galaxy spectrum, T the
optimal template spectrum, P$_G$ the polynomial fit to the galaxy
spectrum, and P$_T$ the polynomial fit to the template. The
quantity in the parentheses is the ``model galaxy'', a spectrum
with the exact shape of the galaxy and high S/N features of the
optimal template.  R refers to the residual noise between the
galaxy spectrum and the model galaxy.

However, to build a true error spectrum we must take into account
the actual noise of the galaxy together with the uncertainty due
to the template mismatch.  We did this by taking the square root
of the galaxy and scaling this spectrum to the average number of
its counts.  Then, we multiplied that quantity by the amount of
noise from the residuals:
\begin{equation}
\label{errorspectrum} {\textrm{E}=\frac{\sqrt\textrm{G}} {\langle
\sqrt\textrm{G} \rangle} \times {\sqrt{\langle \textrm{R$^2$}
\rangle}}}
\end{equation}
Here, E is the error spectrum, $\langle \sqrt{G} \rangle$ is the
average number of counts of the square root of the galaxy, and
$\langle$R$^2$$\rangle$ is the average number of counts of
residuals squared. The error spectrum created in this way is, in a
sense, a spectrum of the exact same shape as the galaxy with the
level of noise which takes into account the Poissonian noise and
the uncertainty due to the mismatch between the templates and the
galaxy spectrum. Additionally, we calculate the uncertainty
associated with the flux calibration.  The program INDEX estimates
the uncertainties associated to flux calibration by measuring the
line-strength indices with all the individual response curves and
computing the standard deviation from these measurements.  In
summary, our uncertainty measurements include the photon noise,
the error due to the mismatch between the galaxy and the template,
and the flux related error.

%%%%%%%%%%%%%%%%%%%%%%%%%%%%%%%%%%%%%%%%%%%%%%%%%%%%%%%%%%%%%%%%%%%%%%%%%%%%%
\section{DETERMINING SIGNAL-TO-NOISE RATIO}
\label{snr}

In our study of both the internal kinematics and stellar
populations of the faint early-type galaxies, we consider only the
galaxies whose average signal-to-noise ratio S/N $\geqslant 15$
(see Paper I). There are 74 galaxies in our central Coma cluster
sample which satisfy this condition.  The S/N which we calculate
for our galaxies is an average value of S/N per pixel

Determining the S/N ratio involved a couple of steps.  A few of
these steps are already described in $\S$ \ref{errors}, where we
also show how one finds the residuals (Equation \ref{residuals})
between the galaxy spectrum and the model galaxy.  We define an
average value of S/N per pixel of each galaxy as:
\begin{equation}
\label{s/n} \textrm{S/N}=\frac{\langle \textrm{G}
\rangle}{\sqrt\textrm{R}^2}
%{\textrm{S/N}=\frac{\langle \textrm{G} \rangle}{\sqrt\textrm{R$^2$}}}
\end{equation}

The value of S/N ratio calculated in this way is an average value
for each galaxy and it includes the mismatch between the actual
galaxy spectrum and that of the model galaxy.

%%%%%%%%%%%%%%%%%%%%%%%%%%%%%%%%%%%%%%%%%%%%%%%%%%%%%%%%%%%%%%%%%%%%%%%%%%%%%%

%*************************************************************************

%%%%%%%%%%%%%%%%%%%%%%%%%%%%%%%%%%%%%%%%%%%%%%%%%%%%%%%%%%%%%%%%%%%%

\clearpage

%%%%%%%%%%%%%%%%%%%%%%%%%%%%%%%%%%%%%%%%%%%%%%%%%%%%%%%%%%%%%%%%%%
%%%%%%%%%%%%%%%%%%%%%%%%%%%%%%%%%%%%%%%%%%%%%%%%%%%%%%%%%%%%%%%%%%
%              TABLES                                            %
%%%%%%%%%%%%%%%%%%%%%%%%%%%%%%%%%%%%%%%%%%%%%%%%%%%%%%%%%%%%%%%%%%
%%%%%%%%%%%%%%%%%%%%%%%%%%%%%%%%%%%%%%%%%%%%%%%%%%%%%%%%%%%%%%%%%%
%  Table of measurements : MASTER table
%%%%%%%%%%%%%%%%%%%%%%%%%%%%%%%%%%%%%%%%%%%%%%%%%%%%%%%%%%%%%%%%%%
\begin{deluxetable}{lrrrrrrrrrrrrrrrrrr}
%lrrrrrrrrrrrrrrrrrr rrrrrrrrrrrrrrrrrrrrrrrrr}
% should be 45 rows
  \tabletypesize{\tiny}
\rotate
  \tablecaption{Index Measurements \label{mastertable1}}
  \tablewidth{0pt}
%  \tablewidth{10in}
  \tablehead{
    \colhead{GMP}&
\colhead{Ca4227}& \colhead{$\delta_{4227}$}& \colhead{G4300}&
\colhead{$\delta_{4300}$}& \colhead{H$\gamma_A$}& \colhead{
$\delta_{H\gamma A}$}& \colhead{H$\gamma_F$}&
\colhead{$\delta_{H\gamma F}$}& \colhead{Fe4383}&
\colhead{$\delta_{4383}$}& \colhead{Ca4455}&
\colhead{$\delta_{4455}$}& \colhead{Fe4531}&
\colhead{$\delta_{4531}$}& \colhead{C$_2$4668}&
\colhead{$\delta_{C_{2}4668}$}& \colhead{H$\beta$}&  \colhead{
$\delta_{H\beta}$}
}
\startdata
%2478&0.972&0.124&4.413&0.214&-3.346&0.246&-0.298&0.147&2.817&0.317&0.343&0.162&2.473&0.234&2.134&0.450&2.359&0.143\\
%2489&1.211&0.077&5.460&0.131&-5.220&0.174&-1.052&0.098&4.814&0.190&1.189&0.097&3.271&0.142&6.442&0.399&2.005&0.091\\
2478 &   0.972&   0.170&   4.413&   0.293&   -3.347&  0.329&   -0.298&  0.198&   2.817&   0.422&   0.343&   0.217&   2.474&   0.313&   2.136&   0.551&   2.352&   0.192 \\
2489 &   1.211&   0.089&   5.460&   0.152&   -5.221&  0.194&   -1.052&  0.110&   4.813&   0.214&   1.189&   0.109&   3.270&   0.159&   6.442&   0.414&   1.994&   0.100	\\
2510 &   1.213&   0.101&   5.217&   0.172&   -5.553&  0.211&   -1.392&  0.123&   5.145&   0.238&   1.231&   0.121&   3.411&   0.176&   6.435&   0.383&   1.911&   0.112	\\
2516 &   1.128&   0.104&   4.811&   0.181&   -4.762&  0.219&   -1.088&  0.127&   4.728&   0.250&   1.285&   0.129&   3.371&   0.187&   6.490&   0.439&   1.571&   0.119	\\
2529 &   1.406&   0.168&   4.619&   0.297&   -3.547&  0.338&   -0.422&  0.204&   3.915&   0.419&   0.919&   0.214&   2.202&   0.315&   3.698&   0.552&   1.800&   0.195	\\
2535 &   1.208&   0.108&   5.517&   0.180&   -5.635&  0.225&   -1.489&  0.131&   4.959&   0.251&   1.206&   0.127&   3.535&   0.183&   6.422&   0.431&   1.902&   0.113	\\
2541 &   1.164&   0.113&   5.427&   0.191&   -5.939&  0.237&   -1.659&  0.138&   5.028&   0.262&   1.252&   0.135&   3.178&   0.196&   6.969&   0.430&   1.490&   0.121	\\
2585 &   0.483&   0.216&   3.690&   0.363&   -2.299&  0.395&   -0.056&  0.243&   2.023&   0.536&   0.440&   0.273&   1.480&   0.401&   5.631&   0.662&   2.642&   0.234	\\
2603 &   0.666&   0.115&   3.827&   0.200&   -0.776&  0.215&   1.385 &  0.126&   2.770&   0.285&   0.976&   0.143&   2.993&   0.211&   4.917&   0.437&   2.727&   0.130	\\
2654 &   1.146&   0.097&   5.689&   0.163&   -5.897&  0.208&   -1.637&  0.119&   5.039&   0.227&   1.218&   0.116&   3.145&   0.169&   6.262&   0.420&   1.672&   0.104	\\
2692 &   1.091&   0.179&   4.449&   0.321&   -2.010&  0.353&   0.967 &  0.209&   3.372&   0.456&   0.370&   0.236&   2.535&   0.342&   3.217&   0.613&   2.151&   0.209	\\
2778 &   1.163&   0.144&   4.968&   0.251&   -4.209&  0.297&   -0.125&  0.170&   4.917&   0.341&   1.155&   0.178&   3.832&   0.253&   6.026&   0.499&   1.937&   0.157	\\
2784 &   0.682&   0.309&   2.370&   0.558&   -3.480&  0.583&   -0.376&  0.355&   4.612&   0.733&   0.120&   0.398&   1.764&   0.584&   5.023&   0.921&   1.198&   0.370	\\
2799 &   1.679&   0.266&   4.730&   0.478&   -1.878&  0.535&   1.170 &  0.316&   4.548&   0.680&   1.028&   0.353&   3.197&   0.510&   4.475&   0.845&   1.218&   0.321	\\
2805 &   1.453&   0.104&   5.237&   0.180&   -5.281&  0.225&   -1.169&  0.128&   4.954&   0.247&   1.277&   0.126&   3.517&   0.182&   7.224&   0.430&   1.728&   0.112	\\
2839 &   1.096&   0.092&   4.799&   0.158&   -4.592&  0.197&   -0.962&  0.112&   4.732&   0.218&   1.324&   0.111&   3.216&   0.162&   7.421&   0.407&   1.987&   0.100	\\
2852 &   1.418&   0.165&   4.861&   0.295&   -4.457&  0.341&   -0.782&  0.204&   4.525&   0.406&   1.809&   0.203&   3.102&   0.301&   5.179&   0.562&   2.116&   0.188	\\
2861 &   1.176&   0.093&   5.606&   0.156&   -5.780&  0.199&   -1.373&  0.114&   5.102&   0.218&   1.461&   0.110&   3.251&   0.161&   7.020&   0.398&   1.830&   0.101	\\
2879 &   1.460&   0.179&   4.964&   0.319&   -3.992&  0.370&   -1.425&  0.230&   1.784&   0.480&   0.504&   0.236&   1.977&   0.350&   2.855&   0.632&   1.703&   0.216	\\
2912 &   1.363&   0.089&   5.341&   0.155&   -5.654&  0.197&   -1.340&  0.113&   5.347&   0.214&   1.372&   0.110&   3.422&   0.159&   6.590&   0.412&   1.873&   0.100	\\
2922 &   1.269&   0.108&   5.369&   0.184&   -5.629&  0.227&   -1.462&  0.132&   5.392&   0.251&   1.435&   0.129&   3.162&   0.187&   7.771&   0.424&   1.640&   0.116	\\
2940 &   1.154&   0.110&   5.389&   0.187&   -5.246&  0.230&   -1.318&  0.134&   4.762&   0.260&   1.349&   0.133&   3.409&   0.192&   6.957&   0.433&   1.718&   0.120	\\
2960 &   1.227&   0.101&   4.709&   0.179&   -4.176&  0.217&   -0.491&  0.124&   4.352&   0.251&   0.855&   0.130&   3.267&   0.188&   4.420&   0.448&   2.105&   0.117	\\
2985 &   0.803&   0.293&   5.232&   0.499&   -3.378&  0.583&   0.225 &  0.347&   3.444&   0.743&   0.187&   0.388&   3.173&   0.554&   4.963&   0.903&   1.544&   0.354	\\
3017 &   1.497&   0.191&   4.559&   0.343&   -3.308&  0.392&   -0.389&  0.239&   3.732&   0.499&   0.261&   0.264&   3.556&   0.369&   3.050&   0.678&   2.528&   0.231	\\
3058 &   1.013&   0.194&   3.291&   0.348&   -2.581&  0.374&   -0.003&  0.226&   3.660&   0.481&   1.243&   0.246&   2.194&   0.373&   2.684&   0.667&   2.041&   0.229	\\
3068 &   0.985&   0.114&   5.739&   0.189&   -6.077&  0.237&   -1.298&  0.137&   5.815&   0.259&   1.420&   0.134&   3.419&   0.196&   5.541&   0.435&   1.569&   0.124	\\
3073 &   1.163&   0.106&   5.594&   0.180&   -6.237&  0.225&   -1.736&  0.132&   5.245&   0.251&   1.186&   0.130&   3.344&   0.189&   7.065&   0.402&   1.738&   0.121	\\
3121 &   1.777&   0.189&   5.167&   0.336&   -4.600&  0.400&   -1.270&  0.245&   5.104&   0.479&   0.587&   0.253&   2.655&   0.366&   3.481&   0.654&   1.521&   0.233	\\
3126 &   0.989&   0.183&   4.408&   0.322&   -3.215&  0.362&   -0.112&  0.218&   5.406&   0.436&   1.202&   0.227&   2.692&   0.338&   3.921&   0.602&   1.784&   0.214	\\
3133 &   1.476&   0.142&   5.774&   0.242&   -6.601&  0.299&   -1.822&  0.177&   5.388&   0.333&   1.202&   0.172&   3.268&   0.251&   5.065&   0.416&   1.702&   0.158	\\
3166 &   1.571&   0.265&   5.255&   0.486&   -3.519&  0.553&   -0.394&  0.336&   1.646&   0.734&   -0.18&3  0.380&   2.322&   0.536&   2.333&   0.868&   2.278&   0.332	\\
3170 &   1.120&   0.116&   5.314&   0.196&   -5.849&  0.239&   -1.567&  0.141&   5.382&   0.273&   1.562&   0.140&   3.250&   0.209&   6.614&   0.371&   1.920&   0.135	\\
3196 &   1.244&   0.223&   4.442&   0.404&   -4.208&  0.462&   -0.778&  0.280&   4.281&   0.572&   0.923&   0.297&   3.618&   0.427&   4.435&   0.738&   1.422&   0.276	\\
3201 &   1.193&   0.105&   5.303&   0.180&   -5.454&  0.224&   -1.444&  0.131&   4.879&   0.253&   1.208&   0.130&   3.032&   0.190&   6.473&   0.441&   2.029&   0.119	\\
3205 &   0.905&   0.193&   5.408&   0.324&   -3.873&  0.377&   -0.450&  0.225&   3.620&   0.455&   1.490&   0.226&   2.700&   0.338&   4.410&   0.614&   1.660&   0.206	\\
3206 &   1.464&   0.137&   5.296&   0.242&   -4.631&  0.290&   -0.891&  0.172&   4.023&   0.349&   1.312&   0.177&   3.399&   0.257&   4.886&   0.520&   1.796&   0.164	\\
3209 &   0.835&   0.317&   5.238&   0.534&   -1.904&  0.604&   1.296 &  0.355&   2.375&   0.814&   -0.18&5  0.437&   0.951&   0.632&   1.016&   1.027&   1.683&   0.385	\\
3213 &   1.084&   0.110&   5.134&   0.186&   -5.236&  0.228&   -1.340&  0.132&   5.206&   0.251&   1.263&   0.129&   3.383&   0.185&   6.073&   0.439&   1.491&   0.114	\\
3222 &   1.123&   0.115&   5.502&   0.195&   -5.706&  0.244&   -1.566&  0.143&   5.095&   0.275&   1.252&   0.142&   3.062&   0.206&   5.270&   0.463&   1.495&   0.130	\\
3254 &   1.370&   0.158&   4.196&   0.280&   -4.703&  0.320&   -1.338&  0.195&   4.735&   0.387&   0.966&   0.198&   3.269&   0.287&   5.554&   0.541&   1.418&   0.186	\\
3269 &   1.025&   0.123&   5.318&   0.210&   -5.354&  0.254&   -1.041&  0.148&   5.229&   0.291&   1.275&   0.151&   2.830&   0.224&   5.656&   0.453&   2.139&   0.141	\\
3292 &   0.344&   0.226&   4.829&   0.377&   -4.908&  0.444&   -1.446&  0.271&   3.790&   0.544&   2.019&   0.263&   3.639&   0.395&   3.535&   0.695&   1.358&   0.253	\\
3296 &   1.185&   0.106&   5.637&   0.176&   -5.663&  0.219&   -1.552&  0.128&   4.764&   0.246&   1.352&   0.124&   3.384&   0.180&   6.478&   0.404&   1.665&   0.110	\\
3312 &   0.637&   0.186&   4.246&   0.312&   -2.162&  0.344&   -0.180&  0.212&   2.473&   0.454&   0.658&   0.227&   2.561&   0.332&   2.610&   0.613&   1.081&   0.207	\\
3313 &   1.135&   0.102&   5.527&   0.174&   -4.536&  0.218&   -1.023&  0.126&   3.648&   0.252&   1.174&   0.125&   2.702&   0.186&   4.584&   0.446&   1.689&   0.114	\\
3336 &   1.441&   0.229&   4.524&   0.394&   -4.474&  0.453&   -1.216&  0.277&   4.021&   0.560&   1.603&   0.282&   2.831&   0.422&   5.661&   0.722&   1.964&   0.265	\\
3339 &   0.849&   0.158&   4.619&   0.271&   -4.800&  0.315&   -1.044&  0.188&   4.596&   0.371&   0.988&   0.190&   2.766&   0.274&   5.594&   0.532&   1.888&   0.166	\\
3352 &   1.170&   0.112&   5.437&   0.191&   -5.742&  0.237&   -1.408&  0.138&   5.412&   0.264&   1.453&   0.136&   3.567&   0.199&   7.382&   0.439&   1.744&   0.127	\\
3367 &   1.013&   0.111&   4.854&   0.188&   -5.013&  0.230&   -1.060&  0.133&   5.127&   0.260&   1.295&   0.135&   3.318&   0.196&   6.870&   0.443&   1.969&   0.123	\\
3400 &   1.183&   0.097&   5.019&   0.167&   -5.027&  0.213&   -1.209&  0.119&   4.980&   0.229&   1.137&   0.119&   3.239&   0.171&   6.900&   0.394&   1.807&   0.104	\\
3438 &   1.662&   0.289&   5.284&   0.514&   -2.202&  0.585&   0.863 &  0.347&   3.395&   0.767&   -0.15&6  0.410&   2.990&   0.575&   3.175&   0.950&   1.772&   0.367	\\
3471 &   1.006&   0.103&   5.329&   0.174&   -4.605&  0.215&   -0.944&  0.125&   4.629&   0.244&   1.456&   0.123&   3.054&   0.181&   5.554&   0.437&   1.425&   0.113	\\
3486 &   0.916&   0.152&   5.492&   0.254&   -4.708&  0.302&   -0.727&  0.178&   3.544&   0.364&   0.922&   0.184&   2.944&   0.266&   4.016&   0.521&   1.979&   0.163	\\
3510 &   1.256&   0.114&   5.608&   0.191&   -5.964&  0.241&   -1.763&  0.140&   5.254&   0.262&   1.425&   0.134&   3.416&   0.193&   7.630&   0.436&   1.308&   0.118	\\
3534 &   0.830&   0.105&   4.722&   0.179&   -3.604&  0.214&   -0.451&  0.124&   3.646&   0.256&   1.151&   0.129&   2.912&   0.190&   3.035&   0.460&   1.842&   0.120	\\
3554 &   0.829&   0.113&   4.704&   0.193&   -4.271&  0.227&   -0.742&  0.134&   4.116&   0.273&   0.922&   0.140&   2.616&   0.205&   4.783&   0.456&   1.797&   0.127	\\
3561 &   1.213&   0.087&   5.079&   0.150&   -5.097&  0.197&   -0.998&  0.107&   5.284&   0.203&   1.492&   0.104&   2.857&   0.152&   6.875&   0.369&   1.807&   0.091	\\
3565 &   0.684&   0.279&   3.927&   0.490&   -2.993&  0.531&   0.313 &  0.318&   3.268&   0.682&   1.311&   0.344&   2.597&   0.515&   4.142&   0.860&   1.244&   0.329	\\
3645 &   1.769&   0.168&   4.870&   0.310&   -3.954&  0.355&   -0.772&  0.212&   3.265&   0.438&   0.768&   0.223&   3.340&   0.320&   6.710&   0.585&   0.843&   0.213	\\
3656 &   1.019&   0.093&   5.413&   0.158&   -4.768&  0.196&   -1.155&  0.114&   3.502&   0.230&   0.823&   0.116&   2.509&   0.170&   3.990&   0.410&   1.520&   0.108	\\
3681 &   1.149&   0.217&   5.780&   0.367&   -4.236&  0.437&   -0.919&  0.266&   2.369&   0.555&   0.472&   0.282&   3.615&   0.392&   4.010&   0.702&   2.497&   0.243	\\
3706 &   1.115&   0.103&   4.704&   0.178&   -4.581&  0.213&   -0.939&  0.124&   4.764&   0.244&   1.426&   0.122&   3.067&   0.181&   6.080&   0.433&   1.982&   0.111	\\
3707 &   1.239&   0.126&   5.112&   0.215&   -4.791&  0.258&   -0.921&  0.151&   4.023&   0.306&   1.076&   0.157&   3.446&   0.224&   5.972&   0.472&   2.018&   0.140	\\
3733 &   1.141&   0.112&   5.110&   0.190&   -5.382&  0.233&   -1.455&  0.136&   5.082&   0.260&   1.394&   0.133&   3.388&   0.194&   7.896&   0.437&   1.316&   0.120	\\
3739 &   1.159&   0.104&   5.695&   0.174&   -6.209&  0.224&   -1.884&  0.129&   5.129&   0.243&   1.377&   0.124&   3.397&   0.180&   6.875&   0.431&   1.552&   0.113	\\
3761 &   1.349&   0.101&   5.465&   0.172&   -5.465&  0.213&   -1.213&  0.123&   5.189&   0.237&   1.405&   0.122&   3.321&   0.176&   6.687&   0.409&   1.855&   0.109	\\
3780 &   1.049&   0.146&   5.442&   0.248&   -4.562&  0.297&   -0.601&  0.175&   3.853&   0.362&   0.618&   0.189&   3.214&   0.269&   4.018&   0.517&   2.108&   0.170	\\
3782 &   1.211&   0.086&   5.410&   0.147&   -5.515&  0.189&   -1.302&  0.107&   4.727&   0.207&   1.216&   0.106&   3.392&   0.152&   6.014&   0.412&   1.733&   0.096	\\
3792 &   1.329&   0.119&   5.471&   0.201&   -6.266&  0.248&   -1.918&  0.147& \nodata& \nodata& \nodata& \nodata& \nodata& \nodata& \nodata& \nodata& \nodata& \nodata	\\
3794 &   1.084&   0.101&   5.417&   0.170&   -5.270&  0.212&   -1.534&  0.124&   4.243&   0.241&   1.113&   0.123&   3.309&   0.176&   5.586&   0.431&   1.518&   0.110	\\
3851 &   1.156&   0.119&   4.507&   0.209&   -4.299&  0.240&   -0.798&  0.142&   4.531&   0.284&   1.303&   0.143&   3.313&   0.209&   6.501&   0.419&   2.082&   0.130	\\
3855 &   1.345&   0.208&   3.140&   0.383&   -1.341&  0.400&   0.213 &  0.245&   3.181&   0.526&   1.103&   0.263&   2.471&   0.390&   4.917&   0.667&   2.335&   0.234	\\
3914 &   1.027&   0.109&   5.167&   0.183&   -5.668&  0.230&   -1.731&  0.134&   4.475&   0.256&   1.291&   0.130&   3.183&   0.189&   6.836&   0.433&   1.663&   0.116	\\
\enddata
\end{deluxetable}

%%%%%%%%%%%%%%%%%%%%%%%%%%%%%%%%%%%%%%%%%%%%%%%%%%%%%%%%%%%%%%%%%%%%%%%%%%%%%%

\begin{deluxetable}{lrrrrrrrrrrrrrrrrrrr}
%lrrrrrrrrrrrrrrrrrr rrrrrrrrrrrrrrrrrrrrrrrrr}
% should be 45 rows
  \tabletypesize{\tiny}
\rotate
  \tablecaption{Index and Model Measurements \label{mastertable2}}
 \tablewidth{0pt}
%  \tablewidth{10in}
  \tablehead{
    \colhead{GMP}&
\colhead{Fe5015}& \colhead{$\delta_{5015}$}& \colhead{Mg$_1$}&
\colhead{$\delta_{Mg_1}$}& \colhead{Mg$_2$}&
\colhead{$\delta_{Mg_2}$}& \colhead{Mg$_b$}&
\colhead{$\delta_{Mg_b}$}& \colhead{Fe5270}&
\colhead{$\delta_{5270}$}& \colhead{Fe5335}&
\colhead{$\delta_{5335}$}& \colhead{Age}&
\colhead{$\delta_{Age}$}& \colhead{$[$Z/H$]$}&
\colhead{$\delta_{[Z/H]}$}& \colhead{$[\alpha$/Fe$]$}&
\colhead{$\delta_{[\alpha/Fe]}$} & \colhead{SNR} } \startdata
%\tableline
%2478&3.629&0.364&0.027&0.013&0.146&0.012&2.966&0.163&2.031&0.199&2.463&0.237&3.8&1.5&-0.108&0.093&0.145&0.075&28\\
%2489&4.799&0.210&0.075&0.012&0.213&0.015&3.963&0.095&2.555&0.112&2.398&0.135&5.5&1.6&0.133&0.076&0.26&0.032&54\\
2478 &     3.630&   0.460&   0.027&   0.013&   0.146&   0.013&   2.966&   0.213&   2.030&   0.258&   2.462&   0.306&   3.8   &  2.1  &   -0.109 & 0.229  & 0.146  & 0.099  & 28 \\
2489 &     4.799&   0.236&   0.075&   0.012&   0.213&   0.015&   3.963&   0.109&   2.555&   0.128&   2.397&   0.155&   5.7   &  1.7  &   0.132  & 0.080  & 0.260  & 0.037  & 54	\\
2510 &     5.258&   0.288&   0.088&   0.012&   0.227&   0.011&   4.075&   0.122&   2.798&   0.148&   2.593&   0.177&   5.7   &  2.2  &   0.183  & 0.130  & 0.198  & 0.043  & 48	\\
2516 &     5.229&   0.272&   0.112&   0.012&   0.254&   0.016&   4.141&   0.131&   2.724&   0.151&   2.565&   0.180& \nodata &\nodata&  \nodata &\nodata &\nodata &\nodata & 46	\\
2529 &     3.202&   0.465&   0.031&   0.013&   0.131&   0.013&   2.913&   0.213&   2.262&   0.257&   2.133&   0.310&   11.6  &  3.4  &   -0.335 & 0.127  & 0.166  & 0.119  & 28	\\
2535 &     4.328&   0.279&   0.091&   0.012&   0.240&   0.014&   4.370&   0.123&   2.685&   0.144&   2.440&   0.173&   4.6   &  2.4  &   0.259  & 0.099  & 0.311  & 0.040  & 47	\\
2541 &     4.954&   0.298&   0.115&   0.012&   0.263&   0.015&   4.692&   0.129&   2.621&   0.154&   2.342&   0.186&   13.7  &  1.3  &   0.063  & 0.054  & 0.350  & 0.044  & 45	\\
2585 &     5.686&   0.535&   0.020&   0.013&   0.116&   0.017&   2.294&   0.270&   1.080&   0.322&   1.509&   0.384&   5.9   &  1.7  &   -0.714 & 0.164  & 0.478  & 0.156  & 23	\\
2603 &     4.874&   0.328&   0.020&   0.012&   0.115&   0.013&   2.355&   0.150&   2.248&   0.179&   2.352&   0.215&   2.0   &  0.8  &  \nodata &\nodata &\nodata &\nodata & 39	\\
2654 &     4.671&   0.254&   0.101&   0.012&   0.236&   0.015&   4.042&   0.113&   2.710&   0.131&   2.435&   0.158&   11.6  &  1.4  &   -0.015 & 0.082  & 0.196  & 0.046  & 51	\\
2692 &     3.400&   0.497&   0.022&   0.013&   0.138&   0.014&   2.969&   0.229&   2.076&   0.276&   2.257&   0.325&   6.2   &  3.1  &   -0.195 & 0.122  & 0.184  & 0.114  & 26	\\
2778 &     5.554&   0.345&   0.062&   0.012&   0.171&   0.017&   3.612&   0.168&   2.477&   0.196&   2.583&   0.230&   7.9   &  4.2  &   -0.021 & 0.185  & 0.135  & 0.068  & 34	\\
2784 &     5.775&   0.801&   0.040&   0.015&   0.156&   0.017&   2.481&   0.413&   0.974&   0.502&   2.564&   0.567& \nodata &\nodata&  \nodata &\nodata &\nodata &\nodata & 15	\\
2799 &     4.356&   0.706&   0.022&   0.014&   0.142&   0.018&   3.265&   0.339&   1.490&   0.409&   1.063&   0.492& \nodata &\nodata&  \nodata &\nodata &\nodata &\nodata & 17	\\
2805 &     4.980&   0.252&   0.083&   0.011&   0.224&   0.016&   3.692&   0.121&   2.661&   0.138&   2.681&   0.164&   11.6  &  2.7  &   -0.039 & 0.076  & 0.089  & 0.051  & 48	\\
2839 &     4.930&   0.225&   0.112&   0.011&   0.254&   0.016&   4.310&   0.109&   2.685&   0.125&   2.583&   0.148&   4.6   &  1.8  &   0.265  & 0.146  & 0.275  & 0.028  & 53	\\
2852 &     4.426&   0.444&   0.049&   0.013&   0.176&   0.015&   3.373&   0.206&   3.219&   0.238&   2.158&   0.291&   3.9   &  3.4  &   0.175  & 0.251  & 0.083  & 0.072  & 27	\\
2861 &     4.801&   0.260&   0.086&   0.012&   0.217&   0.014&   3.662&   0.112&   3.136&   0.130&   2.550&   0.158&   7.9   &  2.4  &   0.101  & 0.097  & 0.043  & 0.050  & 53	\\
2879 &     4.378&   0.501&   0.047&   0.013&   0.159&   0.015&   2.638&   0.242&   1.493&   0.286&   2.695&   0.331& \nodata &\nodata&  \nodata &\nodata &\nodata &\nodata & 23	\\
2912 &     5.392&   0.240&   0.092&   0.012&   0.241&   0.015&   4.289&   0.109&   2.990&   0.127&   2.787&   0.152&   4.3   &  1.4  &   0.378  & 0.070  & 0.198  & 0.040  & 54	\\
2922 &     5.079&   0.283&   0.121&   0.012&   0.270&   0.014&   4.615&   0.125&   2.833&   0.147&   2.751&   0.175&   11.4  &  3.5  &   0.209  & 0.114  & 0.242  & 0.041  & 45	\\
2940 &     4.550&   0.293&   0.101&   0.012&   0.250&   0.015&   4.240&   0.131&   2.548&   0.155&   2.485&   0.185&   11.2  &  2.9  &   0.038  & 0.094  & 0.261  & 0.047  & 45	\\
2960 &     4.880&   0.264&   0.048&   0.011&   0.164&   0.016&   2.988&   0.130&   2.686&   0.148&   2.113&   0.179&   5.3   &  1.0  &   -0.107 & 0.075  & 0.076  & 0.054  & 45	\\
2985 &     2.363&   0.820&   0.034&   0.014&   0.118&   0.019&   1.529&   0.411&   1.624&   0.466&   2.531&   0.540& \nodata &\nodata&  \nodata &\nodata &\nodata &\nodata & 15	\\
3017 &     5.025&   0.540&   0.038&   0.013&   0.137&   0.017&   2.019&   0.275&   2.763&   0.311&   1.376&   0.385&   3.0   &  2.4  &   -0.272 & 0.078  &\nodata &\nodata & 22	\\
3058 &     3.766&   0.526&   0.050&   0.013&   0.165&   0.017&   2.573&   0.261&   2.856&   0.293&   2.882&   0.346&   5.3   &  4.2  &  \nodata &\nodata &\nodata &\nodata & 23	\\
3068 &     4.430&   0.306&   0.073&   0.012&   0.204&   0.014&   3.756&   0.134&   2.810&   0.157&   2.404&   0.191&   12.9  &  2.1  &   -0.084 & 0.056  & 0.132  & 0.064  & 44	\\
3073 &     5.152&   0.308&   0.112&   0.012&   0.266&   0.013&   4.389&   0.133&   2.853&   0.159&   3.016&   0.189&   9.4   &  3.7  &   0.259  & 0.103  & 0.163  & 0.048  & 45	\\
3121 &     4.480&   0.535&   0.038&   0.013&   0.169&   0.016&   2.844&   0.260&   2.056&   0.305&   1.543&   0.370& \nodata &\nodata&  \nodata &\nodata &\nodata &\nodata & 24	\\
3126 &     3.528&   0.504&   0.046&   0.013&   0.160&   0.014&   3.518&   0.230&   3.017&   0.272&   2.148&   0.332&   10.0  &  4.7  &   -0.051 & 0.135  & 0.093  & 0.108  & 26	\\
3133 &     4.733&   0.365&   0.062&   0.012&   0.204&   0.010&   4.174&   0.170&   3.106&   0.209&   2.554&   0.249&   7.7   &  3.9  &   0.224  & 0.157  & 0.165  & 0.053  & 35	\\
3166 &     1.735&   0.787&   0.021&   0.014&   0.095&   0.016&   2.382&   0.376&   2.517&   0.442&   2.745&   0.532&   3.8   &  4.3  &   -0.163 & 0.092  &\nodata &\nodata & 16	\\
3170 &     5.345&   0.327&   0.099&   0.011&   0.243&   0.010&   4.136&   0.151&   2.673&   0.187&   2.696&   0.217&   4.6   &  4.1  &   0.238  & 0.135  & 0.223  & 0.052  & 41	\\
3196 &     4.152&   0.622&   0.063&   0.014&   0.166&   0.017&   3.079&   0.303&   2.229&   0.353&   1.423&   0.436& \nodata &\nodata&  \nodata &\nodata &\nodata &\nodata & 19	\\
3201 &     4.623&   0.286&   0.101&   0.012&   0.226&   0.016&   3.639&   0.135&   2.668&   0.156&   2.813&   0.185&   5.3   &  1.6  &   0.134  & 0.090  & 0.095  & 0.044  & 45	\\
3205 &     4.661&   0.458&   0.035&   0.013&   0.141&   0.016&   2.195&   0.231&   1.958&   0.261&   2.617&   0.306& \nodata &\nodata&  \nodata &\nodata &\nodata &\nodata & 25	\\
3206 &     4.142&   0.386&   0.076&   0.013&   0.197&   0.016&   3.526&   0.183&   2.010&   0.217&   1.670&   0.261&   12.4  &  3.1  &   -0.297 & 0.091  & 0.484  & 0.108  & 33	\\
3209 &     0.648&   0.912&   0.004&   0.015&   0.084&   0.018&   2.050&   0.430&   1.989&   0.510&   2.227&   0.609& \nodata &\nodata&  \nodata &\nodata &\nodata &\nodata & 15	\\
3213 &     4.986&   0.265&   0.113&   0.012&   0.261&   0.015&   4.405&   0.121&   2.509&   0.141&   2.387&   0.169&   14.6  &  3.1  &   -0.028 & 0.092  & 0.313  & 0.049  & 48	\\
3222 &     4.023&   0.316&   0.110&   0.012&   0.250&   0.016&   4.441&   0.143&   2.492&   0.169&   1.884&   0.205&   15.0  &  0.9  &   -0.077 & 0.028  & 0.436  & 0.064  & 41	\\
3254 &     5.096&   0.429&   0.059&   0.013&   0.228&   0.014&   4.223&   0.201&   3.298&   0.233&   2.329&   0.285&   14.9  &  5.1  &   0.049  & 0.159  & 0.141  & 0.066  & 28	\\
3269 &     3.871&   0.356&   0.068&   0.013&   0.200&   0.013&   3.612&   0.157&   2.694&   0.188&   2.323&   0.226&   3.1   &  1.4  &   0.179  & 0.110  & 0.198  & 0.052  & 38	\\
3292 &     3.017&   0.570&   0.054&   0.012&   0.143&   0.017&   1.676&   0.283&   2.259&   0.311&   2.157&   0.367& \nodata &\nodata&  \nodata &\nodata &\nodata &\nodata & 22	\\
3296 &     4.655&   0.283&   0.106&   0.012&   0.257&   0.012&   4.696&   0.116&   2.989&   0.138&   2.730&   0.168&   7.2   &  3.6  &   0.359  & 0.111  & 0.260  & 0.042  & 49	\\
3312 &     3.981&   0.473&   0.019&   0.013&   0.124&   0.015&   2.298&   0.232&   1.715&   0.271&   2.510&   0.321& \nodata &\nodata&  \nodata &\nodata &\nodata &\nodata & 25	\\
3313 &     3.584&   0.263&   0.056&   0.012&   0.171&   0.015&   3.081&   0.126&   2.428&   0.145&   2.102&   0.175&   11.9  &  3.1  &   -0.282 & 0.066  & 0.166  & 0.072  & 49	\\
3336 &     5.271&   0.615&   0.111&   0.014&   0.254&   0.017&   2.998&   0.314&   2.568&   0.347&   1.916&   0.421&   6.8   &  4.9  &   -0.172 & 0.188  & 0.153  & 0.150  & 19	\\
3339 &     3.370&   0.383&   0.054&   0.013&   0.180&   0.016&   3.511&   0.179&   2.842&   0.206&   2.299&   0.249&   8.0   &  3.3  &   -0.016 & 0.124  & 0.102  & 0.073  & 33	\\
3352 &   \nodata& \nodata& \nodata& \nodata& \nodata& \nodata& \nodata& \nodata& \nodata& \nodata& \nodata& \nodata& \nodata &\nodata&  \nodata &\nodata &\nodata &\nodata & 42	\\
3367 &     5.059&   0.281&   0.139&   0.011&   0.273&   0.016&   4.309&   0.137&   2.301&   0.159&   2.459&   0.187&   4.5   &  2.0  &   0.191  & 0.106  & 0.365  & 0.043  & 45	\\
3400 &   \nodata& \nodata& \nodata& \nodata& \nodata& \nodata& \nodata& \nodata& \nodata& \nodata& \nodata& \nodata& \nodata &\nodata&  \nodata &\nodata &\nodata &\nodata & 51	\\
3438 &     3.395&   0.837&   0.023&   0.014&   0.133&   0.019&   2.475&   0.412&   1.007&   0.496&   1.329&   0.578& \nodata &\nodata&  \nodata &\nodata &\nodata &\nodata & 15	\\
3471 &     4.150&   0.266&   0.067&   0.012&   0.193&   0.015&   3.575&   0.123&   2.581&   0.143&   2.074&   0.173& \nodata &\nodata&   -0.209 & 0.025  & 0.240  & 0.057  & 49	\\
3486 &     2.674&   0.400&   0.040&   0.013&   0.150&   0.014&   3.560&   0.177&   2.780&   0.212&   2.749&   0.253&   5.4   &  3.5  &   0.079  & 0.147  & 0.056  & 0.067  & 32	\\
3510 &     4.526&   0.279&   0.135&   0.012&   0.292&   0.015&   4.943&   0.124&   2.958&   0.143&   2.675&   0.172& \nodata &\nodata&   0.219  & 0.033  & 0.285  & 0.035  & 46	\\
3534 &     4.351&   0.274&   0.051&   0.011&   0.165&   0.016&   2.917&   0.134&   2.452&   0.154&   2.551&   0.183&   10.3  &  3.5  &   -0.241 & 0.074  & 0.006  & 0.064  & 43	\\
3554 &     3.509&   0.311&   0.050&   0.012&   0.165&   0.015&   3.165&   0.141&   2.400&   0.166&   2.104&   0.202&   11.0  &  3.9  &   -0.248 & 0.073  & 0.186  & 0.082  & 43	\\
3561 &   \nodata& \nodata& \nodata& \nodata& \nodata& \nodata& \nodata& \nodata& \nodata& \nodata& \nodata& \nodata& \nodata &\nodata&  \nodata &\nodata &\nodata &\nodata & 58	\\
3565 &     6.410&   0.716&   0.066&   0.014&   0.178&   0.017&   3.373&   0.362&   2.085&   0.432&   2.163&   0.514& \nodata &\nodata&  \nodata &\nodata &\nodata &\nodata & 15	\\
3645 &     4.128&   0.471&   0.086&   0.013&   0.231&   0.016&   3.257&   0.232&   2.273&   0.265&   2.182&   0.315& \nodata &\nodata&  \nodata &\nodata &\nodata &\nodata & 25	\\
3656 &     4.062&   0.281&   0.061&   0.012&   0.199&   0.013&   3.605&   0.120&   2.557&   0.143&   2.029&   0.175& \nodata &\nodata&   -0.211 & 0.050  & 0.262  & 0.064  & 53	\\
3681 &     5.864&   0.560&   0.045&   0.013&   0.152&   0.017&   2.884&   0.279&   2.805&   0.324&   2.266&   0.392&   2.4   &  0.7  &   0.126  & 0.066  & 0.038  & 0.091  & 21	\\
3706 &     4.526&   0.267&   0.071&   0.012&   0.204&   0.015&   3.752&   0.122&   2.474&   0.142&   2.449&   0.172&   6.3   &  2.1  &   0.033  & 0.078  & 0.209  & 0.037  & 49	\\
3707 &     4.988&   0.334&   0.079&   0.012&   0.199&   0.015&   3.391&   0.156&   2.542&   0.181&   2.302&   0.217&   5.8   &  3.1  &   -0.053 & 0.129  & 0.150  & 0.058  & 38	\\
3733 &     5.145&   0.277&   0.138&   0.012&   0.299&   0.016&   4.894&   0.128&   2.672&   0.149&   2.542&   0.176& \nodata &\nodata&  \nodata &\nodata &\nodata &\nodata & 45	\\
3739 &     5.143&   0.256&   0.117&   0.012&   0.270&   0.015&   4.505&   0.121&   2.785&   0.140&   2.573&   0.167&   11.4  &  3.5  &   0.123  & 0.096  & 0.254  & 0.037  & 46	\\
3761 &     5.199&   0.277&   0.095&   0.012&   0.244&   0.013&   4.250&   0.119&   2.855&   0.139&   2.707&   0.168&   4.9   &  2.0  &   0.296  & 0.077  & 0.215  & 0.033  & 49	\\
3780 &     3.488&   0.417&   0.045&   0.013&   0.155&   0.013&   3.504&   0.188&   2.240&   0.229&   1.977&   0.277&   5.2   &  2.4  &   -0.104 & 0.186  & 0.321  & 0.092  & 30	\\
3782 &     4.925&   0.223&   0.093&   0.012&   0.237&   0.015&   4.085&   0.105&   2.702&   0.122&   2.319&   0.147&   8.7   &  3.8  &   0.055  & 0.100  & 0.241  & 0.032  & 56	\\
3792 &   \nodata& \nodata& \nodata& \nodata& \nodata& \nodata& \nodata& \nodata& \nodata& \nodata& \nodata& \nodata& \nodata &\nodata&  \nodata &\nodata &\nodata &\nodata & 41	\\
3794 &     4.573&   0.269&   0.109&   0.012&   0.253&   0.014&   4.528&   0.119&   2.643&   0.141&   2.374&   0.170&   13.2  &  1.8  &   0.030  & 0.054  & 0.308  & 0.043  & 47	\\
3851 &     4.167&   0.329&   0.063&   0.012&   0.187&   0.013&   3.627&   0.143&   3.031&   0.170&   2.482&   0.206&   3.2   &  1.5  &   0.261  & 0.111  & 0.118  & 0.051  & 41	\\
3855 &     3.441&   0.541&   0.055&   0.013&   0.151&   0.017&   3.692&   0.257&   1.424&   0.312&   2.714&   0.359&   3.0   &  3.8  &   0.179  & 0.328  & 0.397  & 0.098  & 22	\\
3914 &     4.973&   0.260&   0.112&   0.011&   0.257&   0.016&   4.387&   0.124&   2.793&   0.142&   2.729&   0.168&   11.4  &  2.6  &   0.146  & 0.093  & 0.206  & 0.031  & 46	\\
\enddata
\end{deluxetable}

%%%%%%%%%%%%%%%%%%%%%%%%%%%%%%%%%%%%%%%%%%%%%%%%%%%%%%%%%%%%%%%%%%%%%%%%%%%%%%
%  Comparison Table
%%%%%%%%%%%%%%%%%%%%%%%%%%%%%%%%%%%%%%%%%%%%%%%%%%%%%%%%%%%%%%%%%%
% This table compares the our indices with NFPS,Poggianti and Hectospec.
% including off-grid galaxies.

\begin{deluxetable}{llrrr}
 \tabletypesize{\tiny} \tablecaption{Comparison with other
studies\label{offsetstabl}} \tablewidth{0pt} \tablehead{
\colhead{Index} & \colhead{Ref.} & \colhead{$\langle \Delta I
\rangle$} & \colhead{std} & \colhead{$\sigma_{exp}$} } \startdata
Ca4227     & P01 & $ -0.111 \pm  0.129 $ &  0.755 &  0.364 \\
           & N05 & $ -0.207 \pm  0.047 $ &  0.236 &  0.232 \\
           & S08 & $  0.098 \pm  0.095 $ &  0.426 &  0.241 \\
\tableline
G4300      & P01 & $ -0.217 \pm  0.260 $ &  1.519 &  0.600 \\
           & N05 & $  0.020 \pm  0.094 $ &  0.471 &  0.376 \\
           & S08 & $ -0.145 \pm  0.234 $ &  1.046 &  0.467 \\
\tableline
H$\gamma_A$& P01 & $  0.212  \pm  0.299 $ &  1.745 &  0.663 \\
           & N05 & $   0.105 \pm  0.114$ &   0.569 &  0.515 \\
           & S08 & $  0.359  \pm  0.225 $ &  1.004 &  0.518 \\
\tableline
H$\gamma_F$& P01 & $ -0.040 \pm  0.162 $ &  0.945 &  0.424 \\
           & N05 & $  0.037 \pm  0.064 $ &  0.320 &  0.247 \\
           & S08 & $  0.146 \pm  0.146 $ &  0.653 &  0.303 \\
\tableline
Fe4383     & P01 & $ -0.970 \pm  0.275 $ &  1.604 &  0.846 \\
           & N05 & $  0.154 \pm  0.139 $ &  0.696 &  0.474 \\
           & S08 & $ -0.113 \pm  0.205 $ &  0.916 &  0.627 \\
\tableline
Ca4455     & P01 & $ -0.259  \pm  0.116 $ &  0.678 &  0.425 \\
           &  N05& $  -0.001 \pm  0.038$ &   0.189 &  0.199 \\
           & S08 & $ -0.089  \pm  0.122 $ &  0.547 &  0.309 \\
\tableline
Fe4531     & P01 & $ -0.193  \pm  0.153 $ &  0.890 &  0.616 \\
           &  N05& $  -0.152 \pm  0.066$ &   0.329 &  0.311 \\
           & S08 & $ -0.207  \pm  0.149 $ &  0.668 &  0.473 \\
\tableline
C$_2$4668  & P01 & $ -0.236  \pm  0.345 $ &  2.010 &  0.989 \\
           &  N05& $   0.154 \pm  0.139$ &   0.696 &  0.635 \\
           & S08 & $  0.140  \pm  0.232 $ &  1.037 &  0.838 \\
\tableline
 H$\beta$  & P01 & $  0.121  \pm  0.125 $ &  0.726 &  0.376 \\
           &  N05& $   0.062 \pm  0.041$ &   0.207 &  0.167 \\
           & S08 & $ -0.119  \pm  0.094 $ &  0.419 &  0.297 \\
\tableline
Fe5015     & P01 & $ -0.652  \pm  0.208 $ &  1.216 &  0.780 \\
           &  N05& $  -0.314 \pm  0.102$ &   0.477 &  0.428 \\
           & S08 & $ -0.243  \pm  0.246 $ &  1.099 &  0.936 \\
\tableline
Mg$_1$     & P01 & $ -0.012  \pm  0.006 $ &  0.037 &  0.014 \\
           &  N05& $  -0.023 \pm  0.002$ &   0.010 &  0.014 \\
           & S08 & $ -0.014  \pm  0.003 $ &  0.014 &  0.015 \\
\tableline
Mg$_2$     & P01 & $  0.004  \pm  0.007 $ &  0.041 &  0.017 \\
           &  N05& $  -0.017 \pm  0.002$ &   0.010 &  0.017 \\
           & S08 & $ -0.018  \pm  0.004 $ &  0.018 &  0.016 \\
\tableline
Mg$_b$     & P01 & $  0.015  \pm  0.119 $ &  0.692 &  0.372 \\
           &  N05& $  -0.051 \pm  0.041$ &   0.194 &  0.177 \\
           & S08 & $ -0.016  \pm  0.115 $ &  0.513 &  0.319 \\
\tableline
Fe5270     & P01 & $ -0.178  \pm  0.112 $ &  0.652 &  0.420 \\
           &  N05& $  -0.018 \pm  0.051$ &   0.241 &  0.201 \\
           & S08 & $ -0.334  \pm  0.127 $ &  0.567 &  0.369 \\
\tableline
Fe5335     & P01 & $  0.074  \pm  0.098 $ &  0.571 &  0.480 \\
           &  N05& $  -0.066 \pm  0.050$ &   0.233 &  0.232 \\
           & S08 & $  0.058  \pm  0.146 $ &  0.653 &  0.431 \\

\enddata
\tablecomments{Comparison of line strengths measured in this and other
  studies.  Ref.: reference in the literature where P01 is from
  \citet*{Poggianti01}, N05 is from \citet*{Nelan05} and S08 is from
  \citet*{Smith08}; $\langle \Delta I \rangle$: mean offset between
  our study and other ($I_{here}-I_{other}$) and the error in the
  mean; {\it std} standard deviation of the differences;
  $\sigma_{exp}$: standard deviation expected from the errors.}
\end{deluxetable}
%\clearpage

%%%%%%%%%%%%%%%%%%%%%%%%%%%%%%%%%%%%%%%%%%%%%%%%%%%%%%%%%%%%%%%%%%%%%%%%%%%%%%
%  Linear Fits to low and high
%%%%%%%%%%%%%%%%%%%%%%%%%%%%%%%%%%%%%%%%%%%%%%%%%%%%%%%%%%%%%%%%%%%%%%%%%%%%%%
%this table contains the linear fits to low- and high-sigma galaxies for all indices: Index-sigma.
% it is from /lick_index/nfps/index_sigma_stats_wHg_new.pro 
\begin{deluxetable}{lrrrrrr}
\tabletypesize{\scriptsize}
\tablecaption{$I-\sigma$ Linear Fits \label{linear-table}}
\tablewidth{0pt}
\tablehead{
\colhead{Index} 
&\colhead{Intercept$_{<100}$}
&\colhead{Slope$_{<100}$}
& \colhead{$\rho_{<100}$}
&\colhead{Intercept$_{\geqslant100}$}
&\colhead{Slope$_{\geqslant100}$}
& \colhead{$\rho_{\geqslant100}$}
}
\startdata
Ca4227      &$ 0.826 \pm 0.640$& $ 0.140 \pm 0.352$&  0.063& $ 0.030 \pm 0.590$ & $ 0.494 \pm 0.268$ &  0.186\\   
G4300       &$-0.585 \pm 1.717$& $ 2.974 \pm 0.944$&  0.543& $ 4.748 \pm 1.140$ & $ 0.285 \pm 0.519$ & -0.029\\  
H$\gamma_A$ &$ 7.167 \pm 3.017$& $-6.111 \pm 1.658$& -0.652& $-3.035 \pm 1.852$ & $-1.194 \pm 0.843$ & -0.170\\  
H$\gamma_F$ &$ 4.643 \pm 1.620$& $-2.835 \pm 0.890$& -0.612& $ 0.368 \pm 0.834$ & $-0.834 \pm 0.380$ & -0.290\\  
Fe4383      &$-3.369 \pm 1.731$& $ 4.094 \pm 0.951$&  0.597& $ 4.512 \pm 1.091$ & $ 0.234 \pm 0.497$ &  0.043\\   
Ca4455      &$-0.852 \pm 0.874$& $ 1.014 \pm 0.480$&  0.369& $ 1.774 \pm 0.445$ & $-0.236 \pm 0.203$ & -0.089\\  
Fe4531      &$-0.630 \pm 1.040$& $ 1.962 \pm 0.572$&  0.426& $ 2.773 \pm 0.606$ & $ 0.216 \pm 0.276$ &  0.017\\   
C$_2$4668   &$-3.646 \pm 2.460$& $ 4.622 \pm 1.352$&  0.492& $-3.203 \pm 2.079$ & $ 4.450 \pm 0.947$ &  0.546\\   
H$\beta$    &$ 3.902 \pm 0.939$& $-1.082 \pm 0.516$& -0.375& $ 2.933 \pm 0.503$ & $-0.595 \pm 0.229$ & -0.255\\  
Fe5015      &$ 2.792 \pm 1.921$& $ 0.782 \pm 1.055$&  0.163& $ 4.102 \pm 1.026$ & $ 0.340 \pm 0.469$ &  0.032\\   
Mg$_1$      &$-0.135 \pm 0.040$& $ 0.106 \pm 0.022$&  0.684& $-0.204 \pm 0.036$ & $ 0.141 \pm 0.017$ &  0.702\\   
Mg$_2$      &$-0.136 \pm 0.063$& $ 0.171 \pm 0.035$&  0.652& $-0.143 \pm 0.045$ & $ 0.178 \pm 0.020$ &  0.717\\   
Mg$b$       &$-0.944 \pm 0.981$& $ 2.292 \pm 0.539$&  0.530& $-0.437 \pm 0.743$ & $ 2.120 \pm 0.340$ &  0.667\\   
Fe5270      &$ 0.150 \pm 0.945$& $ 1.303 \pm 0.519$&  0.257& $ 3.235 \pm 0.650$ & $-0.243 \pm 0.297$ & -0.064\\  
Fe5335      &$ 0.985 \pm 0.952$& $ 0.730 \pm 0.523$&  0.103& $ 2.470 \pm 0.640$ & $ 0.021 \pm 0.292$ &  0.157\\   
$\langle$Fe$\rangle$ &$ 0.568 \pm 0.813$& $ 1.017 \pm 0.447$&  0.254& $ 2.853 \pm 0.586$ & $-0.111 \pm 0.268$ & -0.003\\  
$[$MgFe$]\arcmin$ &$-0.256 \pm 0.708$& $ 1.683 \pm 0.389$&  0.513& $ 1.742 \pm 0.536$ & $ 0.726 \pm 0.245$ &  0.387\\   
\enddata
\tablecomments{$I-\sigma$ Linear Fits for low- and high-$\sigma$
  early-type galaxies.  The intercept and slope of the $I-\sigma$
  linear fit, and Spearman Rank correlation coefficient ($\rho$), are
  presented for low- and for high-$\sigma$ galaxies respectively.
  Note that the correlation coefficients are low for galaxies in group
  I implying that for these indices the $I-\sigma$ relations are not
  significant. }
\end{deluxetable}

%%%%%%%%%%%%%%%%%%%%%%%%%%%%%%%%%%%%%%%%%%%%%%%%%%%%%%%%%%%%%%%%%%%%%%%%%%%%%%
%  Group Ia
%%%%%%%%%%%%%%%%%%%%%%%%%%%%%%%%%%%%%%%%%%%%%%%%%%%%%%%%%%%%%%%%%%%%%%%%%%%%%%
%this table contains the statistics for first two columns of I-sigma plots
% it is from /lick_index/nfps/index_sigma_stats.pro and in here it is 'stats.table'
%the titles are Index name, Number of obj. with sigma<100, N>100 km/s, Student's T-test,, Kolmogorov-Smirnof statistics.

\begin{deluxetable}{llllllll}
\tabletypesize{\scriptsize}
\tablecaption{$I-\sigma$ Statistics for Group Ia \label{stats-table}}
\tablewidth{0pt}
\tablehead{
\colhead{Index} & \colhead{N$_{<100}$}
& \colhead{N$_{\geqslant 100}$}
& \colhead{$\bar{I}_{<100}$}
& \colhead{$\bar{I}_{\geqslant 100}$}
& \colhead{std$_{<100}$}
& \colhead{std$_{\geqslant 100}$}
%& \colhead{KS$_d$}
&  \colhead{prob$_{KS}$}
%& \colhead{$\rho_{<100}$}
%& \colhead{$\rho_{\geqslant 100}$}
}
\startdata
  G4300 &  41 & 70 & 4.791 & 5.328 & 0.861 & 0.482 & 1.72E-06\\
 Fe4383 &  41 & 69 & 4.080 & 5.058 & 0.950 & 0.453 & 1.65E-09\\
 Ca4455 &  41 & 69 & 0.986 & 1.252 & 0.413 & 0.186 & 3.59E-04\\
 Fe4531 &  41 & 69 & 2.931 & 3.243 & 0.532 & 0.251 & 2.48E-04\\
 Fe5270 &  41 & 66 & 2.515 & 2.702 & 0.456 & 0.260 & 8.01E-04\\
\enddata
\tablecomments{Statistics for group Ia of $I-\sigma$ plots.  N:
  number of galaxies with $\sigma < 100$ km s$^{-1}$ and with $\sigma
  \geqslant 100$ km s$^{-1}$ excluding the ``off-grid galaxies'';
  $\bar{I}$: average value of the index for low- and high-$\sigma$
  galaxies; std$_{<100}$ and std$_{\geqslant 100}$: standard deviation
  of the points for the low- and high-$\sigma$ galaxies; prob$_{KS}$:
  the KS probability that the two sets are drawn from the same
  distribution.
  % ; $\rho_{<100}$ and $\rho_{\geqslant 100}$: Spearman Rank
  % correlation coefficients for low- and high-$\sigma$ galaxies.
  % KS$_d$: Kolmogorov-Smirnov test value representing the maximum
  % deviation between the distribution of the two sets;
}
\end{deluxetable}
%\clearpage

%%%%%%%%%%%%%%%%%%%%%%%%%%%%%%%%%%%%%%%%%%%%%%%%%%%%%%%%%%%%%%%%%%%%%%%%%%%%%%
%  Group Ib
%%%%%%%%%%%%%%%%%%%%%%%%%%%%%%%%%%%%%%%%%%%%%%%%%%%%%%%%%%%%%%%%%%%%%%%%%%%%%%
%this table contains the lines by fitexy, linear fits when
%weighing in both x and y direction.
%from /scratch/matkovic/dEs_gEs/lick_index/nfps/index_sigma_scatter_corr.table

\begin{deluxetable}{lrrrrrrrrr}    % no Ks test in the table
  \tabletypesize{\scriptsize}
  \tablecaption{$I-\sigma$ linear fits and statistics for index
 group Ib \label{fitexytabl2}}
  \tablewidth{0pt}
  \tablehead{ \colhead{Index}
&\colhead{N}
&\colhead{Intercept}
&\colhead{Slope}
&\colhead{$\rho$}
&\colhead{$\sigma_{Int_{<100}}$}
&\colhead{$\sigma_{Err_{<100}}$}
&\colhead{$\sigma_{Int_{\geqslant100}}$}
&\colhead{$\sigma_{Err_{\geqslant100}}$}
%&\colhead{KS$_d$}
&\colhead{prob$_{KS}$}
  }
  \startdata
  C4227             &111&$0.776 \pm 0.240$&$0.156 \pm 0.116$& 0.156&  0.161 &  0.245 &  0.214&   0.259&   0.333    \\
  Fe5015            &107&$1.814 \pm 0.586$&$1.363 \pm 0.285$& 0.356&  0.582 &  0.364 &  0.000&   0.432&   0.006    \\
  Fe5335            &107&$1.443 \pm 0.310$&$0.483 \pm 0.151$& 0.323&  0.255 &  0.221 &  0.197&   0.173&   0.719    \\
$\langle$Fe$\rangle$&107&$1.526 \pm 0.276$&$0.491 \pm 0.135$& 0.320&  0.285 &  0.099 &  0.364&   0.149&   2.46E-07 \\
%---------------------------------------------------------
 \enddata
 \tablecomments{Statistics for group Ib of $I-\sigma$ plots.  N:
   number of galaxies; Intercept and Slope: linear regression applied
   to all the galaxies; $\rho$: Spearman-Rank correlation coefficient;
   $\sigma_{Int}$: intrinsic scatter for the two sub-samples,
   i.e. standard deviation of residuals between the points and the
   linear fit for each sub-sample of galaxies (low and high-$\sigma$);
   $\sigma_{Err}$: scatter due to the errors; prob$_{KS}$: the KS
   probability that the two sets are drawn from the same distribution.
 }
\end{deluxetable}
%\clearpage

%%%%%%%%%%%%%%%%%%%%%%%%%%%%%%%%%%%%%%%%%%%%%%%%%%%%%%%%%%%%%%%%%%%%
%   Group III
%%%%%%%%%%%%%%%%%%%%%%%%%%%%%%%%%%%%%%%%%%%%%%%%%%%%%%%%%%%%%%%%%%%%
%This table has Linear fits for column III, rms around the line, intrinsic scatter around the line, and Spearman-Rank coefficient (rho).

\begin{deluxetable}{lrrrrrr}
 \tabletypesize{\scriptsize}
  \tablecaption{$I-\sigma$ linear fits and statistics for index group II and Balmer lines \label{fitexytabl3}}
    \tablewidth{0pt}
  \tablehead{
    \colhead{Index}
&\colhead{N}
&\colhead{Intercept}
&\colhead{Slope}
&\colhead{$\rho$} 
&\colhead{$\sigma_{Int}$}
&\colhead{$\sigma_{Err}$}
}
  \startdata
%  C$_2$4668       & 110 & $-3.890 \pm 0.415$ &$ 4.782 \pm 0.197$ &  0.750&  0.972 & 0.524  \\
%  Mg$_1$          & 107 & $-0.177 \pm 0.010$ &$ 0.128 \pm 0.004$ &  0.844&  0.016 & 0.011  \\
%  Mg$_2$          & 107 & $-0.186 \pm 0.012$ &$ 0.197 \pm 0.006$ &  0.871&  0.023 & 0.014  \\
%  Mg$b$           & 107 & $-1.157 \pm 0.134$ &$ 2.435 \pm 0.062$ &  0.854&  0.359 & 0.167  \\
%$[$MgFe$]\arcmin$ & 107 & $ 0.508 \pm 0.081$ &$ 1.290 \pm 0.040$ &  0.729&  0.372 & 0.112  \\
%H$\gamma_A$       & 111 & $ 3.503 \pm 0.229$ &$-4.160 \pm 0.109$ & -0.665&  1.357 & 0.565  \\
%H$\gamma_F$       & 111 & $ 3.563 \pm 0.126$ &$-2.286 \pm 0.059$ & -0.709&  0.684 & 0.297  \\
%H$\beta$          & 110 & $ 3.706 \pm 0.100$ &$-0.943 \pm 0.047$ & -0.502&  0.380 & 0.190  \\
%
  C$_2$4668       & 110 & $-4.249 \pm  0.877$ &$ 5.026 \pm 0.425 $ &   0.756 & 0.801 &   0.559   \\
  Mg$_1$          & 107 & $-0.171 \pm  0.015$ &$ 0.126 \pm 0.007 $ &   0.846 & 0.012 &   0.011   \\
  Mg$_2$          & 107 & $-0.169 \pm  0.021$ &$ 0.190 \pm 0.010 $ &   0.871 & 0.018 &   0.014   \\
  Mg$b$           & 107 & $-1.316 \pm  0.334$ &$ 2.513 \pm 0.163 $ &   0.855 & 0.308 &   0.189   \\
$[$MgFe$]\arcmin$ & 107 & $ 0.359 \pm  0.246$ &$ 1.353 \pm 0.120 $ &   0.730 & 0.351 &   0.128   \\
H$\gamma_A$       & 111 & $ 3.518 \pm  0.969$ &$-4.130 \pm 0.469 $ &  -0.663 & 1.233 &   0.592   \\
H$\gamma_F$       & 111 & $ 3.558 \pm  0.488$ &$-2.263 \pm 0.236 $ &  -0.703 & 0.612 &   0.316   \\
H$\beta$          & 110 & $ 3.271 \pm  0.266$ &$-0.714 \pm 0.129 $ &  -0.488 & 0.314 &   0.214   \\
  \enddata
  \tablecomments{Statistics for group II of $I-\sigma$ plots and
    Balmer lines.  N: number of galaxies; Intercept and slope of the
    linear fit which takes into account both the uncertainties in the
    index and velocity dispersion; $\rho$: Spearman-Rank correlation
    coefficient; $\sigma_{Int}$: intrinsic scatter; $\sigma_{Err}$:
    scatter due to the errors.  }
\end{deluxetable}
\begin{deluxetable}{lrrrrrr}
%\tablecolumns{6}
\tabletypesize{\scriptsize}
\tablecaption{Velocity Dispersions, Ages, Metallicities and $[\alpha$/Fe$]$ for Binned Galaxies \label{tablBins}}
\tablewidth{0pt}  %MUST HAVE THIS HERE ALWAYS!!! ALSO WATCH THE ORDER OF THINGS!!!
\tablehead{
\colhead{Bin} &
\colhead{N} &
\colhead{$\log \sigma$ Range} &
\colhead{$\langle \log \sigma \rangle$}&
\colhead{$\langle \log $Age $\rangle$} &
\colhead{$\langle[$Z/H$] \rangle$} &
\colhead{$\langle[\alpha$/Fe$]\rangle$}
}
\startdata
%1 &$2.291 \leqslant \sigma < 2.127$ &$2.212 \pm 0.045$ &$8.3 \pm 3.8$ &$ 0.156 \pm 0.155$ &$0.247 \pm 0.078$ \\
%2 &$2.127 \leqslant \sigma < 1.963$ &$2.053 \pm 0.055$ &$8.8 \pm 3.9$ &$ 0.060 \pm 0.093$ &$0.190 \pm 0.080$ \\
%3 &$1.963 \leqslant \sigma < 1.799$ &$1.890 \pm 0.046$ &$7.9 \pm 3.9$ &$-0.073 \pm 0.220$ &$0.147 \pm 0.137$ \\
%4 &$1.799 \leqslant \sigma < 1.635$ &$1.744 \pm 0.030$ &$5.6 \pm 3.1$ &$-0.092 \pm 0.148$ &$0.110 \pm 0.116$ \\
%5 &$1.635 \leqslant \sigma < 1.400$ &$1.582 \pm 0.065$ &$5.7 \pm 2.6$ &$-0.206 \pm 0.366$ &$0.113 \pm 0.173$ \\
%%
%This is from: /scratch/matkovic/dEs_gEs/lick_index/indtmk/myplots/make_age_error_avgSPM.pro
%#Name  N  Avg_Age err_avg  AvgMet err_avg  AvgAlf err_avg  AvgSig errSig
%# 0    1    2       3       4       5       6       7       8       9
%#------------------------------------------------------------------------
%1&  5& $2.291 \leqslant \sigma < 2.127$& $0.893 \pm 0.092$& $ 0.150 \pm 0.069$& $0.255 \pm 0.035$& $2.212 \pm 0.020$ \\
%2& 12& $2.127 \leqslant \sigma < 1.963$& $0.880 \pm 0.064$& $ 0.086 \pm 0.027$& $0.202 \pm 0.023$& $2.053 \pm 0.016$ \\
%3&  8& $1.963 \leqslant \sigma < 1.799$& $0.765 \pm 0.102$& $-0.041 \pm 0.078$& $0.179 \pm 0.048$& $1.890 \pm 0.016$ \\
%4& 13& $1.799 \leqslant \sigma < 1.635$& $0.721 \pm 0.066$& $-0.085 \pm 0.043$& $0.162 \pm 0.037$& $1.744 \pm 0.008$ \\
%5&  4& $1.635 \leqslant \sigma < 1.400$& $0.778 \pm 0.094$& $-0.111 \pm 0.147$& $0.143 \pm 0.031$& $1.610 \pm 0.010$ \\
%#------------------------------------------------------------------------
1& 17& $2.291 \leqslant \sigma < 2.127$& $0.905 \pm 0.055$& $ 0.150 \pm  0.039$& $ 0.255 \pm 0.020$& $2.212 \pm 0.011$ \\
2& 12& $2.127 \leqslant \sigma < 1.963$& $0.902 \pm 0.061$& $ 0.078 \pm  0.028$& $ 0.199 \pm 0.024$& $2.053 \pm 0.016$ \\
3&  8& $1.963 \leqslant \sigma < 1.799$& $0.776 \pm 0.102$& $-0.042 \pm  0.079$& $ 0.136 \pm 0.025$& $1.890 \pm 0.016$ \\
4& 13& $1.799 \leqslant \sigma < 1.635$& $0.735 \pm 0.066$& $-0.095 \pm  0.039$& $ 0.110 \pm 0.021$& $1.744 \pm 0.008$ \\
5&  4& $1.635 \leqslant \sigma < 1.400$& $0.793 \pm 0.100$& $-0.118 \pm  0.152$& $ 0.144 \pm 0.031$& $1.610 \pm 0.010$ \\
\enddata
\end{deluxetable}
%\clearpage

%%%%%%%%%%%%%%%%%%%%%%%%%%%%%%%%%%%%%%%%%%%%%%%%%%%%%%%%%%%%%%%%%%%%%%%%%%%%%%
%  Age-sigma, Met-sigma, alpha-sigma
%%%%%%%%%%%%%%%%%%%%%%%%%%%%%%%%%%%%%%%%%%%%%%%%%%%%%%%%%%%%%%%%%%%%%%%%%%%%%
%this table is at /scratch/matkovic/dEs_gEs/lick_index/indtmk/myplots/age_met_alph.pro/tex
\begin{deluxetable}{lrr}
  \tabletypesize{\scriptsize}
  \tablecaption{Model Parameters vs. $\sigma$ Relations \label{tablage_met_sig}}
  \tablewidth{0pt}
  \tablehead{
    \colhead{ } &
\colhead{Intercept} &
\colhead{Slope}
  }
\startdata
%---------------------------------------------------------------------------
%  $\log$ Age vs. $\sigma$ & $  0.25 \pm 0.78 $ & $ 0.32 \pm 0.41$ \\
%   $[$Z/H$]$ vs. $\sigma$ & $ -1.06 \pm 0.77 $ & $ 0.55 \pm 0.39$ \\
%$[\alpha$/H$]$ vs. $\sigma$& $ -0.32 \pm 0.46 $ & $ 0.25 \pm 0.23$ \\
%-----------------------------------------------------------------------rho-
%    $\log$ Age vs. $\sigma$ & $  0.22 \pm 0.34$ & $0.31 \pm 0.18$ \\ %& 0.33 \\
%     $[$Z/H$]$ vs. $\sigma$ & $ -1.00 \pm 0.26$ & $0.53 \pm 0.13$ \\ %& 0.56 \\
%$[\alpha$/H$]$ vs. $\sigma$ & $ -0.13 \pm 0.13$ & $0.17 \pm 0.07$ \\ %& 0.55 \\
%-------------------------------------------------------------------------------
% This is when error in each bin is error in the mean value:
     log Age vs. $\sigma$   &  $ 0.24 \pm 0.29$ & $0.31 \pm 0.15$ \\ 
   $[$Z/H$]$ vs. $\sigma$   &  $-1.01 \pm 0.22$ & $0.53 \pm 0.11$ \\  
$[\alpha$/H$]$ vs. $\sigma$ &  $-0.30 \pm 0.10$ & $0.25 \pm 0.05$ \\  
% This is when the error in each bin is stdev:
%     log Age vs. $\sigma$ &  $   0.35 \pm   0.84 $ & $   0.25 \pm   0.44$ \\ %& 0.31 \\ 
%   $[$Z/H$]$ vs. $\sigma$ &  $  -0.98 \pm   0.71 $ & $   0.51 \pm   0.36$ \\ %& 0.58 \\ 
%[$\alpha$/H] vs. $\sigma$ &  $  -0.18 \pm   0.26 $ & $   0.18 \pm   0.14$ \\ %& 0.56 \\ 
\enddata
\end{deluxetable}
%\clearpage
%\clearpage

%%%%%%%%%%%%%%%%%%%%%%%%%%%%%%%%%%%%%%%%%%%%%%%%%%%%%%%%%%%%%%%%%%%%%%%%%%%%%%
%This table gives relations between age & Z, age & alpha, Z & alpha
% rho is the spearman rank coefficient

\begin{deluxetable}{lrrr}
  \tabletypesize{\scriptsize}
  \tablecaption{Relations Between Age and Metallicity \label{tablage_met}}
  \tablewidth{0pt}
  \tablehead{
    \colhead{$[$Z/H$]=a+b\cdot\log$ Age} & \colhead{Intercept} & \colhead{Slope} & \colhead{$\rho$}
  }
  \startdata
%  High-$\sigma$ & $ 0.795 \pm  0.143 $ & $ -0.723 \pm  0.128  $ &  -0.750  \\
%  Low-$\sigma$ & $ 0.482 \pm  0.105 $ & $ -0.744 \pm  0.119  $ &  -0.594  \\
%unbroadened errors:
High-$\sigma$ & $ 0.860 \pm 0.165 $ & $-0.782 \pm 0.148 $ & -0.803  \\
 Low-$\sigma$ & $ 0.640 \pm 0.177 $ & $-0.886 \pm 0.201 $ & -0.527  \\
  \enddata
\end{deluxetable}

\clearpage
%%%%%%%%%%%%%%%%%%%%%%%%%%%%%%%%%%%%%%%%%%%%%%%%%%%%%%%%%%%%%%%%%%
%              FIGURES                                           %
%%%%%%%%%%%%%%%%%%%%%%%%%%%%%%%%%%%%%%%%%%%%%%%%%%%%%%%%%%%%%%%%%%
% Comparison Plots
%%%%%%%%%%%%%%%%%%%%%%%%%%%%%%%%%%%%%%%%%%%%%%%%%%%%%%%%%%%%%%%%%%
\clearpage

%%%%%%%%%%%%%%%%%%%%%%%%%%%%%%%%%%%%%%%%%%%%%%%%%%%%%%%%%%%%%%%%%%
\clearpage

%%%%%%%%%%%%%%%%%%%%%%%%%%%%%%%%%%%%%%%%%%%%%%%%%%%%%%%%%%%%%%%%%%
\clearpage

%%%%%%%%%%%%%%%%%%%%%%%%%%%%%%%%%%%%%%%%%%%%%%%%%%%%%%%%%%%%%%%%%%
% I-sigma plot
%%%%%%%%%%%%%%%%%%%%%%%%%%%%%%%%%%%%%%%%%%%%%%%%%%%%%%%%%%%%%%%%%%
\clearpage
\clearpage

%%%%%%%%%%%%%%%%%%%%%%%%%%%%%%%%%%%%%%%%%%%%%%%%%%%%%%%%%%%%%%%%%%%
%%   [MgFe]-HgA  AND HgF
%%%%%%%%%%%%%%%%%%%%%%%%%%%%%%%%%%%%%%%%%%%%%%%%%%%%%%%%%%%%%%%%%%%
\clearpage

%%%%%%%%%%%%%%%%%%%%%%%%%%%%%%%%%%%%%%%%%%%%%%%%%%%%%%%%%%%%%%%%%%
% Classification for [MgFe]-Hb plot:

%\begin{figure}
%%%%%\includegraphics[width=3.5in,angle=-90]{MgFe_Hbeta_class.ps}

%  \centerline{\psfig{figure=MgFe_Hbeta_class.ps,width=5truein,angle=-90}}
%  \figcaption{[MgbFe]$\arcmin$ vs. H$\beta$ plot with classifications.
%    The solid circles are galaxies with B/T=0, open circles $0.5< B/T
%    <1.0$, stars have B/T$<0.5$, and asterisks are galaxies for which
%    we do not have classification. {\it Not sure if we should include
%      this figure...}
%    \label{classfig}}
%\end{figure}
%\clearpage
%%%%%%%%%%%%%%%%%%%%%%%%%%%%%%%%%%%%%%%%%%%%%%%%%%%%%%%%%%%%%%%%%%
% The off-grid galaxy section:
%%%%%%%%%%%%%%%%%%%%%%%%%%%%%%%%%%%%%%%%%%%%%%%%%%%%%%%%%%%%%%%%%%
\clearpage

%%%%%%%%%%%%%%%%%%%%%%%%%%%%%%%%%%%%%%%%%%%%%%%%%%%%%%%%%%%%%%%%%%
% /scratch/matkovic/dEs_gEs/lick_index/nfps/Hbeta/outlying_GCs_paper.pro
\clearpage

%%%%%%%%%%%%%%%%%%%%%%%%%%%%%%%%%%%%%%%%%%%%%%%%%%%%%%%%%%%%%%%%%%
%  Age-sigma, Met-sigma, a/fe-sigma
%%%%%%%%%%%%%%%%%%%%%%%%%%%%%%%%%%%%%%%%%%%%%%%%%%%%%%%%%%%%%%%%%%
\clearpage

%%%%%%%%%%%%%%%%%%%%%%%%%%%%%%%%%%%%%%%%%%%%%%%%%%%%%%%%%%%%%%%%%%
%  Age-Z, a/Fe-Age, Z-a/Fe
%%%%%%%%%%%%%%%%%%%%%%%%%%%%%%%%%%%%%%%%%%%%%%%%%%%%%%%%%%%%%%%%%%


\begin{thebibliography}{}

\bibitem[\protect\citeauthoryear{{Bender}, {Burstein}, \& {Faber}}{{Bender}
  et~al.}{1993}]{Bender93}
{Bender}, R., {Burstein}, D.,  \& {Faber}, S.~M. 1993, \apj, 411,
153

\bibitem[\protect\citeauthoryear{{Bernardi} et~al.}{{Bernardi}
  et~al.}{2006}]{Bernardi06}
{Bernardi}, M., {Nichol}, R.~C., {Sheth}, R.~K., {Miller}, C.~J.,
\&
  {Brinkmann}, J. 2006, \aj, 131, 1288

\bibitem[\protect\citeauthoryear{{Bernardi} et~al.}{{Bernardi}
  et~al.}{1998}]{Bernardi98}
{Bernardi}, M., {Renzini}, A., {da Costa}, L.~N., {Wegner}, G.,
{Alonso},
  M.~V., {Pellegrini}, P.~S., {Rit{\'e}}, C.,  \& {Willmer}, C.~N.~A. 1998,
  \apjl, 508, L143

\bibitem[\protect\citeauthoryear{{Brodie} \& {Huchra}}{{Brodie} \&
  {Huchra}}{1991}]{Brodie91}
{Brodie}, J.~P.,  \& {Huchra}, J.~P. 1991, \apj, 379, 157

\bibitem[\protect\citeauthoryear{{Bruzual} \& {Charlot}}{{Bruzual} \&
  {Charlot}}{2003}]{Bruzual03}
{Bruzual}, G.,  \& {Charlot}, S. 2003, \mnras, 344, 1000

\bibitem[\protect\citeauthoryear{{Burstein} et~al.}{{Burstein}
  et~al.}{1984}]{Burstein84}
{Burstein}, D., {Faber}, S.~M., {Gaskell}, C.~M.,  \& {Krumm}, N.
1984, \apj,
  287, 586

\bibitem[\protect\citeauthoryear{{Caldwell}}{{Caldwell}}{1983}]{Caldwell83b}
{Caldwell}, N. 1983, \aj, 88, 804

\bibitem[\protect\citeauthoryear{{Caldwell}, {Rose}, \& {Concannon}}{{Caldwell}
  et~al.}{2003}]{Caldwell03}
{Caldwell}, N., {Rose}, J.~A.,  \& {Concannon}, K.~D. 2003, \aj,
125, 2891

\bibitem[\protect\citeauthoryear{{Caldwell} et~al.}{{Caldwell}
  et~al.}{1993}]{Caldwell93}
{Caldwell}, N., {Rose}, J.~A., {Sharples}, R.~M., {Ellis}, R.~S.,
\& {Bower},
  R.~G. 1993, \aj, 106, 473

\bibitem[\protect\citeauthoryear{{Cardiel}}{{Cardiel}}{1999}]{Cardiel99}
{Cardiel}, N. 1999, Ph.D. thesis, Universidad Complutense de
Madrid, Spain,
  (1999)

\bibitem[\protect\citeauthoryear{{Cenarro} et~al.}{{Cenarro}
  et~al.}{2007}]{Cenarro07}
{Cenarro}, A.~J., {Beasley}, M.~A., {Strader}, J., {Brodie},
J.~P.,  \&
  {Forbes}, D.~A. 2007, \aj, 134, 391

\bibitem[\protect\citeauthoryear{{Chiosi} \& {Carraro}}{{Chiosi} \&
  {Carraro}}{2002}]{Chiosi02}
{Chiosi}, C.,  \& {Carraro}, G. 2002, \mnras, 335, 335

\bibitem[\protect\citeauthoryear{{Colless} et~al.}{{Colless}
  et~al.}{1999}]{Colless99}
{Colless}, M., {Burstein}, D., {Davies}, R.~L., {McMahan}, R.~K.,
{Saglia},
  R.~P.,  \& {Wegner}, G. 1999, \mnras, 303, 813

\bibitem[\protect\citeauthoryear{{Colless} \& {Dunn}}{{Colless} \&
  {Dunn}}{1996}]{Colless96}
{Colless}, M.,  \& {Dunn}, A.~M. 1996, \apj, 458, 435

\bibitem[\protect\citeauthoryear{{Concannon}, {Rose}, \&
  {Caldwell}}{{Concannon} et~al.}{2000}]{Concannon00}
{Concannon}, K.~D., {Rose}, J.~A.,  \& {Caldwell}, N. 2000, \apjl,
536, L19

\bibitem[\protect\citeauthoryear{{Denicol{\'o}} et~al.}{{Denicol{\'o}}
  et~al.}{2005}]{Denicolo05b}
{Denicol{\'o}}, G., {Terlevich}, R., {Terlevich}, E., {Forbes},
D.~A.,  \&
  {Terlevich}, A. 2005, \mnras, 358, 813

\bibitem[\protect\citeauthoryear{{Dressler}}{{Dressler}}{1984}]{Dressler84}
{Dressler}, A. 1984, \apj, 281, 512

\bibitem[\protect\citeauthoryear{{Ferguson} \& {Binggeli}}{{Ferguson} \&
  {Binggeli}}{1994}]{Ferguson94}
{Ferguson}, H.~C.,  \& {Binggeli}, B. 1994, \aapr, 6, 67

\bibitem[\protect\citeauthoryear{{Ferreras}, {Charlot}, \& {Silk}}{{Ferreras}
  et~al.}{1999}]{Ferreras99}
{Ferreras}, I., {Charlot}, S.,  \& {Silk}, J. 1999, \apj, 521, 81

\bibitem[\protect\citeauthoryear{{Fisher}, {Franx}, \& {Illingworth}}{{Fisher}
  et~al.}{1995}]{Fisher95}
{Fisher}, D., {Franx}, M.,  \& {Illingworth}, G. 1995, \apj, 448,
119

\bibitem[\protect\citeauthoryear{{Forbes}, {Ponman}, \& {Brown}}{{Forbes}
  et~al.}{1998}]{Forbes98}
{Forbes}, D.~A., {Ponman}, T.~J.,  \& {Brown}, R.~J.~N. 1998,
\apjl, 508, L43

\bibitem[\protect\citeauthoryear{{Gonz\'alez}}{{Gonz\'alez}}{1993}]{Gonzalez93}
{Gonz\'alez}, J.~D.~J. 1993, Ph.D. thesis, AA(Univ.~California,
Santa Cruz.)

\bibitem[\protect\citeauthoryear{{Gorgas}, {Efstathiou}, \& {Aragon
  Salamanca}}{{Gorgas} et~al.}{1990}]{Gorgas90}
{Gorgas}, J., {Efstathiou}, G.,  \& {Aragon Salamanca}, A. 1990,
\mnras, 245,
  217

\bibitem[\protect\citeauthoryear{{Gorgas} et~al.}{{Gorgas}
  et~al.}{1993}]{Gorgas93}
{Gorgas}, J., {Faber}, S.~M., {Burstein}, D., {Gonzalez}, J.~J.,
{Courteau},
  S.,  \& {Prosser}, C. 1993, \apjs, 86, 153

\bibitem[\protect\citeauthoryear{{Gorgas} et~al.}{{Gorgas}
  et~al.}{1997}]{Gorgas97}
{Gorgas}, J., {Pedraz}, S., {Guzman}, R., {Cardiel}, N.,  \&
{Gonzalez}, J.~J.
  1997, \apjl, 481, L19

\bibitem[\protect\citeauthoryear{{Graham} \& {Guzm{\'a}n}}{{Graham} \&
  {Guzm{\'a}n}}{2003}]{GG03}
{Graham}, A.~W.,  \& {Guzm{\'a}n}, R. 2003, \aj, 125, 2936

\bibitem[\protect\citeauthoryear{{Greggio}}{{Greggio}}{1997}]{Greggio97}
{Greggio}, L. 1997, \mnras, 285, 151

\bibitem[\protect\citeauthoryear{{Guzman} et~al.}{{Guzman}
  et~al.}{1992}]{Guzman92}
{Guzman}, R., {Lucey}, J.~R., {Carter}, D.,  \& {Terlevich}, R.~J.
1992,
  \mnras, 257, 187

\bibitem[\protect\citeauthoryear{{Hammer} et~al.}{{Hammer}
  et~al.}{2001}]{Hammer01}
{Hammer}, F., {Gruel}, N., {Thuan}, T.~X., {Flores}, H.,  \&
{Infante}, L.
  2001, \apj, 550, 570

\bibitem[\protect\citeauthoryear{{Jorgensen}}{{Jorgensen}}{1997}]{Jorgensen97}
{Jorgensen}, I. 1997, \mnras, 288, 161

\bibitem[\protect\citeauthoryear{{J{\o}rgensen}}{{J{\o}rgensen}}{1999}]{Jorgen%
sen99} {J{\o}rgensen}, I. 1999, \mnras, 306, 607

\bibitem[\protect\citeauthoryear{{Jorgensen}, {Franx}, \&
  {Kjaergaard}}{{Jorgensen} et~al.}{1996}]{Jorgensen96}
{Jorgensen}, I., {Franx}, M.,  \& {Kjaergaard}, P. 1996, \mnras,
280, 167

\bibitem[\protect\citeauthoryear{{Kawata}}{{Kawata}}{2001}]{Kawata01}
{Kawata}, D. 2001, \apj, 558, 598

\bibitem[\protect\citeauthoryear{{Kelson} et~al.}{{Kelson}
  et~al.}{2006}]{Kelson06}
{Kelson}, D.~D., {Illingworth}, G.~D., {Franx}, M.,  \& {van
Dokkum}, P.~G.
  2006, \apj, 653, 159

\bibitem[\protect\citeauthoryear{{Korn}, {Maraston}, \& {Thomas}}{{Korn}
  et~al.}{2005}]{Korn05}
{Korn}, A.~J., {Maraston}, C.,  \& {Thomas}, D. 2005, \aap, 438,
685

\bibitem[\protect\citeauthoryear{{Kuntschner}}{{Kuntschner}}{2000}]{Kuntschner%
00} {Kuntschner}, H. 2000, \mnras, 315, 184

\bibitem[\protect\citeauthoryear{{Kuntschner} \& {Davies}}{{Kuntschner} \&
  {Davies}}{1998}]{Kuntschner98}
{Kuntschner}, H.,  \& {Davies}, R.~L. 1998, \mnras, 295, L29

\bibitem[\protect\citeauthoryear{{Kuntschner} et~al.}{{Kuntschner}
  et~al.}{2001}]{Kuntschner01}
{Kuntschner}, H., {Lucey}, J.~R., {Smith}, R.~J., {Hudson}, M.~J.,
\&
  {Davies}, R.~L. 2001, \mnras, 323, 615

\bibitem[\protect\citeauthoryear{{Kuntschner} et~al.}{{Kuntschner}
  et~al.}{2002}]{Kuntschner02}
{Kuntschner}, H., {Smith}, R.~J., {Colless}, M., {Davies}, R.~L.,
{Kaldare},
  R.,  \& {Vazdekis}, A. 2002, \mnras, 337, 172

\bibitem[\protect\citeauthoryear{{Maraston}}{{Maraston}}{1998}]{Maraston98}
{Maraston}, C. 1998, \mnras, 300, 872

\bibitem[\protect\citeauthoryear{{Matkovi{\'c}} \& {Guzm{\'a}n}}{{Matkovi{\'c}}
  \& {Guzm{\'a}n}}{2005}]{Matkovic05}
{Matkovi{\'c}}, A.,  \& {Guzm{\'a}n}, R. 2005, \mnras, 362, 289

\bibitem[\protect\citeauthoryear{{Matteucci}}{{Matteucci}}{1994}]{Matteucci94}
{Matteucci}, F. 1994, \aap, 288, 57

\bibitem[\protect\citeauthoryear{{Mehlert} et~al.}{{Mehlert}
  et~al.}{2003}]{Mehlert03}
{Mehlert}, D., {Thomas}, D., {Saglia}, R.~P., {Bender}, R.,  \&
{Wegner}, G.
  2003, \aap, 407, 423

\bibitem[\protect\citeauthoryear{{Michielsen} et~al.}{{Michielsen}
  et~al.}{2007}]{Michielsen07}
{Michielsen}, D., {Boselli}, A., {Conselice}, C.~J., {Toloba}, E.,
{Whiley},
  I.~M., {Aragon-Salamanca}, A., {Balcells}, M., {Cardiel}, N., {Cenarro},
  A.~J., {Gorgas}, J., {Peletier}, R.~F.,  \& {Vazdekis}, A. 2007, ArXiv
  e-prints, 712

\bibitem[\protect\citeauthoryear{{Nelan} et~al.}{{Nelan}
  et~al.}{2005}]{Nelan05}
{Nelan}, J.~E., {Smith}, R.~J., {Hudson}, M.~J., {Wegner}, G.~A.,
{Lucey},
  J.~R., {Moore}, S.~A.~W., {Quinney}, S.~J.,  \& {Suntzeff}, N.~B. 2005, \apj,
  632, 137

\bibitem[\protect\citeauthoryear{{Pedraz} et~al.}{{Pedraz}
  et~al.}{1998}]{Pedraz98}
{Pedraz}, S, {Gorgas}, J, {Cardiel}, N., \& {Guzm\'an}, R. 1998, \apss,
  632, 137

\bibitem[\protect\citeauthoryear{{Poggianti} et~al.}{{Poggianti}
  et~al.}{2001a}]{Poggianti01}
{Poggianti}, B.~M., {Bridges}, T.~J., {Carter}, D., {Mobasher},
B., {Doi}, M.,
  {Iye}, M., {Kashikawa}, N., {Komiyama}, Y., {Okamura}, S., {Sekiguchi}, M.,
  {Shimasaku}, K., {Yagi}, M.,  \& {Yasuda}, N. 2001a, \apj, 563, 118

\bibitem[\protect\citeauthoryear{{Poggianti} et~al.}{{Poggianti}
  et~al.}{2001b}]{Poggianti01a}
{Poggianti}, B.~M., {Bridges}, T.~J., {Carter}, D., {Mobasher},
B., {Doi}, M.,
  {Iye}, M., {Kashikawa}, N., {Komiyama}, Y., {Okamura}, S., {Sekiguchi}, M.,
  {Shimasaku}, K., {Yagi}, M.,  \& {Yasuda}, N. 2001b, \apj, 563, 118

\bibitem[\protect\citeauthoryear{{Proctor} et~al.}{{Proctor}
  et~al.}{2004}]{Proctor04a}
{Proctor}, R.~N., {Forbes}, D.~A., {Hau}, G.~K.~T., {Beasley},
M.~A., {De
  Silva}, G.~M., {Contreras}, R.,  \& {Terlevich}, A.~I. 2004, \mnras, 349,
  1381

\bibitem[\protect\citeauthoryear{{Proctor} \& {Sansom}}{{Proctor} \&
  {Sansom}}{2002}]{Proctor02}
{Proctor}, R.~N.,  \& {Sansom}, A.~E. 2002, \mnras, 333, 517

\bibitem[\protect\citeauthoryear{{Rakos} \& {Schombert}}{{Rakos} \&
  {Schombert}}{2004}]{Rakos04}
{Rakos}, K.,  \& {Schombert}, J. 2004, \aj, 127, 1502

\bibitem[\protect\citeauthoryear{{S{\'a}nchez-Bl{\'a}zquez}
  et~al.}{{S{\'a}nchez-Bl{\'a}zquez} et~al.}{2006a}]{Sanchez06a}
{S{\'a}nchez-Bl{\'a}zquez}, P., {Gorgas}, J., {Cardiel}, N.,  \&
  {Gonz{\'a}lez}, J.~J. 2006a, \aap, 457, 787

\bibitem[\protect\citeauthoryear{{S{\'a}nchez-Bl{\'a}zquez}
  et~al.}{{S{\'a}nchez-Bl{\'a}zquez} et~al.}{2006b}]{Sanchez06b}
{S{\'a}nchez-Bl{\'a}zquez}, P., {Gorgas}, J., {Cardiel}, N.,  \&
  {Gonz{\'a}lez}, J.~J. 2006b, \aap, 457, 809

\bibitem[\protect\citeauthoryear{{Smith} et~al.}{{Smith}
et~al.}{2008}]{Smith08} {Smith}, R.~J., {Marzke}, R.~O.,
{Hornschemeier}, A.~E., {Bridges}, T.~J., {Hudson}, M.~J.,{Miller},
N.~A., {Lucey}, J.~R., {V{\'a}zquez}, G.~A.,{Carter}, D., 2008,
\mnras,386, L96

\bibitem[\protect\citeauthoryear{{Terlevich} \& {Forbes}}{{Terlevich} \&
  {Forbes}}{2002}]{Terlevich02}
{Terlevich}, A.~I.,  \& {Forbes}, D.~A. 2002, \mnras, 330, 547

\bibitem[\protect\citeauthoryear{{Terlevich} et~al.}{{Terlevich}
  et~al.}{1999}]{Terlevich99}
{Terlevich}, A.~I., {Kuntschner}, H., {Bower}, R.~G., {Caldwell},
N.,  \&
  {Sharples}, R.~M. 1999, \mnras, 310, 445

\bibitem[\protect\citeauthoryear{{Terlevich} et~al.}{{Terlevich}
  et~al.}{1981}]{Terlevich81}
{Terlevich}, R., {Davies}, R.~L., {Faber}, S.~M.,  \& {Burstein},
D. 1981,
  \mnras, 196, 381

\bibitem[\protect\citeauthoryear{{Thomas}, {Maraston}, \& {Bender}}{{Thomas}
  et~al.}{2002}]{TMB02}
{Thomas}, D., {Maraston}, C.,  \& {Bender}, R. 2002, \apss, 281,
371

\bibitem[\protect\citeauthoryear{{Thomas}, {Maraston}, \& {Bender}}{{Thomas}
  et~al.}{2003}]{TMB03}
{Thomas}, D., {Maraston}, C.,  \& {Bender}, R. 2003, \mnras, 343,
279

\bibitem[\protect\citeauthoryear{{Thomas} et~al.}{{Thomas}
  et~al.}{2005}]{Thomas05}
{Thomas}, D., {Maraston}, C., {Bender}, R.,  \& {Mendes de
Oliveira}, C. 2005,
  \apj, 621, 673

\bibitem[\protect\citeauthoryear{{Thomas}, {Maraston}, \& {Korn}}{{Thomas}
  et~al.}{2004}]{TMK04}
{Thomas}, D., {Maraston}, C.,  \& {Korn}, A. 2004, \mnras, 351,
L19

\bibitem[\protect\citeauthoryear{{Trager} et~al.}{{Trager}
  et~al.}{2000a}]{Trager00b}
{Trager}, S.~C., {Faber}, S.~M., {Worthey}, G.,  \&
{Gonz{\'a}lez}, J.~J.
  2000a, \aj, 120, 165

\bibitem[\protect\citeauthoryear{{Trager} et~al.}{{Trager}
  et~al.}{2000b}]{Trager00a}
{Trager}, S.~C., {Faber}, S.~M., {Worthey}, G.,  \&
{Gonz{\'a}lez}, J.~J.
  2000b, \aj, 119, 1645

\bibitem[\protect\citeauthoryear{{Trager} et~al.}{{Trager}
  et~al.}{1998}]{Trager98}
{Trager}, S.~C., {Worthey}, G., {Faber}, S.~M., {Burstein}, D.,
\& {Gonzalez},
  J.~J. 1998, \apjs, 116, 1

\bibitem[\protect\citeauthoryear{{Tripicco} \& {Bell}}{{Tripicco} \&
  {Bell}}{1995}]{Tripi95}
{Tripicco}, M.~J.,  \& {Bell}, R.~A. 1995, \aj, 110, 3035

\bibitem[\protect\citeauthoryear{{Worthey}}{{Worthey}}{1994}]{Worthey94b}
{Worthey}, G. 1994, \apjs, 95, 107

\bibitem[\protect\citeauthoryear{{Worthey} \& {Collobert}}{{Worthey} \&
  {Collobert}}{2003}]{Worthey03}
{Worthey}, G.,  \& {Collobert}, M. 2003, \apj, 586, 17

\bibitem[\protect\citeauthoryear{{Worthey}, {Faber}, \& {Gonzalez}}{{Worthey}
  et~al.}{1992}]{Worthey92}
{Worthey}, G., {Faber}, S.~M.,  \& {Gonzalez}, J.~J. 1992, \apj,
398, 69

\bibitem[\protect\citeauthoryear{{Worthey} et~al.}{{Worthey}
  et~al.}{1994}]{Worthey94a}
{Worthey}, G., {Faber}, S.~M., {Gonzalez}, J.~J.,  \& {Burstein},
D. 1994,
  \apjs, 94, 687

\bibitem[\protect\citeauthoryear{{Worthey} \& {Ottaviani}}{{Worthey} \&
  {Ottaviani}}{1997}]{Worthey97}
{Worthey}, G.,  \& {Ottaviani}, D.~L. 1997, \apjs, 111, 377

\end{thebibliography}
\end{document}